\documentclass[preprint2]{aastex}
\usepackage{hyperref}
\hypersetup{colorlinks=true,citecolor=blue}
\usepackage[usenames]{color}
\usepackage{natbib}
\newcommand{\kms}{\mbox{km s$^{-1}$}}

\slugcomment{Draft version June, 2017}

\shorttitle{Disentangling the Circumnuclear Environs of Centaurus A III. }
\shortauthors{D. Espada et al.}

\begin{document}

\title{Disentangling the Circumnuclear Environs of Centaurus A: \\
 III. An Inner Molecular Ring, Nuclear Shocks and the CO to warm H$_2$ interface}

\author{
D. Espada\altaffilmark{1,2,3}, 
S. Matsushita\altaffilmark{4},
R. E. Miura\altaffilmark{1},
F. P. Israel\altaffilmark{5},
N. Neumayer\altaffilmark{6},
S. Martin\altaffilmark{3,7}, 
C. Henkel\altaffilmark{8,9}, 
T. Izumi\altaffilmark{10},
D. Iono\altaffilmark{1,2},
S. Aalto\altaffilmark{11},
J. Ott\altaffilmark{12},
A. B. Peck\altaffilmark{13},
A. C. Quillen\altaffilmark{14}, and
K. Kohno\altaffilmark{10}
}
\altaffiltext{1}{National Astronomical Observatory of Japan, 2-21-1 Osawa, Mitaka, Tokyo 181-8588, Japan}
\altaffiltext{2}{The Graduate University for Advanced Studies (SOKENDAI), 2-21-1 Osawa, Mitaka, Tokyo, 181-0015, Japan}
\altaffiltext{3}{Joint ALMA Observatory, Alonso de C{\'o}rdova, 3107, Vitacura, Santiago 763-0355, Chile}
\altaffiltext{4}{Academia Sinica, Institute of Astronomy \& Astrophysics, P.O. Box 23-141, Taipei 10617, Taiwan} 
\altaffiltext{5}{Sterrewacht Leiden, Leiden University, PO Box 9513, 2300 RA, Leiden, The Netherlands}
\altaffiltext{6}{Max Planck Institute for Astronomy (MPIA), K{\"o}nigstuhl 17, D-69121 Heidelberg, Germany}
\altaffiltext{7}{European Southern Observatory, Alonso de C{\'o}rdova 3107, Vitacura, Santiago, Chile}
\altaffiltext{8}{Max-Planck-Institut f{\"u}r Radioastronomie, Auf dem H{\"u}gel 69, 53121, Bonn, Germany}
\altaffiltext{9}{Dept. of Astronomy, King Abdulaziz University, PO Box 80203, 21589 Jeddah, Saudi Arabia}
\altaffiltext{10}{Institute of Astronomy, School of Science, The University of Tokyo, 2-21-1 Osawa, Mitaka, Tokyo 181-0015, Japan}
\altaffiltext{11}{Department of Earth and Space Sciences, Chalmers University, Sweden}
\altaffiltext{12}{National Radio Astronomy Observatory, Socorro, NM, USA}
\altaffiltext{13}{Gemini Observatory, 670 N'Aohoku Pl, Hilo 96720-2700, Hawaii, HI, USA}
\altaffiltext{14}{Department of Physics and Astronomy, University of Rochester, Rochester, NY 14627, USA}

\begin{abstract}

We present the distribution and kinematics of the molecular gas in the circumnuclear disk (CND, 400 pc $\times$ 200 pc)  of Centaurus\,A  with resolutions of $\sim$5\,pc (0\farcs3) and shed light onto the mechanism feeding the Active Galactic Nucleus (AGN) using CO(3--2), HCO$^+$(4--3), HCN(4--3), and CO(6--5) observations obtained with ALMA. 
Multiple filaments or streamers of tens to a hundred parsec scale exist within the CND, which form a ring-like structure with an unprojected diameter of 9\arcsec $\times$ 6\arcsec (162\,pc $\times$ 108\,pc) and a position angle $PA$ $\simeq$ 155\arcdeg . 
 Inside the nuclear ring, there are two leading and straight filamentary structures with lengths of about 30--60\,pc at $PA$ $\simeq$ 120\arcdeg\ on opposite sides of the AGN, with a rotational symmetry of 180\arcdeg\ and steeper position-velocity diagrams, which are interpreted as nuclear shocks due to non-circular motions. Along the filaments, and unlike other nearby AGNs, several dense molecular clumps present low HCN/HCO$^+$(4--3) ratios ($\lesssim$0.5).
The filaments abruptly end in the probed transitions at $r$ $\simeq$ 20\,pc from the AGN, but previous near-IR H$_2$(J=1--0)S(1) maps show that they continue in an even warmer gas phase (T$\sim$1000\,K), winding up in the form of nuclear spirals, and forming an inner ring structure with another set of symmetric filaments along the N--S direction and within $r$ $\simeq$ 10\,pc. 
The molecular gas is governed primarily by non-circular motions, being the successive shock fronts at different scales where loss of angular momentum occurs, a mechanism which may feed efficiently powerful radio galaxies down to parsec scales.

\end{abstract}


\section{Introduction}
\label{introduction}

Active Galactic Nuclei (AGNs) are thought to be powered by accretion onto Super Massive Black Holes (SMBHs) and their luminosities require large mass accretion rates. However, a {crucial} outstanding problem is to identify the mechanisms that drive gas from external regions toward the nuclei of these galaxies, removing its angular momentum in order to trigger nuclear activity and feed the SMBHs that reside there. 
On the other hand, AGNs will have an effect on the surrounding molecular gas, from positive feedback such as gas compressed by jets or winds to negative feedback where outflows also drive gas away from the nuclear region \citep[for reviews see][]{2012ARA&A..50..455F,2015ARA&A..53..115K}.

In powerful radio-galaxies the properties of the circumnuclear gas from hundreds of parsec scale down to the accretion disk are this far poorly understood. Their luminosities, which may exceed 10$^{46}$\,erg\,s$^{-1}$, require mass accretion rates of $>$ 1 M$_\sun$\,yr$^{-1}$.
Radio galaxies are radio-loud active galaxies, usually of elliptical type. Their large scale synchrotron jets are presumably powered by the accretion of gas onto SMBHs, which are fueled by reservoirs of neutral and ionized gas in the host galaxy.  Some of these radio galaxies possess dust lanes containing large amounts of material comprising different gas phases of the interstellar medium (e.g. \citealt{2002ApJS..139..411A}).  
Powerful radio sources are rare in the local Universe, and thus the lack of high resolution observations have so far inhibited studies of the properties of the molecular gas in their nuclear regions.

\citet{1989Natur.338...45S} proposed that funneling gas to the centers of galaxies may occur from large to small scales by successive dynamical instabilities, also known as the \emph{bars within bars} mechanism, in which a primary bar may drive the gas inwards forming a circumnuclear disk which then becomes unstable again to form a decoupled nuclear bar.
Although many observational and numerical studies of spiral galaxies have been carried out \citep[e.g.][]{1990ApJ...363..391P,1993A&A...277...27F,1997A&AS..125..479J,2004ApJ...617L.115E}, a similar scenario may also be present in elliptical galaxies, where material accreted from external sources is not supported rotationally. This gas would go into Keplerian orbits close to the center and may form a gaseous bar which funnels gas to radii close to their SMBHs \citep{1989Natur.338...45S}.

Centaurus A (Cen\,A) is a radio source associated with the giant elliptical NGC\,5128 at a distance of only D $\simeq$ 3.8\,Mpc (\citealt{2010PASA...27..457H}, where 1\arcsec = 18\,pc). Although characterized by a more modest bolometric luminosity ($\sim$ 2 $\times$ 10$^{43}$\,erg\,s$^{-1}$, \citealt{1998A&ARv...8..237I}) and accretion rate ($\dot{M}_{Bondi}$ = 6.4 $\times$ 10$^{-4}$\,M$_\odot$\,yr$^{-1}$ and a Bondi efficiency of $\sim$ 0.23\%, \citealt{2004ApJ...612..786E})
than other powerful radio-galaxies, it is by far the nearest and possibly best studied one  \citep[for reviews, see ][]{1998A&ARv...8..237I,2010PASA...27..463M}. Because of its proximity it is the best target among the class of powerful radio galaxies for detailed studies of the feeding mechanisms of an active galactic nucleus (AGN) and feedback on the surrounding molecular gas. Centaurus A represents a particularly interesting case of an elliptical galaxy that was replenished recently (a few 10$^8$\,yr) by gas from an external source { \citep[e.g.][]{2010A&A...515A..67S}}. 

\begin{figure*}
\begin{center}
\includegraphics[width=14cm]{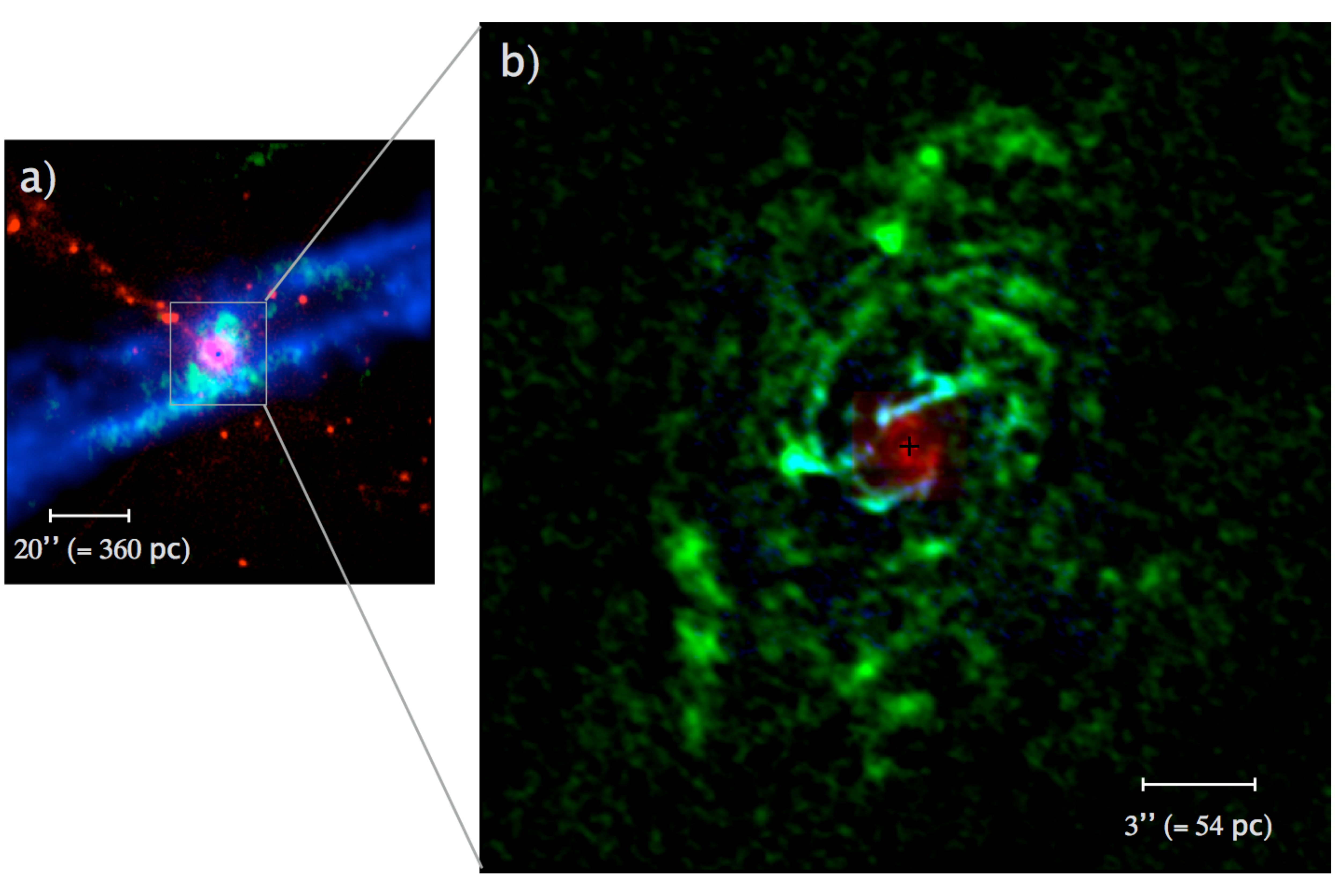}
\end{center}
\caption{Kiloparsec to parsec scale view of the molecular disk of Cen~A (NGC\,5128): {\bf a)} Integrated CO(2--1) emission map (green) observed using the Submillimeter Array (SMA, \citealt{2009ApJ...695..116E}). The (white) rectangle encompasses the molecular gas in the circumnuclear disk (CND, $r$ $<$ 200 pc) in the form of a disk/torus just perpendicular to the X-ray/radio jet (red, Chandra/ACIS-I, \citealt{2002ApJ...569...54K}), and represents the area covered by the CO(3--2) observations as shown in panel $b$. A more extended molecular gas component in form of spiral arms \citep{2012ApJ...756L..10E} is seen to be coextensive with a parallelogram structure previously observed in dust emission along $PA$ = 120$\arcdeg$ (blue, 8\,$\mu$m Spitzer/IRAC, \citealt{2006ApJ...645.1092Q}).  b) Composite image of the CND of Cen~A including the ALMA CO(3--2) (green) and CO(6--5) (blue) integrated intensity maps, as presented in this Paper, as well as VLT/SINFONI H$_2$ (1--0) S(1) integrated intensity map \citep[red,][]{2007ApJ...671.1329N}. The ALMA CO(3--2) and CO(6--5) maps cover a field of view of 24\arcsec\ and 12\arcsec , and have a resolution of $\sim$0\farcs3 (or 5\,pc) resolution.  The distribution of molecular hydrogen as traced by the H$_2$ line is mostly contained within a field of view of 3\arcsec\ (54\,pc). The cross sign in the center of the image shows the AGN position at $RA$\ = 13${\rm ^h}$25${\rm ^m}$27.${\rm^s}$615 ; $Dec$ =  --43${\rm ^o}$01$\arcmin$08\farcs805 \citep{1998AJ....116..516M}. 
\label{fig-1}}
\end{figure*}

Projected toward the inner several hundred parsec of Cen\,A, the following components have been identified (see Fig.\,1): 
 i) \emph{Molecular gas at large radii} ($>$ 1\,kpc) as traced by CO transitions \citep[e.g.][]{1987ApJ...322L..73P,1990ApJ...365..522E,1993A&A...270L..13R,2001A&A...371..865L,2009ApJ...695..116E} and corresponding to kpc scale spiral features \citep{2012ApJ...756L..10E}. This component is associated with the dust lane and is seen to be coextensive with other components of the interstellar medium, such as 
 H$\alpha$ (e.g. \citealt{1992ApJ...387..503N}), near infrarred \citep{1993ApJ...412..550Q}, submillimeter continuum \citep[e.g.][]{1993MNRAS.260..844H,2002ApJ...565..131L},
and mid-IR continuum emission \citep[e.g.][]{1999A&A...341..667M,2006ApJ...645.1092Q};
ii) \emph{A circumnuclear gaseous disk (CND) or torus} of $\sim$ 400\,pc extent ($\sim$ 24\arcsec) and a position angle $PA$ = $155\arcdeg$, perpendicular to the inner jet, at least in projection \citep{2009ApJ...695..116E} (see Fig.\,1a). The estimated total gas mass in this component is 8.4 $\times$ 10$^7$\,M$_\odot$ \citep{2014A&A...562A..96I}. The CO emission line is brightest at the edges of the disk at the NW and SE of the AGN, and \citet{2009ApJ...695..116E} reported a possible gap of CO(2--1) emission in the inner $r$\,$<$\,80\,pc (see also Fig.\,\ref{fig-1}a). ALMA science verification CO(2--1) observations, also with tens of parsec resolution showed similar results \citep{2013ASPC..476...69E};
iii) \emph{A nuclear disk} ($\sim$ 40\,pc { diameter}, or 2\arcsec) containing ionized and molecular gas presumably feeding the nuclear massive object \citep[e.g.][]{2001ApJ...549..915M,2006A&A...448..921M}.  While ionized gas shows an elongated distribution along the jet and is likely related to it, the molecular hydrogen as traced by H$_2$(J=1--0) S(1) seems to be mostly unperturbed within an irregular nuclear disk-like structure (see Fig.\,\ref{fig-1}b), and with an S-shaped velocity field \citep{2007ApJ...671.1329N};
and iv) \emph{Absorption lines} toward the continuum source detected in \ion{H}{1} (e.g. \citealt{1970ApJ...161L...9R,1983ApJ...264L..37V,2002ApJ...564..696S,2010ApJ...720..666E}) and molecular lines (e.g. \citealt{1990A&A...227..342I,1997A&A...324...51W,1999ApJ...516..769E,2010ApJ...720..666E}).

A {fundamental} question regarding this object, though, is how the gas in the nuclear disk at parsec scales is replenished by molecular gas in the { CND. Previous CO(2--1) observations} seemed to indicate that there was a lack of emission (Fig.~\ref{fig-1}, \citealt{2009ApJ...695..116E}), which suggested a lack of molecular gas { in the center} or a sudden change in the physical properties of the gas.
Historically observing the higher transitions of CO or other dense gas tracers has been very challenging technically, and the problem is compounded for Cen~A due to its location in the southern sky ($Dec$ = -42\arcdeg). ALMA finally provides the missing data, and in this Paper we present CO(3--2), HCO$^+$(4--3), HCN(4--3), and CO(6--5) emission line maps with 5\,pc resolution towards the CND.
These form an excellent suite of transition lines to probe the warm and dense molecular gas. The critical densities of CO(3--2), HCO$^+$(4--3), CO(6--5)  and  HCN(4--3) are 8.4 $\times$ 10$^3$ cm$^{-3}$, 1.0 $\times$ 10$^6$ cm$^{-3}$,  1.8 $\times$ 10$^6$ cm$^{-3}$, and  8.5 $\times$ 10$^6$ cm$^{-3}$, respectively.

The paper is organized as follows:  We introduce our ALMA observations and data reduction in \S~\ref{observationReduction}. In \S~\ref{result} we focus on the identification of the different components, answer whether there is a gap in the molecular gas, and study their major physical properties. We interpret these results in order to shed light on the following questions: 
i) what are the mechanisms that drive the gas from kiloparsec to parsec scales (in \S\,\ref{secfeeding}),
ii) what is the geometry between the CND and the jet (in \S\,\ref{secjet}),
iii) how do the molecular gas properties of the CND compare to those in numerical simulations of the multi-phase ISM around SMBHs (in \S\,\ref{secnumerical}), 
and iv) with our molecular line maps at hand, what are the chemical properties of the molecular gas close to the AGN (in \S~\ref{secratio}). 
Finally we summarize our results in \S~\ref{conclusion}.

\section{Observations and Data Reduction}
\label{observationReduction}

We present single pointing observations towards the center of the molecular CND of Cen~A, observed as part of project 2012.1.00225.S (PI D. Espada). 
The chosen angular resolution was $\sim$0\farcs3 (5\,pc) for the band 7 and 9 observations. The angular resolution obtained in band 7 and 9 is almost a factor of a hundred better in area than the resolution obtained in previous SMA or ALMA CO(2--1) maps (Espada et al. 2009, 2012, 2013), and comparable to the warm H$_2$ map resolution (2\,pc) by \citet{2007ApJ...671.1329N} using VLT/SINFONI.

Observations took place in 2014 July and April for the band 7 and 9 observations, with 33 and 34 antennas, respectively, and 
unprojected baseline lengths spanning  20 -- 650\,m and 20 -- 558\,m.
The maximum angular scale (defined as 0.6$\lambda$/$L_{\rm min}$ where $\lambda$ is the wavelength and $L_{\rm min}$ the minimum baseline)  for which we recover most of the flux is 5\arcsec\ (90\,pc) for band 7, and 3\arcsec\ (54\,pc) for band 9, respectively. 

The CO(3--2)  ($\nu_{\rm rest}$ = 345.796~GHz) emission line was observed in the lower sideband simultaneously with
HCO$^+$(4--3) ($\nu_{\rm rest}$ = 356.734~GHz) and HCN(4--3) ($\nu_{\rm rest}$ = 354.505~GHz) in the upper sideband.  
CO(6--5) ($\nu_{\rm rest}$ = 691.473~GHz) was also observed in the lower side band simultaneously with 
HCN(8--7) ($\nu_{\rm rest}$ = 708.877~GHz), but the latter was not detected in emission. 
One execution of 2 hours was carried out in band 7, and three executions of 2 hours each for Band 9 in different Local Sidereal Time (LST) ranges. The set-ups of our interferometric observations are summarized in Table~\ref{table1}.
The systemic velocity of Cen~A is $V_{sys}$ = 541.6\,\kms \citep{2010ApJ...720..666E}.

The field of view is characterized by a Half Power Beam Width (HPBW) for the primary beam of an ALMA 12m antenna of 16\farcs9 (304\,pc) in band 7 and 8\farcs4 (151\,pc) in band 9.
The bandwidth chosen was 937.5\,MHz  for band 7 and 1.875\,GHz for band 9 ({ both corresponding to} $\sim$800\,\kms), and the channel width of 244\,kHz and 488\,kHz for spectral windows at bands 7 and 9 (or $\sim$ 0.4\,\kms), respectively. Enough line-free channels are present to subtract the continuum emission. 

The CASA package (\citealt{2007ASPC..376..127M}) was used for data reduction. The CASA version used was 4.2.1 (r29048). Each execution was calibrated separately by the East Asian ALMA Regional Center (EA-ARC) and followed the standard ALMA calibration scheme. A priori calibration tables were created for Water Vapor Radiometer phase correction and atmospheric calibration. Then bandpass, gain (amplitude, phase), and flux calibrations were applied. 
Titan and 3C279 were chosen for absolute flux and bandpass calibration, respectively.  An initial gain calibration was performed using J1321-4342 or J1427-4206,  at angular distances of $1\arcdeg$ and $11\arcdeg$ from the target.  Self-calibration was then done by us in phase and amplitude using the bright and compact continuum source of Cen~A, and only using line-free channels.

The fluxes obtained in band 7 at 345.6\,GHz on 2014 July 7 were 6.46\,Jy for 3C279 and 2.42\,Jy for J1427-4206.  This is in very good agreement with fluxes in the ALMA source catalogue database at similar dates:  at 343.48\,GHz the fluxes were 6.60\,Jy for 3C279 and 2.38\,Jy for 1427-4206, on 2014 July 6 and June 30, respectively.
The fluxes obtained in band 9 at 691.1\,GHz on 2014 April 14 were 4.5\,Jy for 3C279 and 2.3\,Jy for J1427-4206, and all three executions show similar results. These fluxes are also consistent with the measurements in 
 the ALMA source catalogue in that band (691.1\,GHz) on 2014 25 April, 4.35-4.99\,Jy for 3C279, and 1.96\,Jy for J1427-4206. Note that J1321-4342 was not monitored as frequently by the ALMA project and it is therefore not useful for a comparison. We thus estimate absolute flux uncertainties of the order of 5\% for band 7 and 10--20\% for band 9.

Once calibration for each dataset was completed, we concatenated all executions and proceeded to imaging also using CASA with task \verb!CLEAN!. First, we imaged the continuum emission using line-free channels. We cleaned the continuum image using the multi frequency synthesis technique (MFS), and using the weighting scheme Briggs with a robust parameter 0.5. The obtained noise levels are 2.2\,mJy/beam and 16.6\,mJy/beam for the 350 and 698\,GHz continuum maps, respectively.
To image the emission lines, and prior to the cleaning, we used task \verb!UVCONTSUB! to substract the continuum using as baseline the line-free channels and an order of 1 in the fitting polynomial function. Since the continuum emission is so strong in this source, we paid special attention that continuum subtraction was done correctly by checking for possible artifacts in the line-free channels. The weighting scheme for imaging of the lines we used was also Briggs and robust parameter 0.5. In the channel maps we binned the data to 10\,\kms , which is enough to resolve spectrally the $\sim$400\,\kms\  velocity width of the circumnuclear gas. 
 Note that velocities are expressed throughout this paper with respect to the kinematic Local Standard of Rest (LSRK) using the radio convention.
											
 The angular resolution expressed as the Full Width at Half Maximum (FWHM) is $0\farcs36 \times 0\farcs29$ ($6.5 \times 5.2$\,pc) with a major axis $PA$ = $70\arcdeg$ in our band 7 observations, and $0\farcs23 \times 0\farcs16$ ($4.1 \times 2.9$\,pc) with a major axis $PA$ = $48\arcdeg$ in our band 9 observations.  The mean rms noise levels are 1.3\,mJy\,beam$^{-1}$ and 5.8\,mJy\,beam$^{-1}$ for the Briggs weighted band 7 and band 9 channel maps with a channel width of 10~\kms .
The channel maps presentation was conducted in MIRIAD \citep{1995ASPC...77..433S}.  
The task \verb!IMMOMENTS! in CASA was used to calculate the integrated intensity maps, intensity-weighted velocity field and velocity dispersion distributions (the latter two clipped at 5 times the rms noise).

\section{Results}
\label{result}

We present in this section the results obtained from the 350 and 698\,GHz continuum maps, as well as CO(3--2), HCO$^+$(4--3), HCN(4-3), and CO(6--5) emission spectral line maps.

\subsection{350 and 698\,GHz Continuum Emission}

The continuum emission was found to be unresolved at 350\,GHz and 698\,GHz. 
The peak coordinates for the 350\,GHz and 698\,GHz continuum maps were calculated using a two dimensional fit and are very close to each other. To our angular resolution, the 350\,GHz and 698\,GHz peaks are identical and located at $RA$ : 13$^{\rm h}$25$^{\rm m}$27.$^{\rm s}$615, $Dec$: -43\arcdeg01\arcmin08\farcs805. The uncertainty is at most one tenth of the synthesized beam.
This is in perfect agreement with the position of the AGN found by \citet{1998AJ....116..516M} using Very Long Baseline Interferometry (VLBI). We will keep this position as our reference for the AGN location.
On 2014 July 7 the flux for the 350\,GHz continuum was measured to be 8.0 $\pm$ 0.1\,Jy.
The 698\,GHz continuum flux was found to be 7.7 $\pm$ 0.1\,Jy on 2014 April 14.

\subsection{CO(3--2) Emission}
\label{subsect:co3-2emissionline}

\begin{figure*}
\centering
\includegraphics[width=9cm]{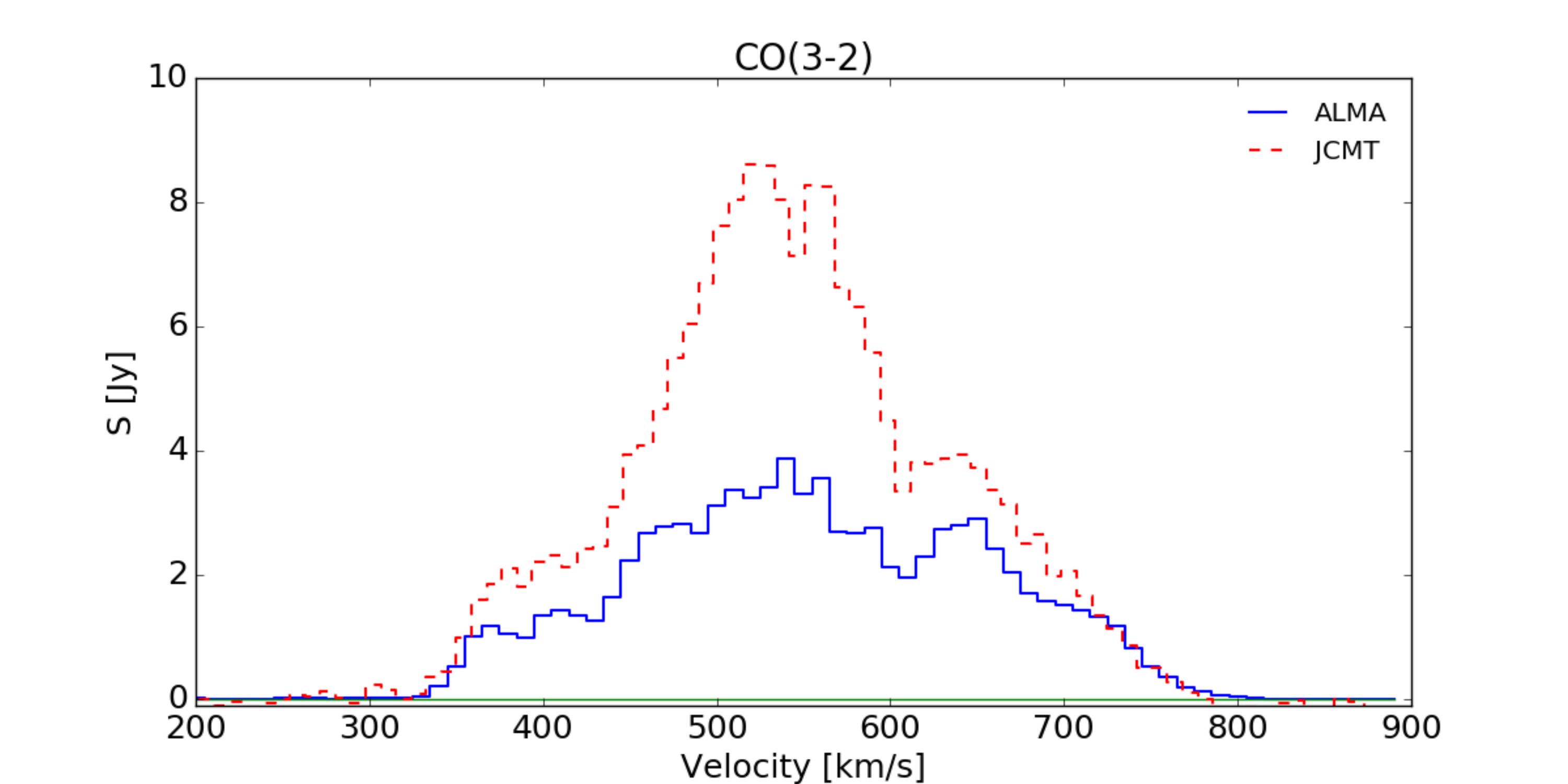}
\includegraphics[width=9cm]{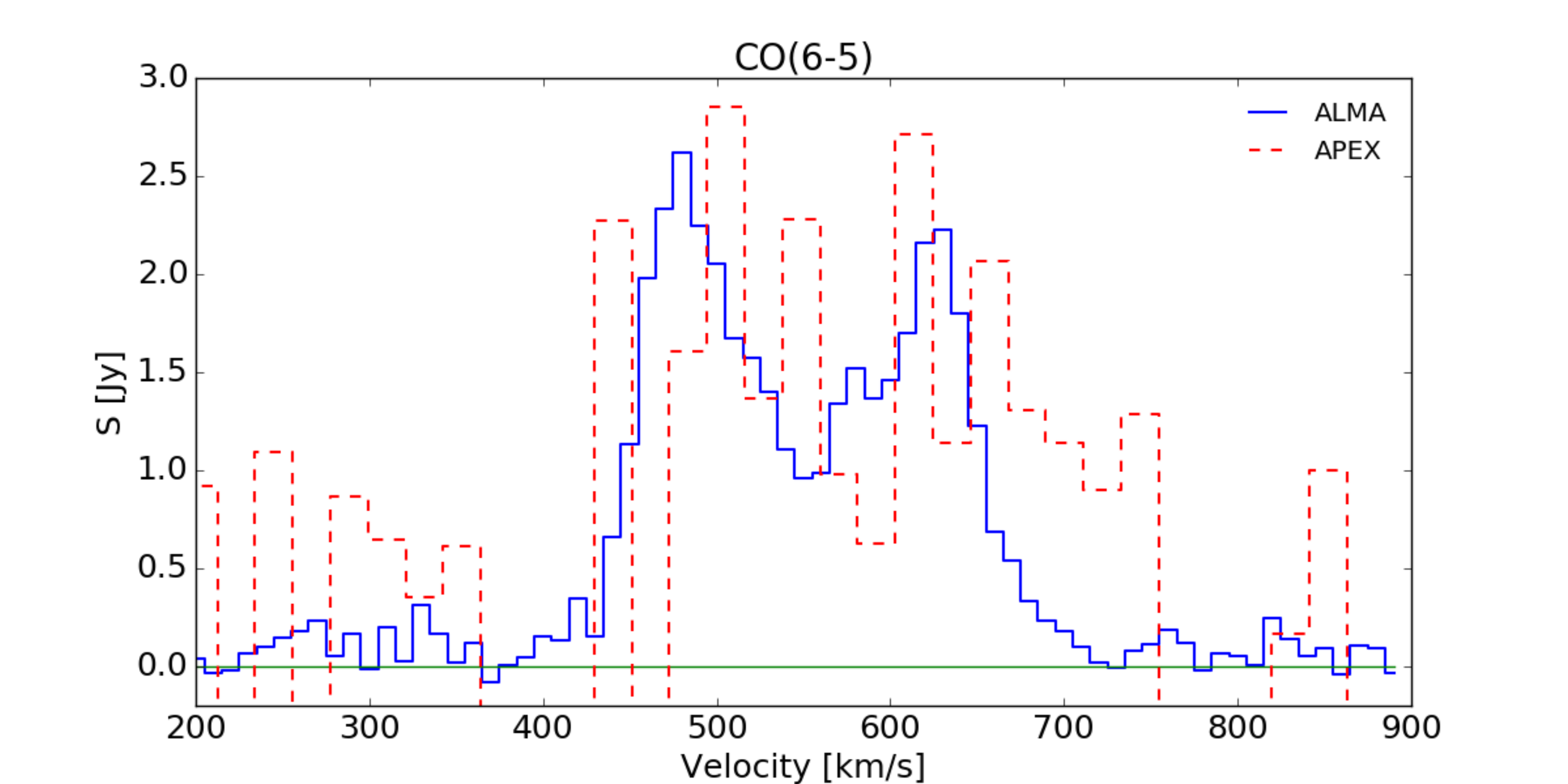}
\caption{CO(3--2) and CO(6--5) spectra. The y-axis is in Jy, from -0.1 to 10\,Jy and -0.2 to 3.0\,Jy. The x-axis shows radio LSR velocity in units of km\,s$^{-1}$ from 200 to 900 km\,s$^{-1}$. Note that these two spectra include Cen~A emission from detected regions in the data cubes  (see Figs.~\ref{fig2} and \ref{fig6} for CO(3--2) and CO(6--5), respectively) and exclude the spatially unresolved region exhibiting absorption at the center of the galaxy. The dashed lines show JCMT CO(3--2) and APEX CO(6--5) spectra from \citet{2014A&A...562A..96I} for comparison.
\label{fig1}}
\end{figure*}

In Fig.~\ref{fig1}, we present the CO(3--2) spectrum integrated over detected regions in the inner 24\arcsec\ and excluding the center of the image where absorption lines are found towards the (unresolved) continuum emission in the velocity range from 540 to 620\,\kms . The profile is relatively flat and shows two or probably three peaks, at about 480\,\kms , 550\,\kms and 650\,\kms . High red and blue-shifted velocity components are also seen at $V$\,$<$\,420\,\kms\ and $V$\,$>$\,680\,\kms, which { were not as clearly seen in} previously obtained CO interferometric data due to a lack of angular resolution and sensitivity. In order to illustrate where missing flux is important, we also plot in Fig.~\ref{fig1}  the 15\,m James Clerk Maxwell Telescope (JCMT) CO(3--2) profile \citep{2014A&A...562A..96I} for comparison. { Note that the ALMA CO(3--2) spectrum presented in Fig.~\ref{fig1} was obtained after correcting the CO(3--2) data by the different beam response of the JCMT (14\arcsec\ HPBW) in order to be able to compare.}

The integrated flux measurement of the ALMA CO(3--2) line is 1102 $\pm$ 6\,Jy\,\kms , with no primary beam correction, and 1552  $\pm$ 8 Jy\,\kms\  with primary beam correction. Error estimates correspond to the nominal value, but note that uncertainties are larger due to absolute flux uncertainties and filtered flux. 
Correcting by the different beam response of the JCMT we obtain 880\,Jy\,\kms.  
This is to be compared with the flux of 1600\,Jy\,\kms\ obtained directly from the JCMT CO(3--2) spectrum in Fig.~\ref{fig1}. The absorption line flux was estimated as 220\,Jy\,\kms\ \citep{2014A&A...562A..96I}. 
Therefore our map recovers $\sim$50\% of the flux.  From Fig.~\ref{fig1} we can see that we are missing emission preferentially around the systemic velocity. The extended emission that { arises from the disk on large spatial scales is likely one of the most important contributors to the missing flux due to the lack of short-spacings.}

\begin{figure*}
\begin{center}
\includegraphics[width=12cm]{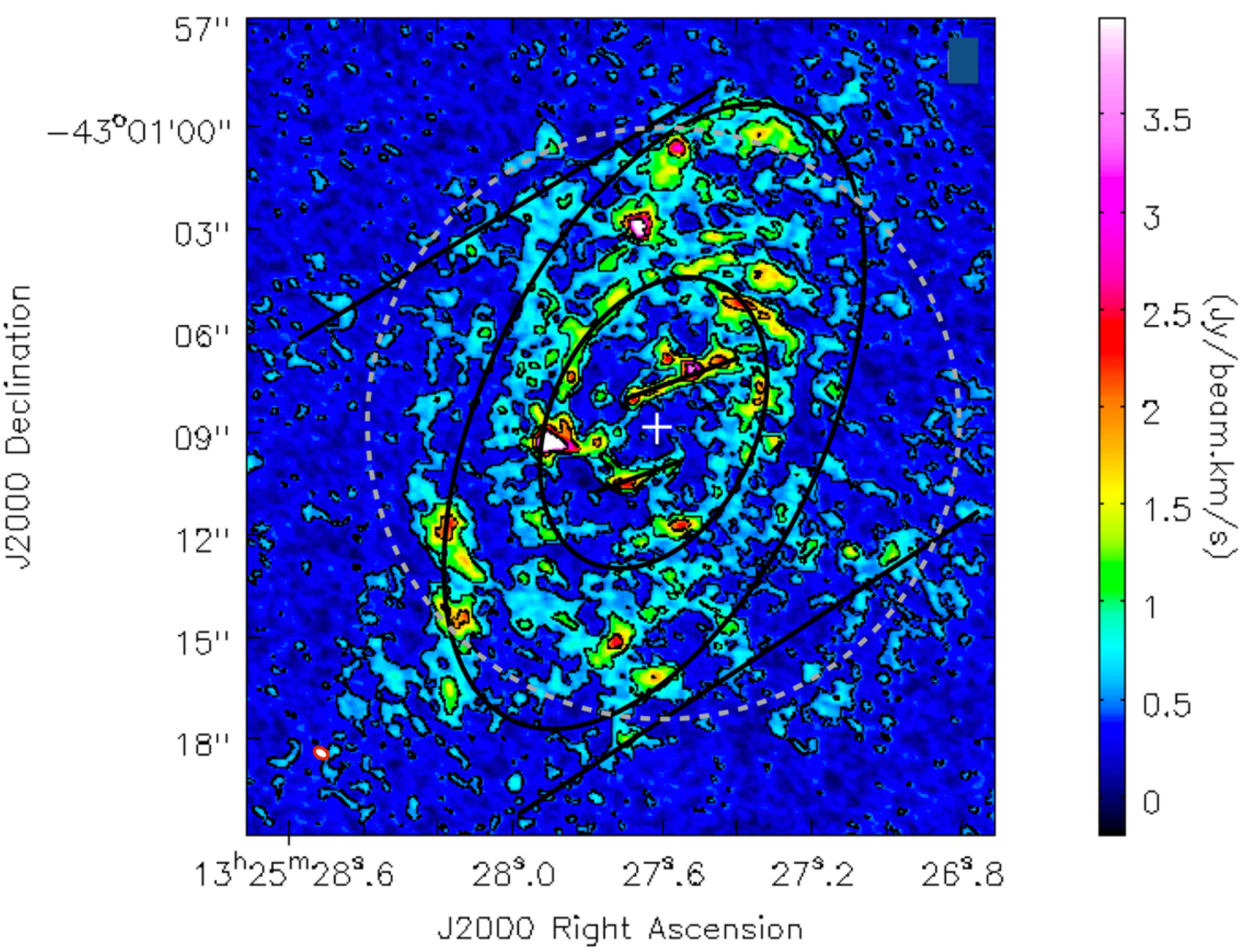}
\end{center}
\caption{ CO(3--2) integrated intensity map of the CND of Cen~A. 
Contour levels are at 0.44, 0.9, 1.8 and 3\,Jy\,beam$^{-1}$\,\kms.  The synthesized beam ($0\farcs36 \times 0\farcs29$, PA=70.44\arcdeg) is indicated by an ellipse in the lower left corner of the plot, and the colour scale is shown beside the plot, also in Jy\,beam$^{-1}$\,\kms units.  
The cross in the center of the image shows the position of the AGN, the (grey) dotted line the primary beam at 345\,GHz (HPBW = 16\farcs9), the two (black) ellipses the CND and the nuclear ring, with semimajor {axes} of  10\arcsec $\times$ 5\arcsec and 4\farcs5 $\times$ 3\arcsec, respectively, and both with a $PA$ = 155\arcdeg. The two lines outside the CND correspond to gas at larger radii seen in projection associated to the spiral arms and parallelogram features aligned along $PA$ = 120\arcdeg\ and at velocities close to the systemic velocity. The two parallel lines inside the inner ring correspond to two nuclear filaments (loci of shocks) with lengths of 4\arcsec\ (NW {of the} AGN) and 3\arcsec\ (SE {of the} AGN), with $PA$ = 110\arcdeg\ and with an offset between them of 2\arcsec .
\label{fig4}}
\end{figure*}

Figs.~\ref{fig-1}b and \ref{fig4} show the CO(3--2) integrated intensity map.
The extent of the CND is represented by an ellipse with major and minor axes of $20\arcsec\ \times 10\arcsec$, or 360\,pc $\times$ 180\,pc in linear scale (without taking into account any projection effect), and a position angle of $PA$ $=155\arcdeg$. Note that there is emission beyond the primary beam of the central pointing at 345\,GHz. 
Fig.~\ref{fig4b} shows a simplified scheme that illustrates the main molecular components within the CND that are discussed in this Paper {(see \S\,\ref{subsect:components}}).

In Fig.~\ref{fig2}, we show the CO(3--2) channel maps over the CND of Cen~A. The map covers the inner 24\arcsec\ (432\,pc) and a velocity interval from 315 to 790\,km~s$^{-1}$ in 20~\kms\ bins. Note that no primary beam correction was performed in these maps.
Fig.~\ref{fig3} shows the enclosed 12\arcsec\ area to highlight the different inner components within the CND. 
The CO(3--2) channel maps show blue-shifted ($\sim$315 -- 495~km~s$^{-1}$) emission on the SE and red-shifted ($\sim$615 -- 790~km~s$^{-1}$) emission on the NW side. The Full Width at Zero Intensity (FWZI) is $\Delta V$ $\simeq$ 475 $\pm$ 7 \kms.  
 
  The higher velocity components in the spectrum (see Fig.~\ref{fig1}) at $V$\,$<$\,420\,\kms\ and $V$\,$>$\,680\,\kms\ are located at the edges of the CND.
The molecular gas at large radii (likely $\geq$ 1\,kpc, although seen in projection in the field of view) associated with the kpc scale spiral features (see \S\,\ref{introduction}, emission around the parallel lines in Figs.\,\ref{fig4} and \ref{fig2}) can be discerned as very extended components in the velocity range from 495\,\kms\ to 635\,\kms : from 495\,\kms\ to 535\,\kms\ towards the NE with respect to the center, and from 515\,\kms\ to 635\,\kms\ towards the SW. Note that these extended components are likely the most affected by missing flux in this interferometric experiment.

There are multiple filaments within the previously discovered CND that are apparent in the maps we present in this Paper. The longest feature, extending more than 15\arcsec\ (or 270\,pc) is located just at the eastern edge of the CND (see for example channel at $V$= 495\,\kms\ in Fig.~\ref{fig2}). Note that some filamentary structures may in part be caused by projection effects, especially in the channels from 495\,\kms\ to 595\,\kms\ (close to the systemic velocity), where emission from the CND and from gas expected to arise from larger galactocentric radii (r$>$1\,kpc) might be projected along a similar line of sight.
Multiple streamers on scales of tens to one hundred parsec scale also exist within the CND, which form a ring-like structure (nuclear ring) with a diameter of 9\arcsec\  $\times$ 6\arcsec\ (160\,pc $\times$ 108\,pc ), and a $PA$ = 155\arcdeg. The nuclear ring can be discerned in the map showing the integrated emission (Fig.~\ref{fig4}), and in channels from 415\,\kms\ to 675\,\kms\ (Fig.~\ref{fig2}). 
Inside the ring there are two filaments elongated along a position angle PA=120\arcdeg\ and separated by 2\arcsec\ to the SE and NW from the AGN location.
More details on the different components are provided in \S\,\ref{subsect:components}.

Fig.~\ref{fig5} presents the intensity-weighted velocity field (left panel) and the velocity dispersion maps (right).  
 The receding side of the CND is to the NW and the approaching side to the SE. As discussed in \citet{2009ApJ...695..116E} there are multiple components along the line of sight and the velocity field should be taken with caution when it overlaps with projected emission arising from gas at larger radii. This is most important at the edges of the CND to the NW and SE. There are deviations with respect to a simple circularly rotating coplanar disk because the velocity field shows an S-shape distribution (see \S\,\ref{secfeeding}).
 The velocity dispersion map shows that there is a large range of values within the CND, from a few \kms\ to tens of \kms. Note that in this plot we excluded the 6 channels at velocities $<$ 60\,\kms\ from the systemic velocity (i.e. 545\,\kms) to avoid contamination by the molecular gas component at $r$ $\geq$ 1\,kpc seen in projection. The velocity width of this component is much narrower ($\sim$140\,\kms) than the CND in the field of view, so we are effectively removing most of its contribution. In general the molecular gas has low velocity dispersions of $\sim$ 5--10\,\kms, but especially in the outermost parts of the CND to the NW and SE we see that the velocity dispersion increases, and there are velocity dispersions of up to 30--40\,\kms . Other regions such as the nuclear filaments have larger velocity dispersions as well.
 The large velocity dispersions there become apparent in the Position--Velocity (P--V) diagrams, which are described next.

P--V diagrams along several cuts are shown in Fig.~\ref{fig-pv}.  Figs.~\ref{fig-pv}a, b and c show the P--V diagrams along $PA$ = $60\arcdeg$, $120\arcdeg$, $150\arcdeg$, respectively, all of them with widths of 6\arcsec. These P--V diagrams  cross and are centered at the AGN position.
From Fig.~\ref{fig-pv}c we estimate that the CND velocity gradient is typically $\Delta V / \Delta r$ = 1.8\,\kms\,pc$^{-1}$ (or 32\,\kms\,arcsec$^{-1}$). At the edges of the CND, at about -8\arcsec\ and 8\arcsec\ from the center, we can see in the P--V diagrams along $PA$ = 120$\arcdeg$ and 150$\arcdeg$ that there is a large radial velocity drop of $\sim$100\,\kms\ towards { more systemic values} (see for example the enclosed region by red boxes in the 150$\arcdeg$ P--V diagram) both in the NW and SE.
This is to be compared with \citet{2009ApJ...695..116E}, where the velocity gradient in the CND (up to $r$ $\simeq$ 200\,pc) was given as 1.2\,\kms\,pc$^{-1}$, and 0.2\,\kms\,pc$^{-1}$ in the outer parts of the disk located at $r$ $>$ 400\,pc and out to a few kiloparsec. This velocity gradient for the CND agrees well with that quoted in this Paper, although it is slightly lower mainly due to beam smearing in the previous SMA observations.

Apparent in all P--V diagrams is that the nuclear filaments (within the inner $r$ = 36\,pc, or 2\arcsec) inside the CND have an even steeper velocity gradient,  which can be characterized by $\Delta V / \Delta r$ = 3.4\,\kms\,pc$^{-1}$ (or 62\,\kms\,arcsec$^{-1}$). See for example the enclosed regions by the (red) boxes in Fig.~\ref{fig-pv}a and the (blue) box in Fig.~\ref{fig-pv}c for reference.

Fig.~\ref{fig18} (left panel) shows the P--V diagrams obtained from CO(3--2) data compared with the compilation performed by \citet{2009ApJ...695..116E}, including CO(2--1) data, H$\alpha$ data \citep{1992ApJ...387..503N} and single dish CO(3--2) data \citep{2001A&A...371..865L}.  In the inner region ($r$ $<$ 0\farcm2, see the right panel) one can distinguish the peculiar velocities of the two nuclear filaments, and in the outer regions ($r$ $\sim$ 0\farcm14) the sudden drop in radial velocity of $\sim$100\,\kms . Data points for both receding and approaching sides are shown in these plots and they show good agreement. Note that a rotation curve obtained from P--V diagrams has large uncertainties when non-circular motions are present \citep{1999ApJS..124..403S}. Instead an average across different directions is closer to the right rotation curve. Therefore we confirm that the estimated rotation curve presented in \citet{2009ApJ...695..116E} is reasonably acceptable.

The decomposition of the CO(3--2) emission for different components mentioned in this section is further explained in \S\,\ref{subsect:components} and the main parameters are summarized in Table~\ref{table2}.

\subsection{CO(6-5) Emission}
\label{subsect:dustlane}

Fig.~\ref{fig1} shows the ALMA CO(6--5) spectrum integrated over the detected regions,  where we have excluded (as for the CO(3--2) spectrum) the absorption line towards the central continuum emission. We also show the APEX (Atacama Pathfinder EXperiment) CO(6--5) profile from \citet{2014A&A...562A..96I} for comparison.
Note that within this field of view we mostly probe the molecular gas located in the ring and nuclear filaments.
We can distinguish more clearly than in the CO(3--2) spectrum a double peak structure.  
 The high velocity components in the CO(3--2) spectrum found at $V$\,$<$\,420\,\kms\ and $V$\,$>$\,680\,\kms\ are not present in the CO(6--5) spectrum partly because these are located in the external components of the CND, which are attenuated because they are located outside the primary beam HPBW.

In order to estimate how affected the CO(6--5) map is by the missing zero spacing problem, we compare the fluxes with that of the single dish measurement. 
The total flux measurement of the ALMA CO(6--5) data is 521$\pm$ 4\,Jy\,\kms, with no primary beam correction.
Again, the error estimates correspond to the nominal value without taking into account absolute flux uncertainties and filtered flux.
We compare this directly with the flux obtained from observations from APEX since that antenna has a similar response as the { individual} ALMA 12m antennas. The flux was found to be 787 $\pm$ 187\,Jy\,\kms\ \citep{2014A&A...562A..96I}.   This indicates that $\sim$ 30\% of the flux is missing within this field of view, and reflects the compact nature of the CO(6--5) emission. Note that although the APEX spectrum is good enough to provide an integrated flux, it is probably too noisy for a channel to channel comparison. The 30\% is an upper limit because the flux peaks are essentially equivalent within the uncertainties, and the flux difference arises mostly from the velocity range 650 -- 750\,\kms, which may indicate that the APEX pointing was slightly offset towards the NW.

In Fig.~\ref{fig6} we show the channel maps of the CO(6--5) emission line covering the inner 12\arcsec\ (same size as the CO(3--2) channel maps in Fig.~\ref{fig3}) and the velocity interval from 315 to 795\,\kms\ in 20\,\kms\ bins. The blue-shifted emission starts in channel 425\,\kms\ at the SE of the AGN and the redshifted side ends at 705\,km~s$^{-1}$ to the NW. 
The morphology is very similar to that of the CO(3--2) maps in this field of view.
The two nuclear filamentary structures are clearly visible, as well as other filaments connected to them to the N and S, and also, although weak (partly because they are close to the edge of the primary beam), some regions along the ring are detected.

Figs.~\ref{fig7} and \ref{fig8} show the  CO(6--5)  integrated intensity, velocity field and velocity dispersion maps.  
Note that the resolution is slightly better (factor of three in area) in the CO(6--5) maps than in the CO(3--2) maps. Also the confusion caused by the multiple components seen in CO(3--2) is not present in the CO(6--5) maps, and we can discern more clearly the structure and kinematics of the CND within its inner 8\arcsec , and in particular the distribution of the nuclear filaments. Large velocity dispersions of $\sim$ 10 -- 20\,\kms\ can be found in some parts of the nuclear filaments.

The filaments to both sides of the AGN are composed by several Giant Molecular Clouds (GMCs) that are resolved with our synthesized beam. The closest {projected distance of these clouds} to the AGN is 1\arcsec , or $\sim$ 20\,pc. They form two nearly straight filaments which are also kinematically distinct, i.e. deviations of typically 50 km\,s$^{-1}$ can be seen from the rotational velocity of the CND at a given radius. However, we confirm that both filaments are seen to be slightly asymmetric with respect to each other. First, the GMCs are brighter by a factor of two in the NW filament than in the SE filament. 
Also, as in the CO(3--2) maps, while the NW filament is perfectly aligned, the one in the SE is slightly curved { in the lower contours at the distant SE end}.  
These asymmetries between the two sides indicate that the molecular gas is not well-settled. 
 CO(6--5) emission also exists to the E and W of the AGN and slightly farther from the AGN than the filaments, which connect the nuclear ring with the nuclear filaments. The eastern connecting point contains substantially more molecular gas than its western counterpart. 
The highest velocity dispersion regions are in the connecting points to the E and W of the AGN, and at the end of the filamentary structures as they approach the AGN.

The CO(6--5) P--V diagrams are displayed in Fig.\,\ref{fig-pv2}: a) a $PA$ = 115$^{\rm o}$ cut along the SE filament with a a length of 3\farcs5 and a width of 1\farcs6, b) a $PA$ = 115$^{\rm o}$ cut along the NW filament with a length of 3\farcs7 and a width of 1\farcs9, and  c) a $PA$ = 115$^{\rm o}$ cut with a length of 9\arcsec\ and a width of 4\arcsec, containing this time both nuclear filaments as well as the connecting points to the E and W of the AGN. The area probed in Fig.\,\,\ref{fig-pv2}c is essentially equivalent to that of the inner 9\arcsec\ of the CO(3--2) P--V diagram in Fig.\,\ref{fig-pv}c ($PA$ = 120\arcdeg), and basically they are in agreement. 

Fig.~\ref{fig-pv3} shows a zoom of the P--V diagram of CO(6--5) at $PA$ = 115$^{\rm o}$ of the NW filament. 
 Along this filament, and farther outside from the nucleus, there is a plateau in the P--V diagram of about 20\,pc in length. As indicated in \S.\,3.2, closer to the nucleus the velocity gradient becomes steeper, $\Delta V$/$\Delta r$ = 3.4\,km\,s$^{-1}$ pc$^{-1}$. A similar plateau and velocity gradient is also seen in the SE nuclear filament. 
At the closest distance from the AGN (1\farcs2, or 22\,pc) along the NW nuclear filament the P--V diagram may show a second plateau in velocity and there is a signature of an even steeper velocity gradient of $\Delta V$/$\Delta r$ $\simeq$ 12\,km\,s$^{-1}$\,pc$^{-1}$  at about 0\farcs5 from it, from 570\,\kms\ to 520\,\kms.

\subsection{CO(6--5)/CO(3--2) Line Ratios}
\label{colineratio}

{
The difference in the critical densities of the J=3--2 and 6--5 CO molecular transitions is more than two orders of magnitude (see \S\,\ref{introduction}), and they are located at $T$ = 33 and 166\,K above the ground state, respectively.}
In this subsection, we calculate CO(6--5)/CO(3--2) line ratios to find out if there is any large gradient in the physical conditions of the molecular gas.

In order to reduce the effect of the different angular scales that are recovered by the CO(3--2) and CO(6--5) observations, we first performed imaging in these two datasets using CASA selecting visibilities for uv distances larger than 40 k$\lambda$ in task \verb!CLEAN! keyword \verb!uvrange!, i.e. the minimum uv distance in the CO(6--5) observations. The resulting maximum recoverable scales 
of the two datasets are similar and equal to 3\arcsec (or 54\,pc).  
 We then obtained primary beam corrected maps and convolve the two images to the same resolution of 0\farcs3 (or 5\,pc) with a Gaussian kernel using CASA task \verb!IMSMOOTH!. Absolute flux calibration uncertainties are as explained in \S\,\ref{observationReduction}. 

CO(6--5) to CO(3--2) line ratios, $R_{65/32}$, for regions where CO(6--5) and CO(3--2) moment 0 maps exceeded 3$\sigma$, are shown in Fig.~\ref{lineratio}. 
R$_{65/32}$ spans from 0.2 to 1.4 using main brightness $T_{mb}$ units (or 0.9 to 5.9 using flux units), with a mean value of 0.9 (respective 3.5).
Along the two inner filaments $R_{65/32}$ (i.e. NW and SE) there is a gradual increase towards the AGN by a factor of 2--3. It is closest to the nucleus where the largest values are found, where velocity dispersions were also substantially larger (i.e. close to 20\,\kms ).

Although unresolved, the average $R_{65/32}$ over the whole CND can also be inferred from the analysis by \citet{2014A&A...562A..96I}. 
Obtaining integrated line intensities with Gaussian fits, and assuming the ratio normalized to a response of a 22\arcsec\ beam and corrections by absorption line loss, $R_{65/32}$ = 0.4 (or $\sim$ 1.85 in flux units). This is a factor of two lower than the average over all regions in our $R_{65/32}$ map. It is unlikely that the different factor found in interferometric and single dish experiments is due to flux loss effects.
For further discussion, please refer to \S\,\ref{secratio}.

\subsection{HCO$^+$(4--3), HCN(4--3) Emission and HCN/HCO$^+$(4--3) Line Ratios}
\label{hcohcn}
In Figs.~\ref{fig9} and \ref{fig10} we show the channel maps of HCO$^+$(4--3) and HCN(4--3) covering the inner 12\arcsec\  (same size as the CO(3--2) and CO(6--5) channel maps in Figs.~\ref{fig3} and \ref{fig6} for comparison) and the velocity interval from 415 to 695\,\kms\ and from 415 to 615\,\kms\ in 20\,\kms\ bins, respectively. 
 Due to limitations in the spectral setup, the HCN(4--3) line was only partially covered in our observations so the interval is cut beyond 615\,km~s$^{-1}$. Although the field of view for these two transitions is almost identical to that of CO(3--2), namely 16\arcsec , we did not detect any emission in the region outside 12\arcsec\ from the center.
Similarly to the CO(6--5) transition, 
the blue-shifted emission in these two dense gas tracers starts to the SE at the 435\,\kms\ channel, and then continues to the redshifted side to the NW, ending by the 635\,\kms\ channel. Essentially 
the HCO$^+$(4--3) line is detected along the nuclear filaments. HCN(4--3) is detected or tentatively detected (above 3$\sigma$ and in several channels) in just four regions and in all of them HCO$^+$(4--3) is also detected.
For a similar rms, $\sigma$ = 1.34 mJy\,beam$^{-1}$, HCO$^+$(4--3) emission is found to have a higher S/N than the HCN(4--3) line. Note that the data of these two lines were taken simultaneously in two different spectral windows and calibrated in an identical manner, so the difference in amplitude cannot be explained in principle by absolute flux calibration uncertainties. 

The average HCN(4--3) to HCO$^+$(4--3) intensity ratio in our maps is $R_{HCN/HCO^+}$ $\simeq$ 0.5, but there are regions where the ratio is even lower, e.g. $R_{HCN/HCO^+}$ $\simeq$ 0.3. 
To illustrate the low $R_{HCN/HCO^+}$, we present in Fig.~\ref{fig20} HCO$^+$(4--3) and HCN(4--3) spectra towards four different positions as indicated in the channel maps. The positions are: 1) $RA$ = 13${\rm ^h}$25${\rm ^m}$27.${\rm^s}$905, $Dec$ = -43${\rm ^o}$01$\arcmin$09\farcs286 (58\,pc from the nucleus), 2) $RA$\ = 13${\rm ^h}$25${\rm ^m}$27${\rm^s}$68, $Dec$ = -43${\rm ^o}$01$\arcmin$07\farcs952 (20\,pc from the nucleus), 3) $RA$\ = 13${\rm ^h}$25${\rm ^m}$27${\rm^s}$.664, $Dec$ = -43${\rm ^o}$01$\arcmin$10\farcs473 (32\,pc from the nucleus), and 4) $RA$\ = 13${\rm ^h}$25${\rm ^m}$27.${\rm^s}$615, $Dec$ =  --43${\rm ^o}$01$\arcmin$08\farcs805 (center, in absorption).

We fitted single Gaussian profiles to the detected HCN(4--3) and HCO$^+$(4--3) lines, where the central velocity and width of the HCN(4--3) profile was fixed to that measured toward the brighter HCO$^+$(4--3). The fit parameters are given in Table~\ref{table4}.
The ratios for the emission lines are $R_{HCN/HCO^+}$=0.37 $\pm$ 0.08 in position 1, $R_{HCN/HCO^+}$ = 0.40 $\pm$ 0.1 in position 2, and $R_{HCN/HCO^+}$ $<$ 0.56 in position 3. For the absorption lines towards the AGN the ratio is $R_{HCN/HCO^+}$ = 0.2 but note that it is likely material far from the center and just seen in projection \citep[e.g.][]{2010ApJ...720..666E}.

High $R_{HCN/HCO^+}$ ratios are often claimed to distinguish AGN over starburst conditions \citep[e.g.][]{2001ASPC..249..672K,2008ApJ...677..262K,2016ApJ...818...42I}, but this does not hold for the nuclear regions of Cen~A even though it is a well known AGN. This is further discussed in \S\,\ref{secratio}.

\subsection{Molecular Gas Components within the CND}
\label{subsect:components}

\begin{figure*}
\begin{center}
\includegraphics[width=12cm]{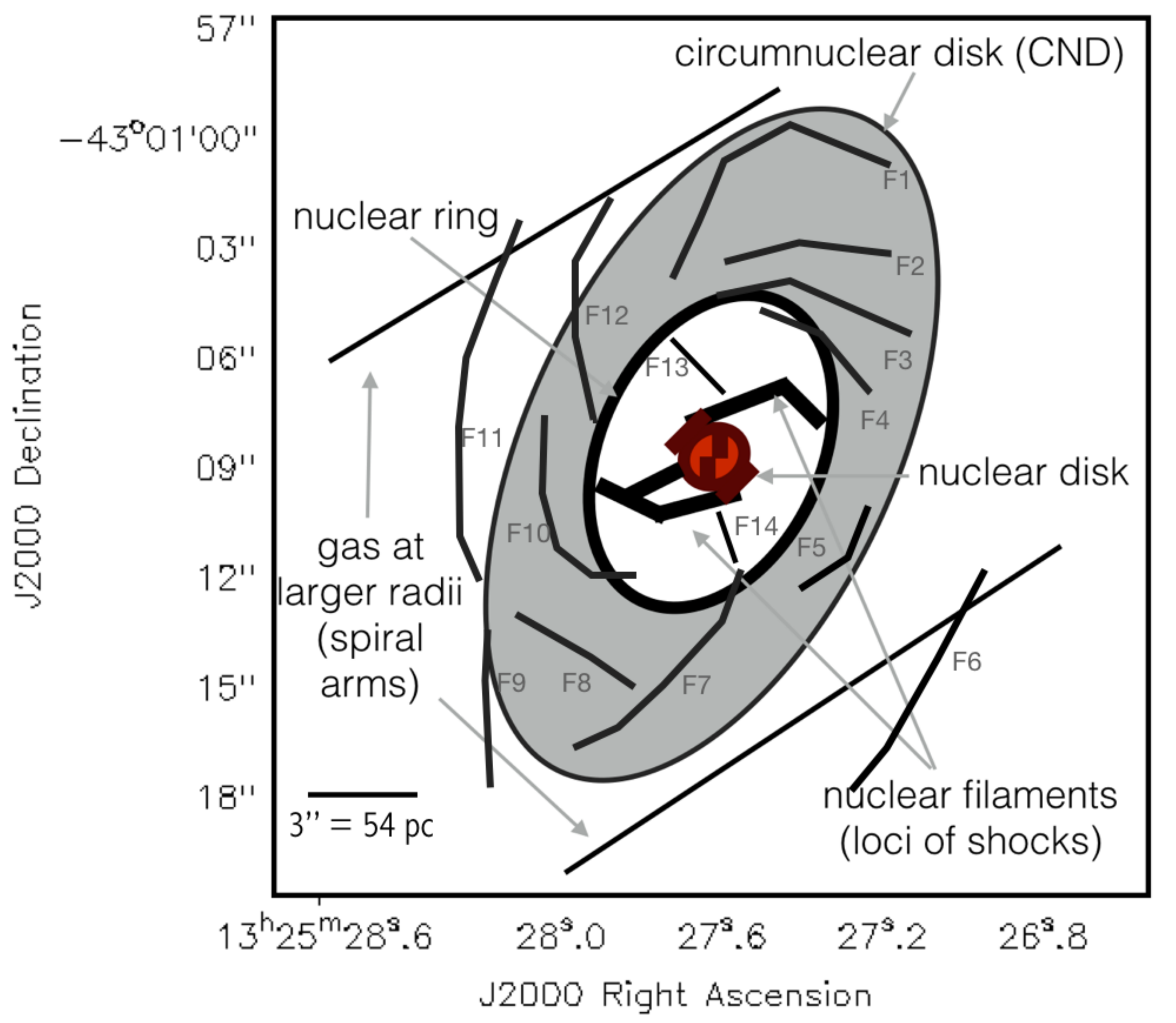}
\end{center}
\caption{
Scheme showing the main molecular components that are present in the CND of Cen~A. The field of view is as in Fig.\,\ref{fig4}. 
Regions marked in black/gray indicate those traced by the transitions presented in this Paper, and in red those traced by the the 1--0 S(1) H$_2$ line in \citet{2007ApJ...671.1329N}. The main filamentary structures within the CND as seen in the channel maps are also indicated as black lines. An identifier for each filament with format F$XX$ is provided. \label{fig4b}}
\end{figure*}

\begin{figure*}
\includegraphics[width=16cm]{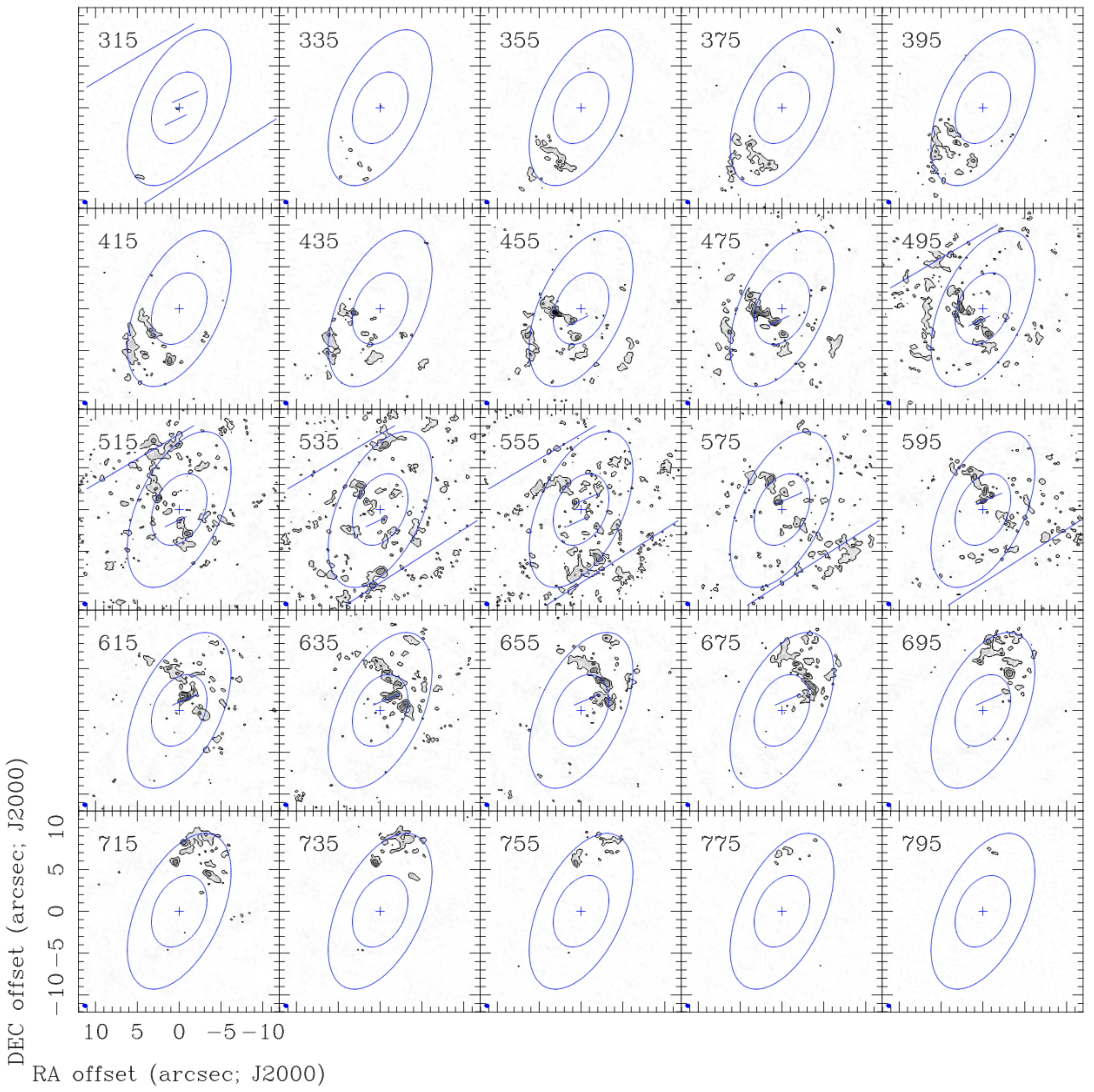}
\caption{{ CO(3--2) channel maps of the CND of Cen~A} in the LSR velocity range V = $315 - 795$\,\kms\ in 20\,\kms\ bins. The size of the maps is 24\arcsec. The velocities are shown in the upper left corner and the synthesized beam is shown at the bottom left corner of each panel.  The rms noise of an individual channel is 0.92\,mJy\,beam$^{-1}$.  The contour levels are 5, 25, and 100 $\sigma$.
The cross sign shows the galaxy's AGN position at $RA$\ = 13${\rm ^h}$25${\rm ^m}$27.${\rm^s}$615 ; $Dec$ =  --43${\rm ^o}$01$\arcmin$08\farcs805 \citep{1998AJ....116..516M}. 
See Fig.~\ref{fig4b} for a description of the symbols representing the main molecular components of the CND of Cen\,A.
\label{fig2}}
\end{figure*}

\begin{figure*}
\includegraphics[width=16cm]{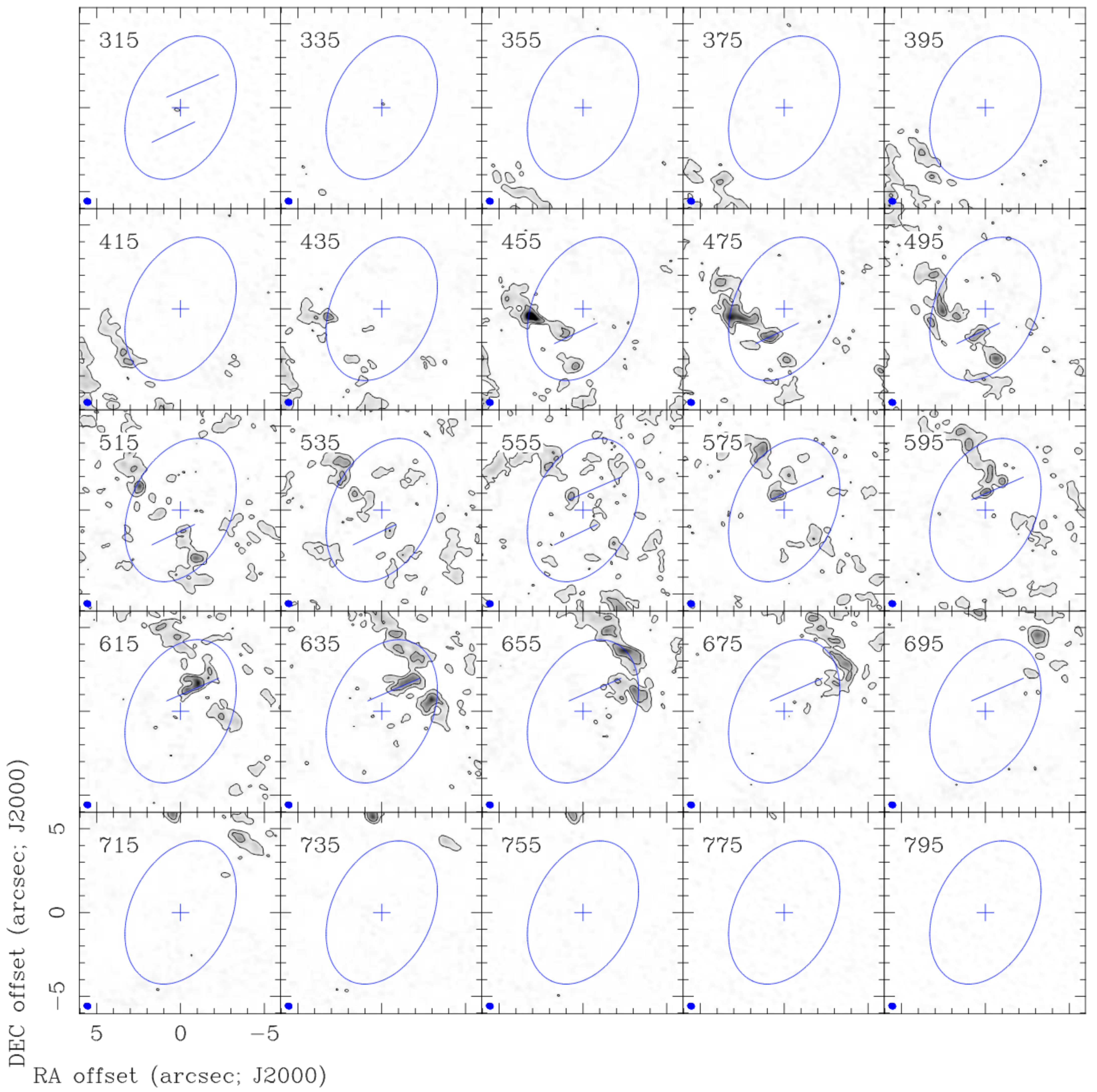}
\caption{As Fig.~\ref{fig2} but the size of the maps is 12\arcsec\ to highlight the inner molecular components. 
\label{fig3}}
\end{figure*}

\begin{figure*}
\begin{center}
\includegraphics[width=8cm]{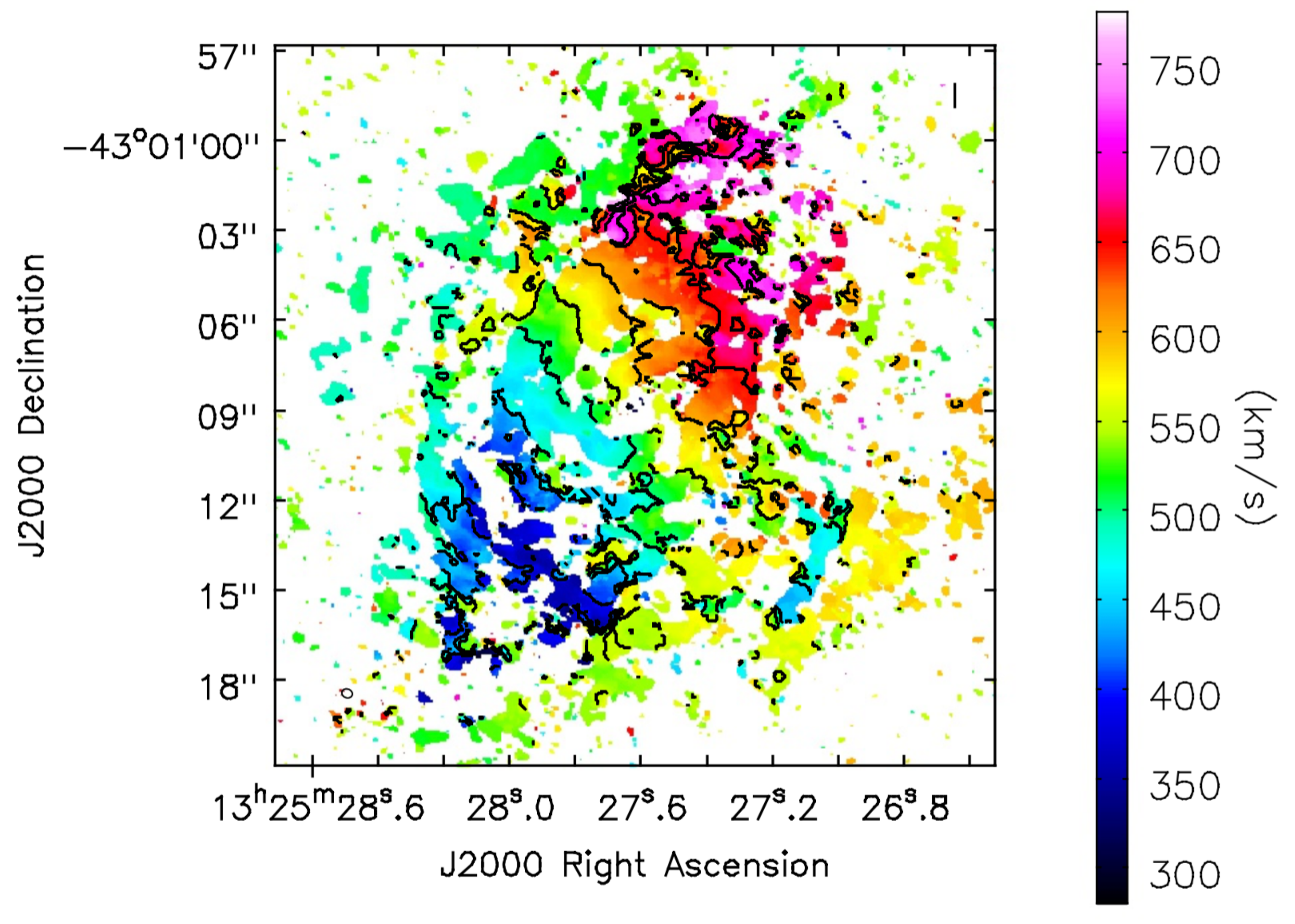}
\includegraphics[width=8cm]{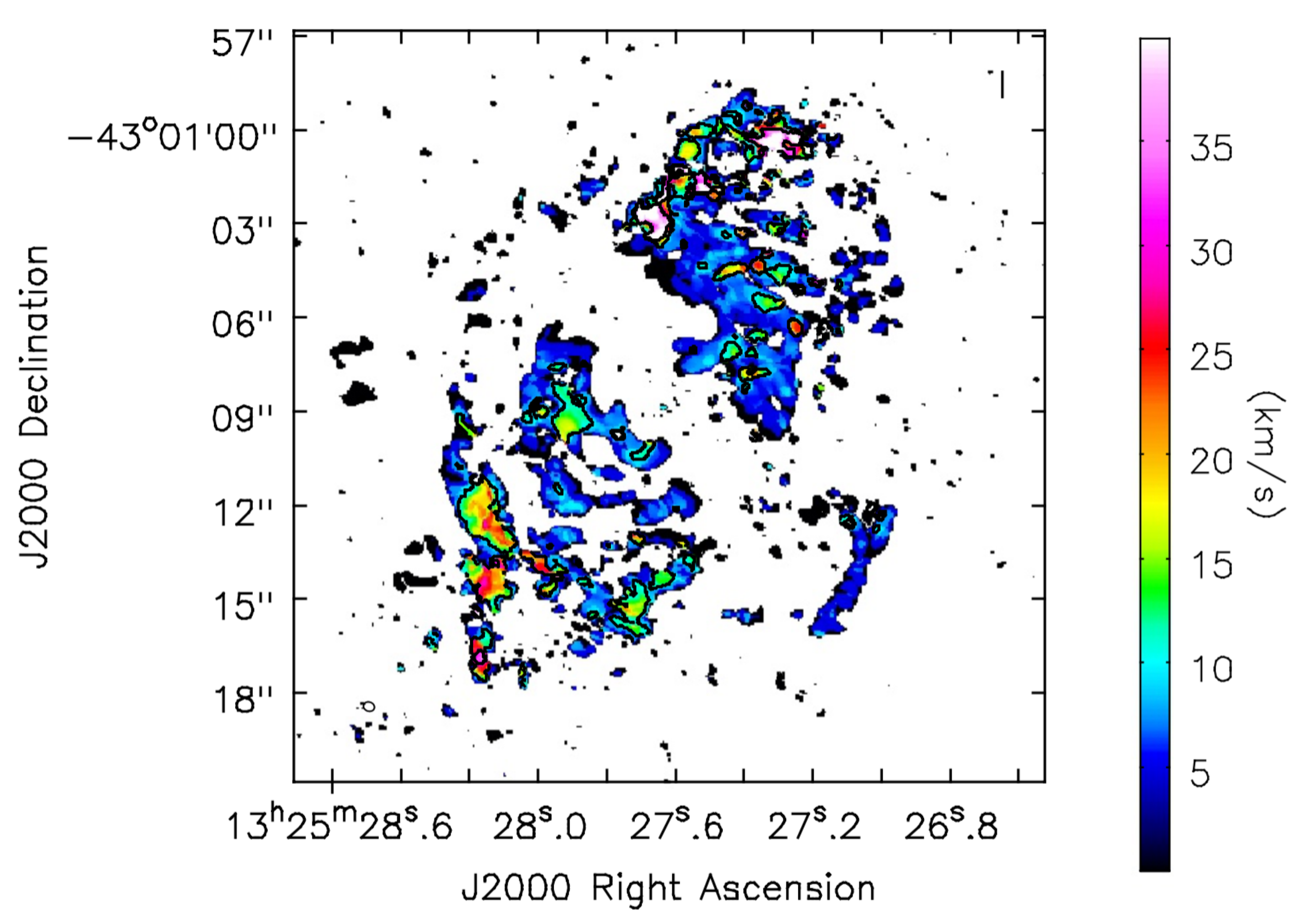}
\end{center}
\caption{ CO(3--2) (intensity weighted) velocity field and velocity dispersion maps of the CND of Cen~A.  In the velocity field map, contours are placed every 50\,km~s$^{-1}$, from $350 - 700\,$km~s$^{-1}$.  The color scale ranges from 280\,\kms\ up to 780 \kms . 
In the velocity dispersion map, we place contours at 10\,\kms\ and 30\,\kms, and the color scale ranges from 0 to 40\,\kms. We excluded channels $<$ 50\,\kms\ from the systemic velocity ($V_{sys}$ = 541.6\,\kms) to avoid contamination by the external molecular gas component. 
The synthesized beam {($0\farcs36 \times 0\farcs29$)} is shown at the bottom left corner of each panel. 
\label{fig5}}
\end{figure*}

\begin{figure*}
\begin{center}
\includegraphics[width=16cm]{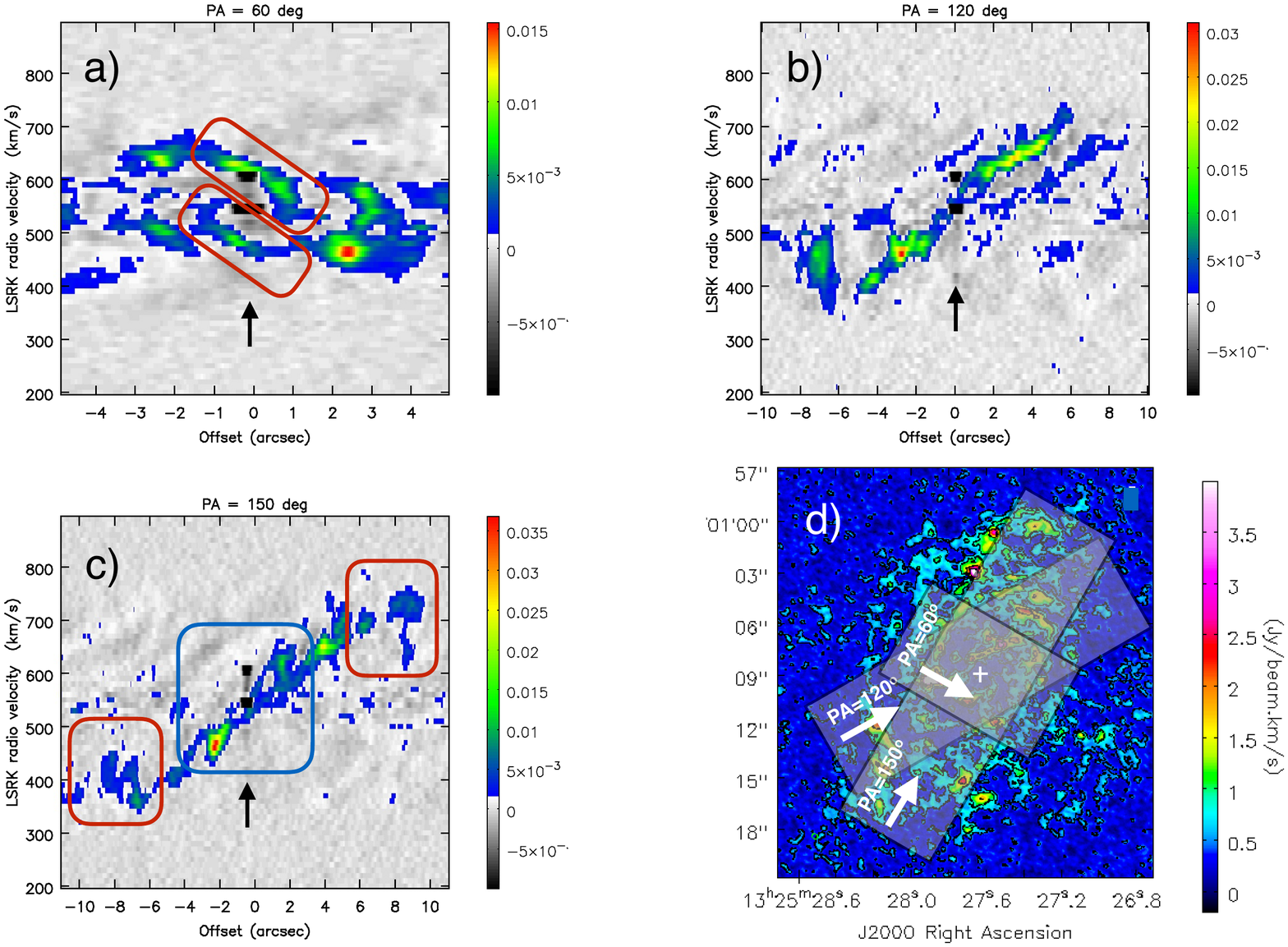}
\end{center}
\caption{ CO(3--2) Position - Velocity (P--V) diagrams: a) $PA$ = 60$^{\rm o}$, with a length of 10\arcsec\ and a width of 6\arcsec , b) $PA$ = 120$^{\rm o}$, with a length of 20\arcsec\ and a width of 6\arcsec , and c) $PA$ = 150$^{\rm o}$, with a length of 22\arcsec\ and a width of 6\arcsec.  Plot d) shows the P--V diagrams along these cuts over the moment 0 map for reference. The white arrows show the direction for each P--V diagram.
In plot a) the red boxes indicate the nuclear filaments. In plot c) the blue box highlights the nuclear ring and filaments, and the red box the edges of the CND. The black arrow symbols indicate the AGN position. 
\label{fig-pv}}
\end{figure*}

\begin{figure*}
\begin{center}
\includegraphics[width=8.cm]{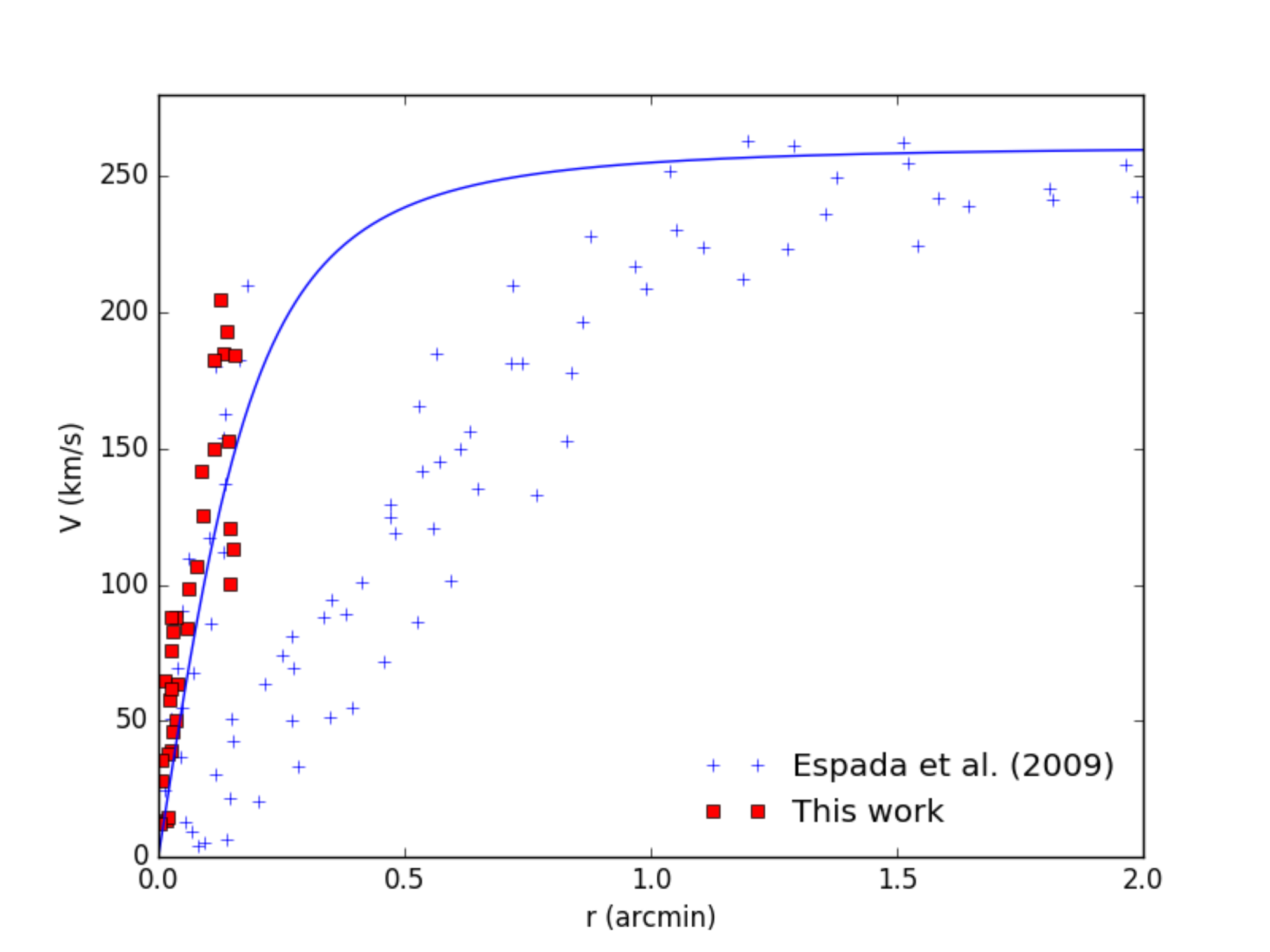}
\includegraphics[width=8.cm]{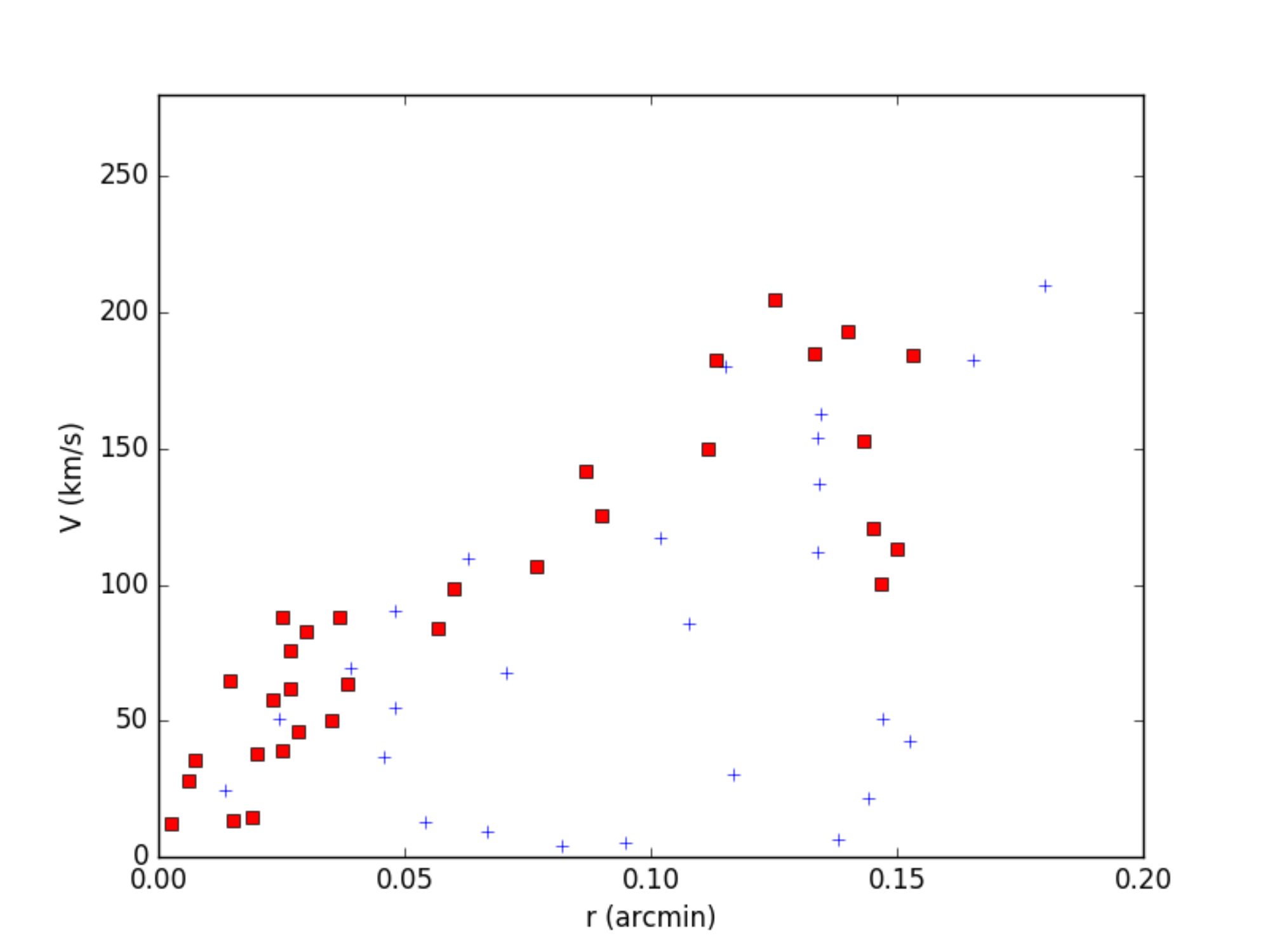}
\caption{Left) P--V diagram as obtained from CO(3--2) data at $PA$ = 150$^{\rm o}$ (red squares) compared with the compilation performed by \citet{2009ApJ...695..116E} (blue crosses, see for reference fig.\,7 of that publication) using CO(2--1) \citep{2009ApJ...695..116E}, H$\alpha$ \citep{1992ApJ...387..503N} and CO(3--2) \citep{2001A&A...371..865L}. { Velocities are provided with respect to the systemic one}. The solid line indicates the rotation curve estimated by \citet{2009ApJ...695..116E} using an axisymmetric logarithmic potential $\Phi_0(r) = 0.5 \times \log(a + r^2/b)$ with $a = 1.0$ and $b = 0.05$. Right) Same as the left panel but zooming into that portion of the plot corresponding to the CND. 
\label{fig18}}
\end{center}
\end{figure*}

\begin{figure*}
\includegraphics[width=16cm]{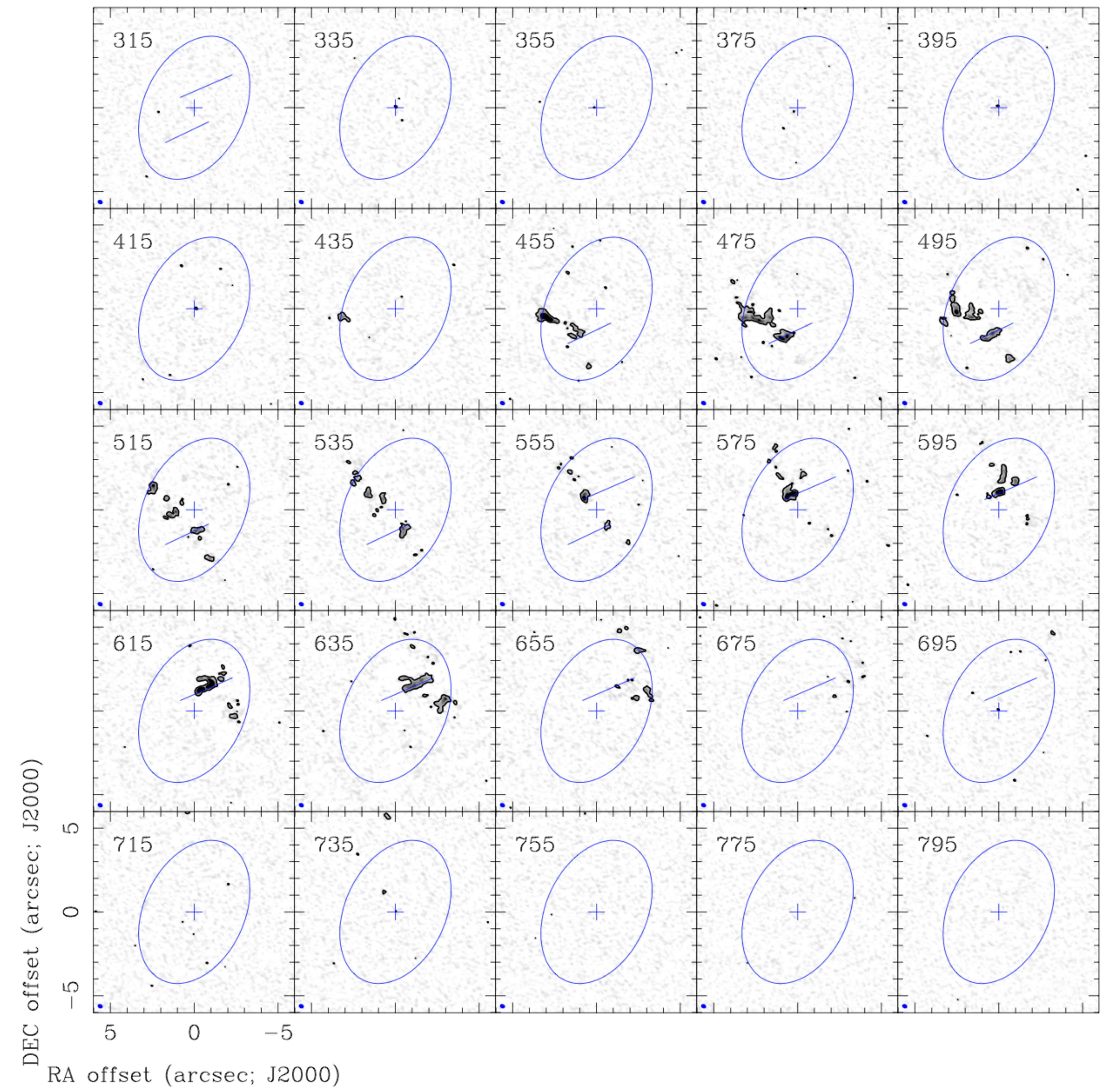}
\caption{{ CO(6--5) channel maps of the CND of Cen~A} in the LSR velocity range V = $315 - 795$\,\kms\ in 20\,\kms\ bins. The size of the maps is 12\arcsec. The velocities are shown in the upper left corner and the synthesized beam at the bottom left corner of each panel.  The rms noise of individual channels is 4.1\,mJy\,beam$^{-1}$.  The contour levels are 5, and 25 $\sigma$.
The cross sign shows the position of the AGN: $RA$ = 13${\rm ^h}$25${\rm ^m}$27.${\rm^s}$615 ; $Dec$ =  --43${\rm ^o}$01$\arcmin$08\farcs805. 
See Fig.~\ref{fig4b} for a description of the symbols representing the main molecular components of the CND of Cen\,A.
\label{fig6}}
\end{figure*}

\begin{figure*}
\begin{center}
\includegraphics[width=14cm]{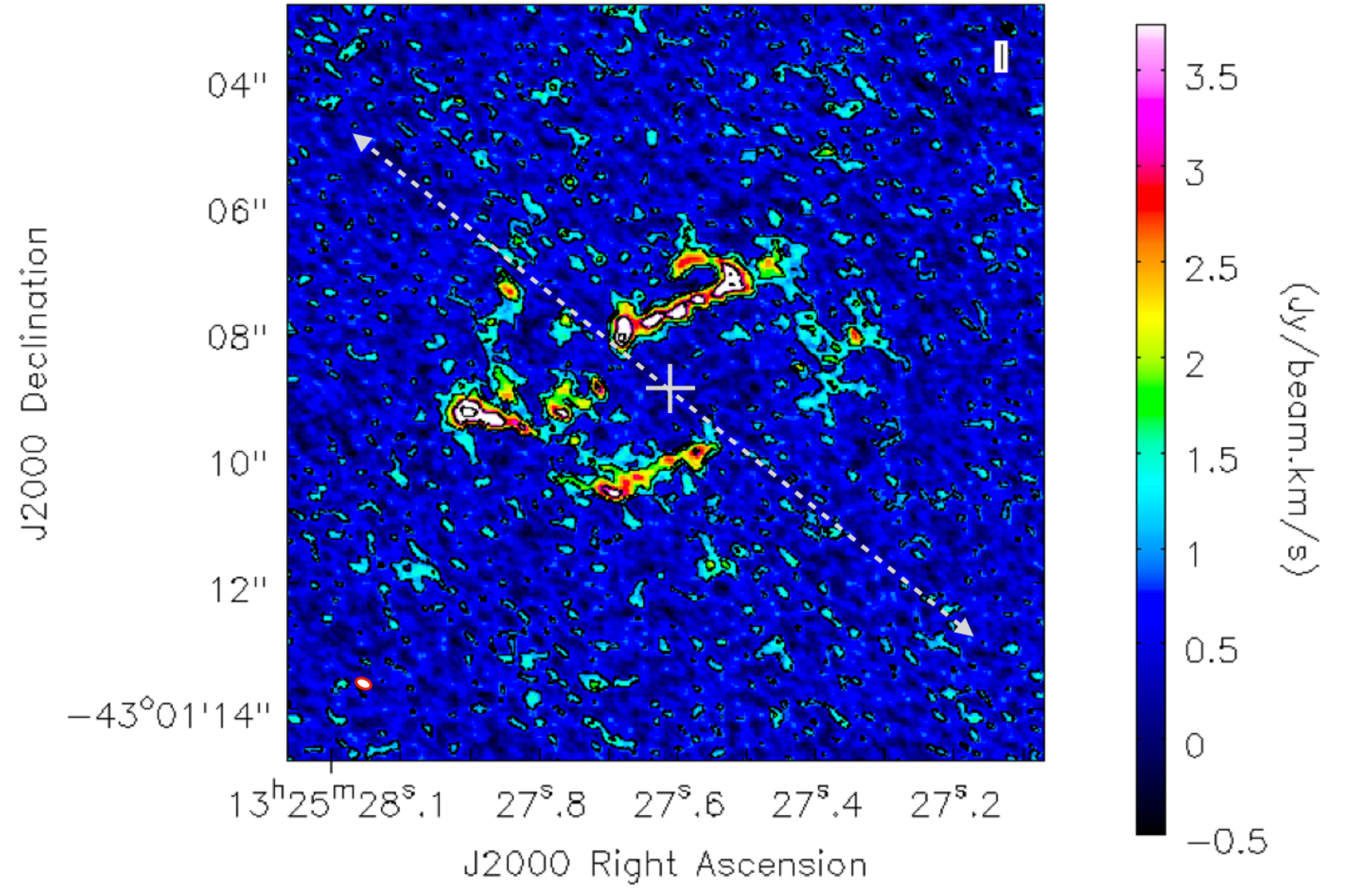}
\end{center}
\caption{CO(6--5) integrated intensity map of the CND of Cen~A, with the color scale ranging from -0.5 to 3.6\,Jy\,beam$^{-1}$\,\kms. The size of the map is 12\arcsec\ and the Half Power Beam Width (HPBW) is 8\arcsec .
Contour levels are at 1.614, 3.2, 5\,Jy\,beam$^{-1}$\,\kms .  The synthesized beam ($0\farcs23 \times 0\farcs17$, PA=64.1\arcdeg) is indicated by an ellipse at the bottom left corner of the plot. The cross sign shows the position of the AGN: $RA$ = 13${\rm ^h}$25${\rm ^m}$27.${\rm^s}$615 ; $Dec$ =  --43${\rm ^o}$01$\arcmin$08\farcs805. The dashed line indicates the jet direction at $PA$ = 51\arcdeg.
\label{fig7}}
\end{figure*}

\begin{figure*}
\includegraphics[width=9cm]{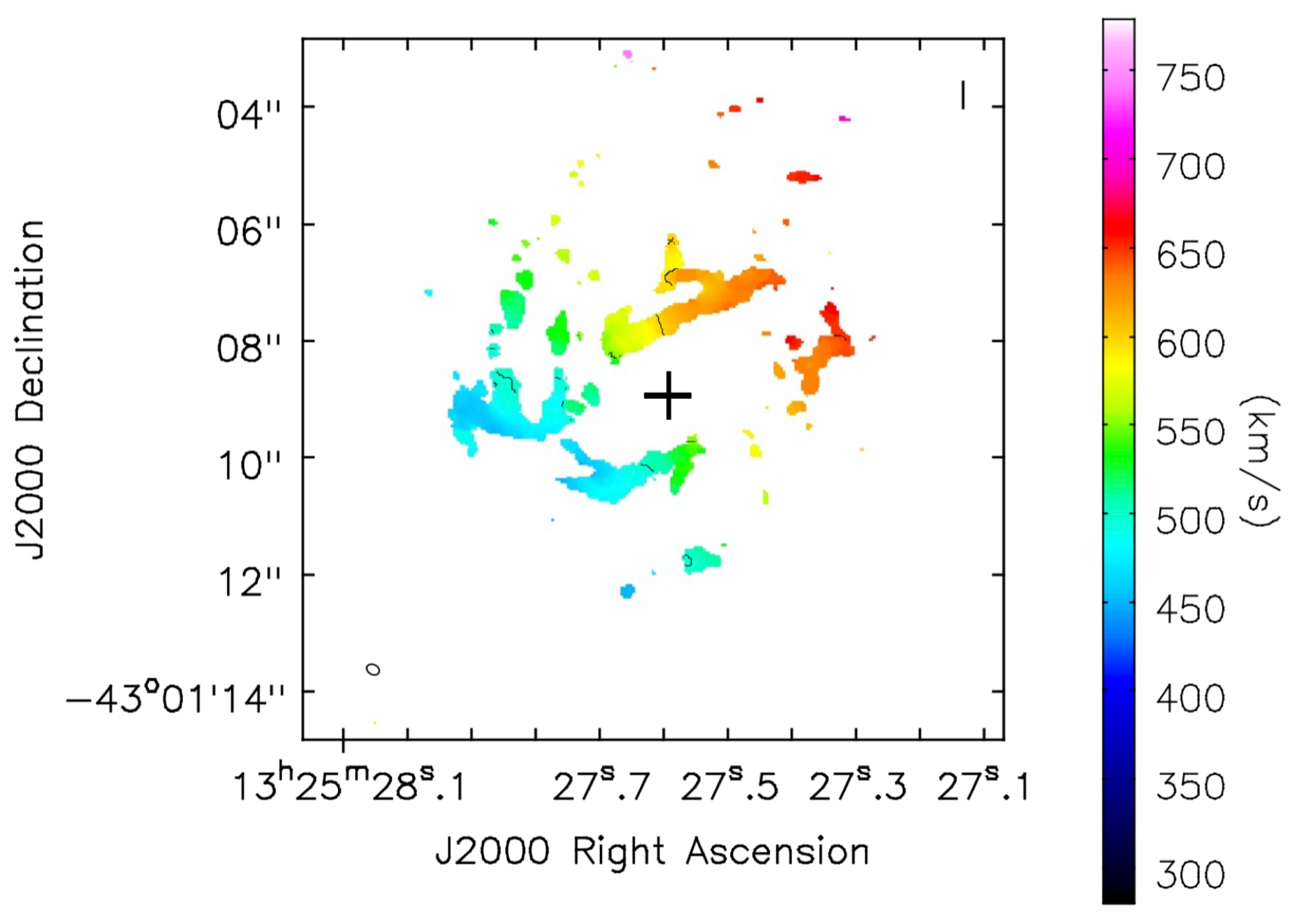}
\includegraphics[width=9cm]{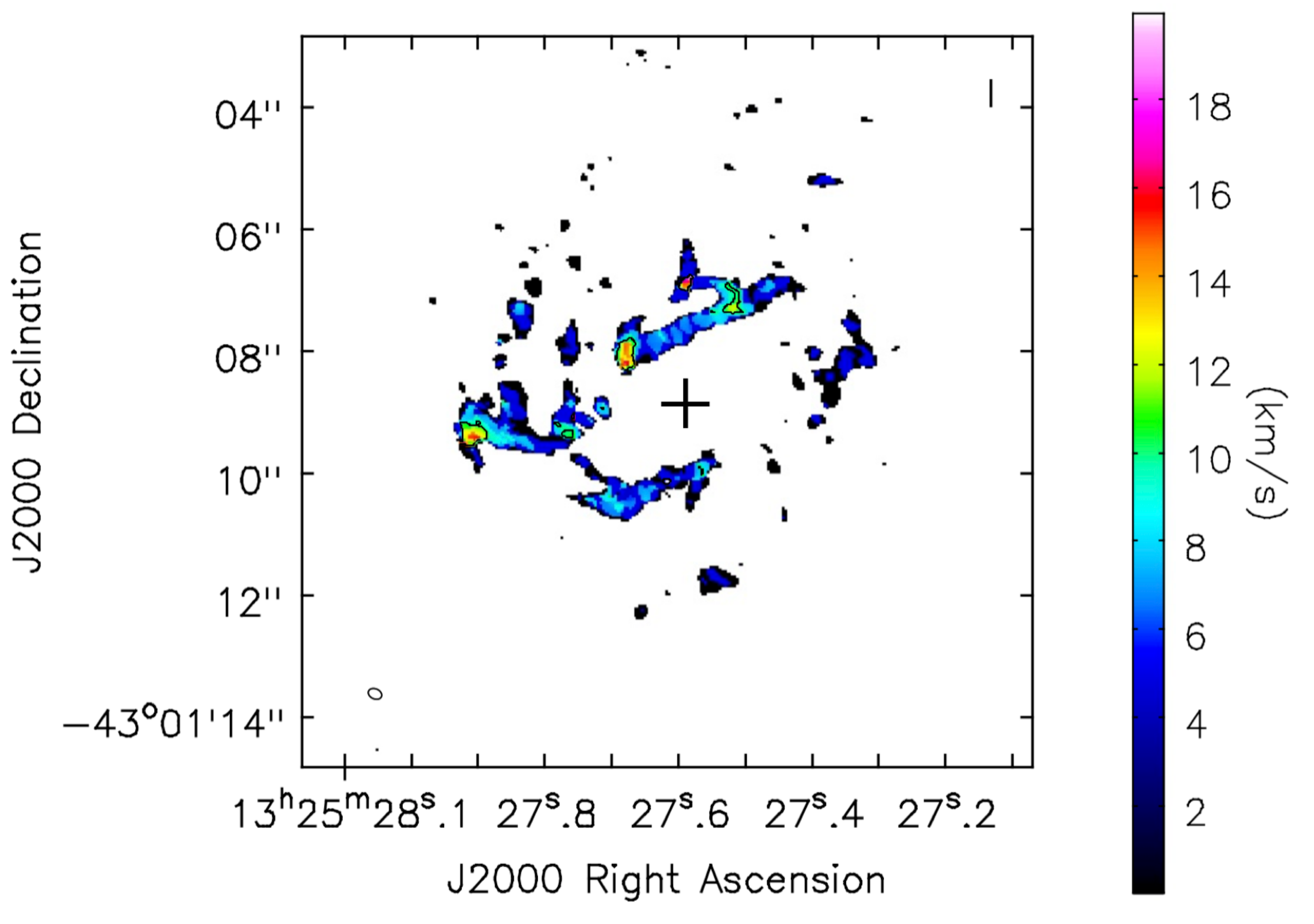}
\caption{ CO(6--5) (intensity weighted) velocity field and velocity dispersion maps of the CND of Cen~A.  In the velocity field map contours are placed every 50\,\kms , from $350 - 700$\,\kms.  The color scale ranges from 280\,\kms\ up to 780\,\kms\ as in the CO(3--2) map in Fig.~\ref{fig5}. In the velocity dispersion map, we place contours at 5\,\kms\ and 15\,\kms, and the color scale ranges from 0 to 20\,\kms. 
The size of the synthesized beam is shown in both panels at the bottom left corner.  
The cross sign shows the position of the AGN: $RA$ = 13${\rm ^h}$25${\rm ^m}$27.${\rm^s}$615 ; $Dec$ =  --43${\rm ^o}$01$\arcmin$08\farcs805. 
\label{fig8}}
\end{figure*}

\begin{figure*}
\begin{center}
\includegraphics[width=18cm]{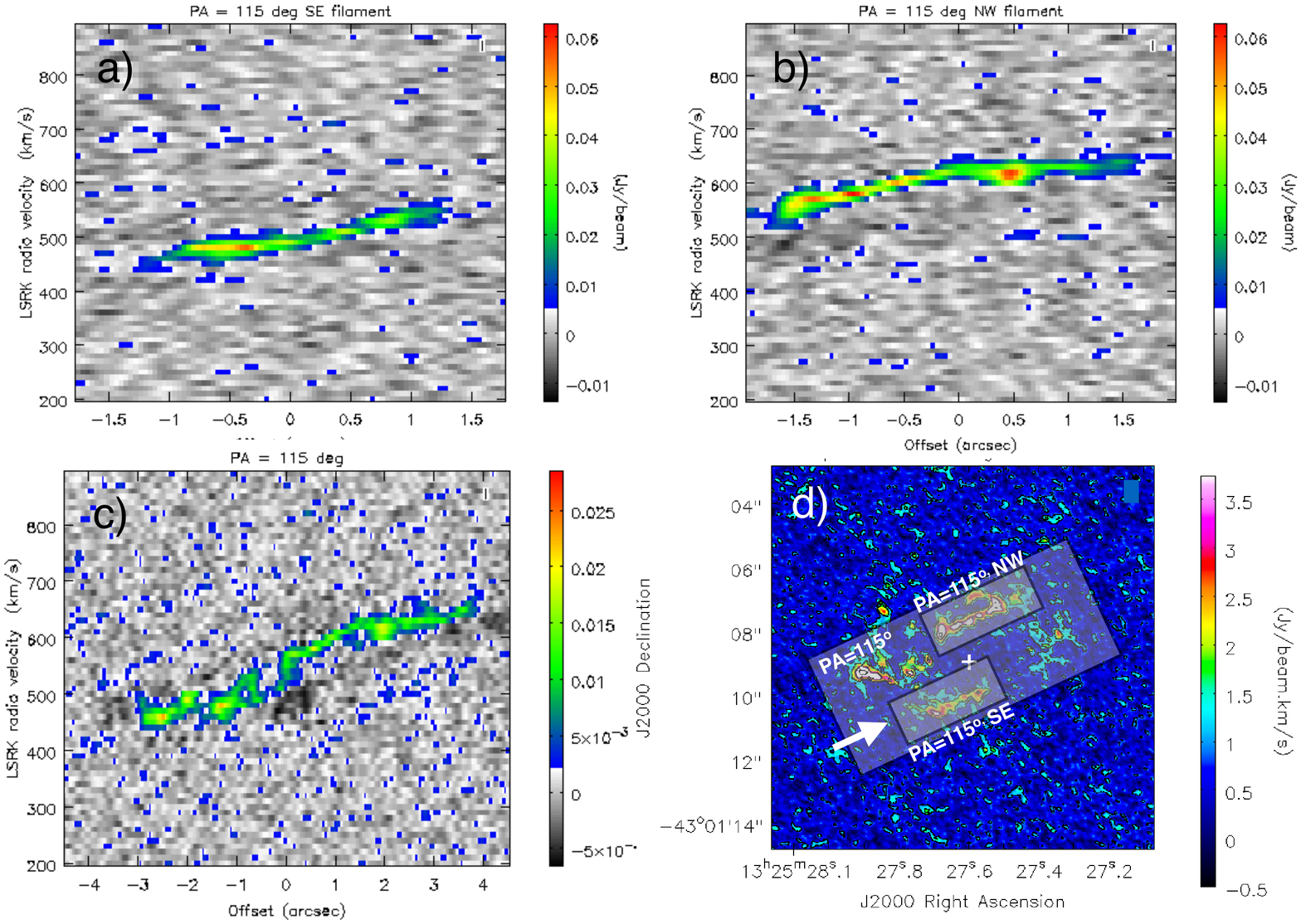}
\end{center}
\caption{  CO(6--5) Position - Velocity (P--V) diagrams: a) $PA$ = 115$^{\rm o}$ along the SE filament with a length of 3\farcs5 and a width of 1\farcs6, b) $PA$ = 115$^{\rm o}$ along the NW filament with a length of 3\farcs7 and a width of 1\farcs9, c) $PA$ = 115$^{\rm o}$, containing the previous two filaments, with a length of 9\arcsec\ and a width of 4\arcsec . Panel d) shows the location of the cuts of the P--V diagrams over the moment 0 CO(6--5) map for reference. The (white) arrow symbol indicates the direction used in the P--V diagrams. 
 \label{fig-pv2}}
\end{figure*}

\begin{figure*}
\begin{center}
\includegraphics[width=10cm]{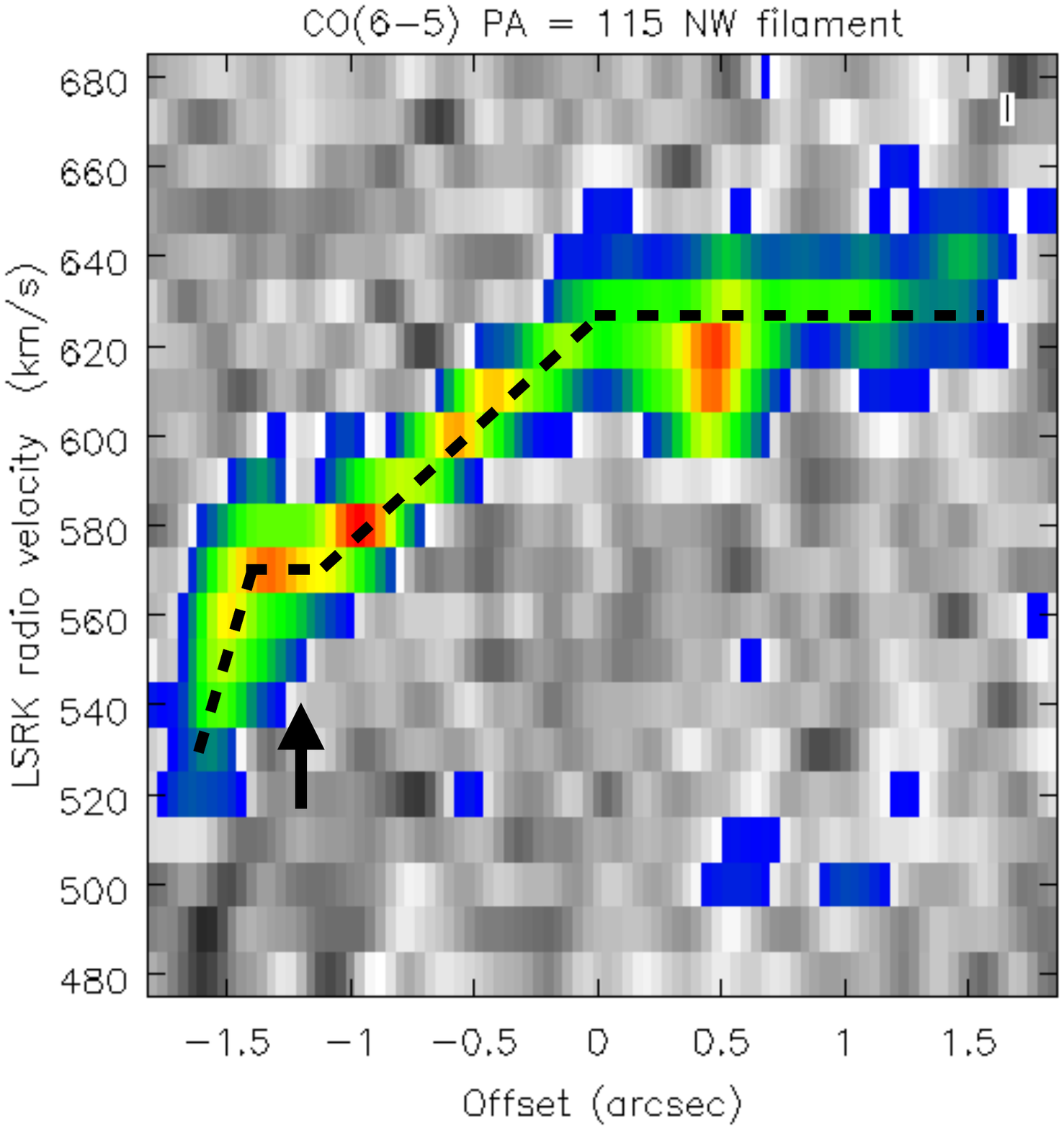}
\end{center}
\caption{ CO(6--5) Position - Velocity (P--V) diagram at $PA$ = 115$^{\rm o}$ along the NW filament with a length of 3\farcs7 and a width of 1\farcs9, as in Fig.~\ref{fig-pv2} b) but with the velocity axis from 480 to 680\,\kms . The dashed lines show the velocity plateau and large velocity gradients along the filament. The (black) arrow symbol indicates the closest location to the AGN ($\sim$1\farcs2, or 22\,pc) and systemic velocity.
\label{fig-pv3}}
\end{figure*}

\begin{figure*}
\begin{center}
\includegraphics[width=9cm]{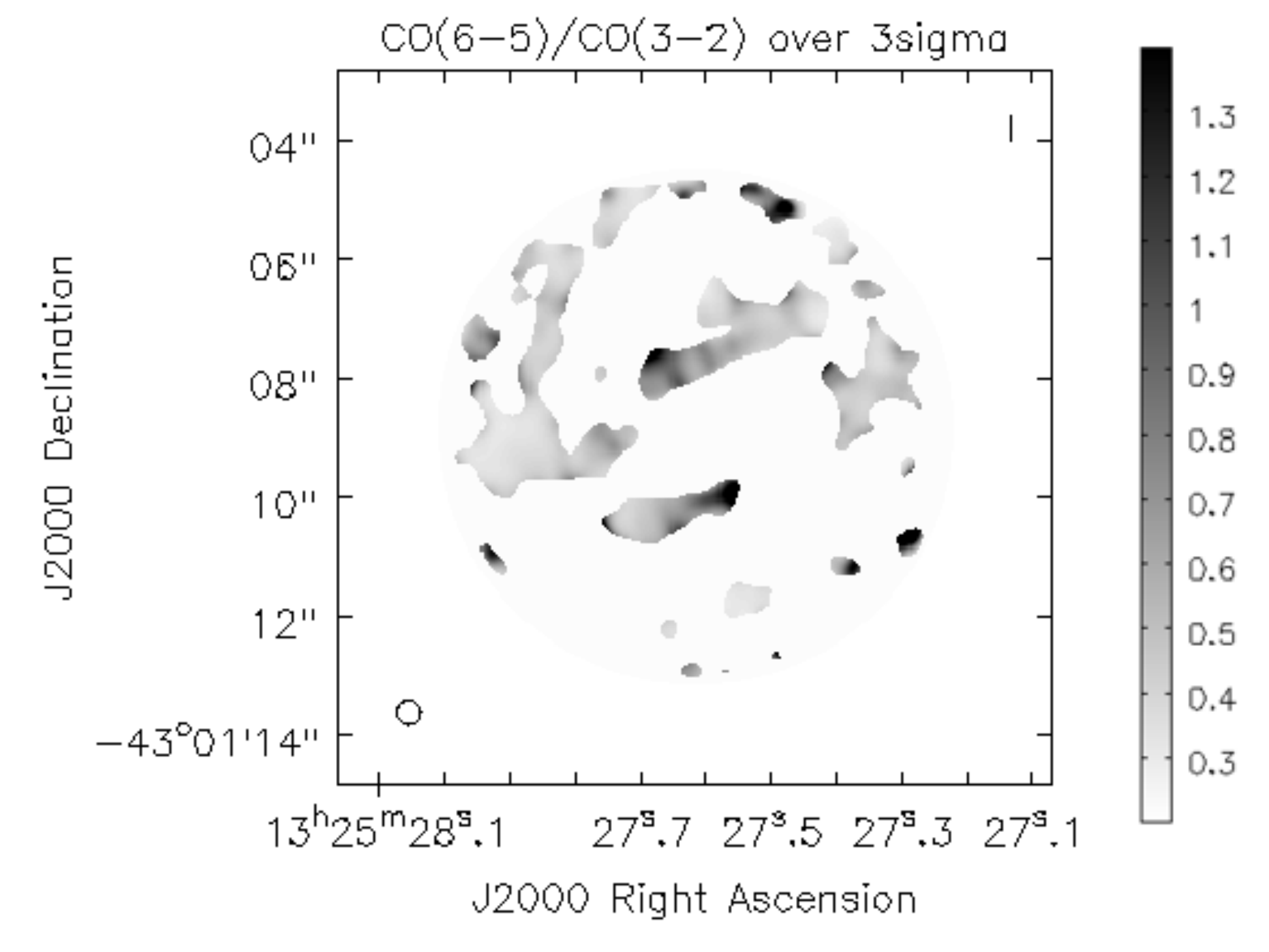}
\end{center}
\caption{ CO(6--5) / CO(3--2) line ratio for values of the CO(6--5) and CO(3--2) maps exceeding 3$\sigma$, with an angular resolution of 0\farcs4 (see bottom left corner).  The grey scale spans from 0.2 to 1.4 in T$_{mb}$ units (or 0.9 to 5.9 in flux units).
\label{lineratio}}
\end{figure*}

\begin{figure*}
\includegraphics[width=16cm]{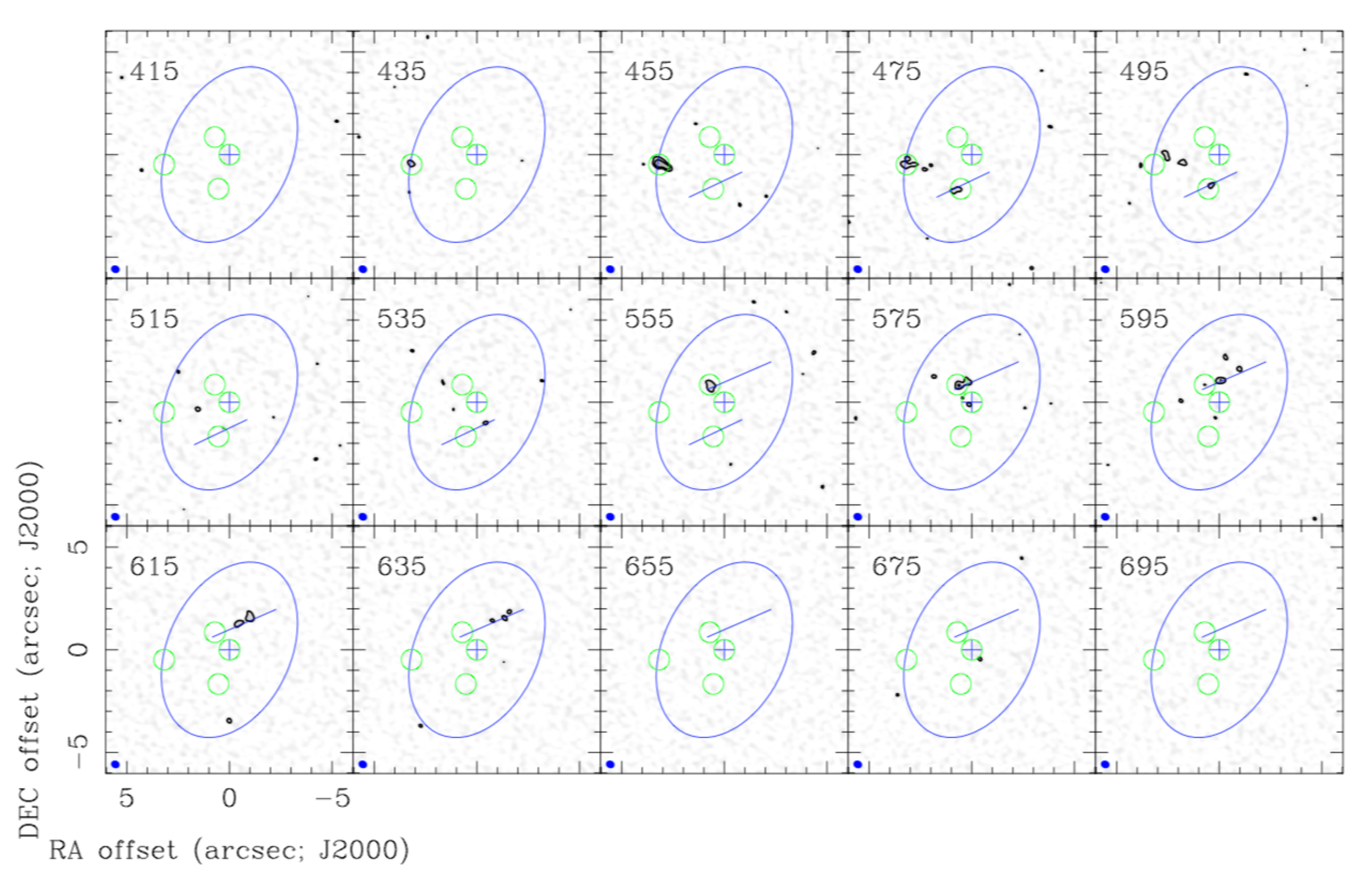}
\caption{{ HCO$^{\rm +}$(4--3) channel maps} of the CND of Cen~A in the LSR velocity range V = $415 - 695$~\kms\ in 20\,km~s$^{-1}$ bins. The size of the maps is 12\arcsec. The velocities are shown in the upper left corner and the synthesized beam at the bottom left corner of each panel. The rms noise of individual channels is 1.3\,mJy\,beam$^{-1}$.
The contour levels are 3.5, and 7$\sigma$.
The cross sign shows the position of the AGN: $RA$\ = 13${\rm ^h}$25${\rm ^m}$27.${\rm^s}$615 ; $Dec$ =  --43${\rm ^o}$01$\arcmin$08\farcs805. 
 The (green) circles show the regions used to calculate the HCO$^+$/HCN(4-3) ratio (\S~\ref{hcohcn}) and spectra are presented in Fig.~\ref{fig20}.
 See Fig.~\ref{fig4b} for a description of the symbols representing the main molecular components of the CND of Cen\,A.
\label{fig9}}
\end{figure*}

\begin{figure*}
\includegraphics[width=16cm]{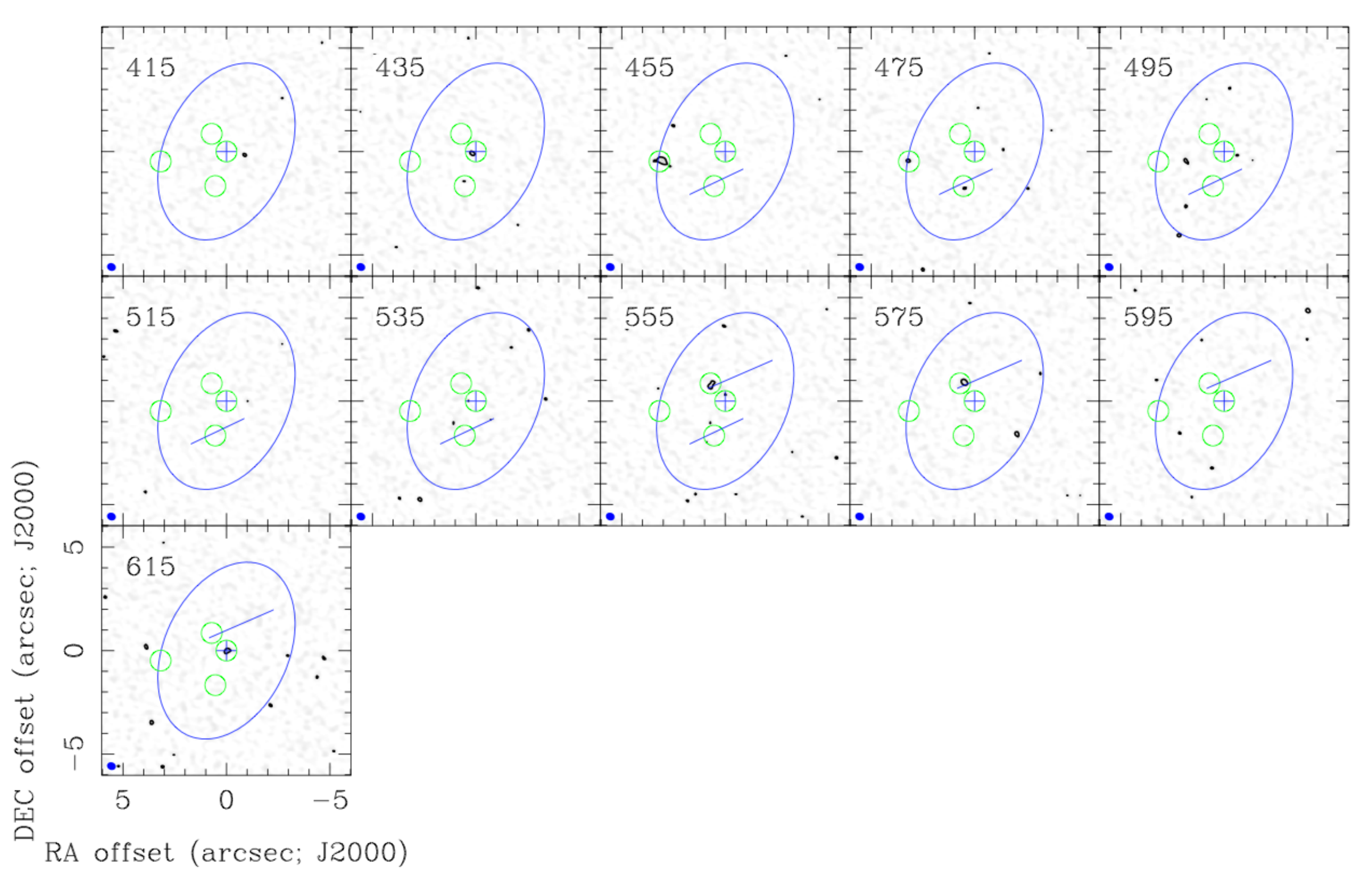}
\caption{{HCN(4--3) channel maps of the CND of Cen~A} in the LSR velocity range V = $415 - 615$\,\kms\ in 20\,km\,s$^{-1}$ bins. The size of the maps is 12\arcsec.  The velocities are shown in the upper left corner and the synthesized beam at the bottom left corner of each panel. The rms noise of individual channels is 1.3\,mJy\,beam$^{-1}$.  
The contour levels are 3.5, and 7$\sigma$.
The cross sign shows the position of the AGN: $RA$\ = 13${\rm ^h}$25${\rm ^m}$27.${\rm^s}$615 ; $Dec$ =  --43${\rm ^o}$01$\arcmin$08\farcs805 . 
 The (green) circles show the regions used to calculate the HCO$^+$/HCN(4-3) ratio (\S~\ref{hcohcn}) and spectra are presented in Fig.~\ref{fig20}.
 See Fig.~\ref{fig4b} for a description of the symbols representing the main molecular components of the CND of Cen\,A.
\label{fig10}}
\end{figure*}

\begin{figure*}
\begin{center}
\includegraphics[width=8cm]{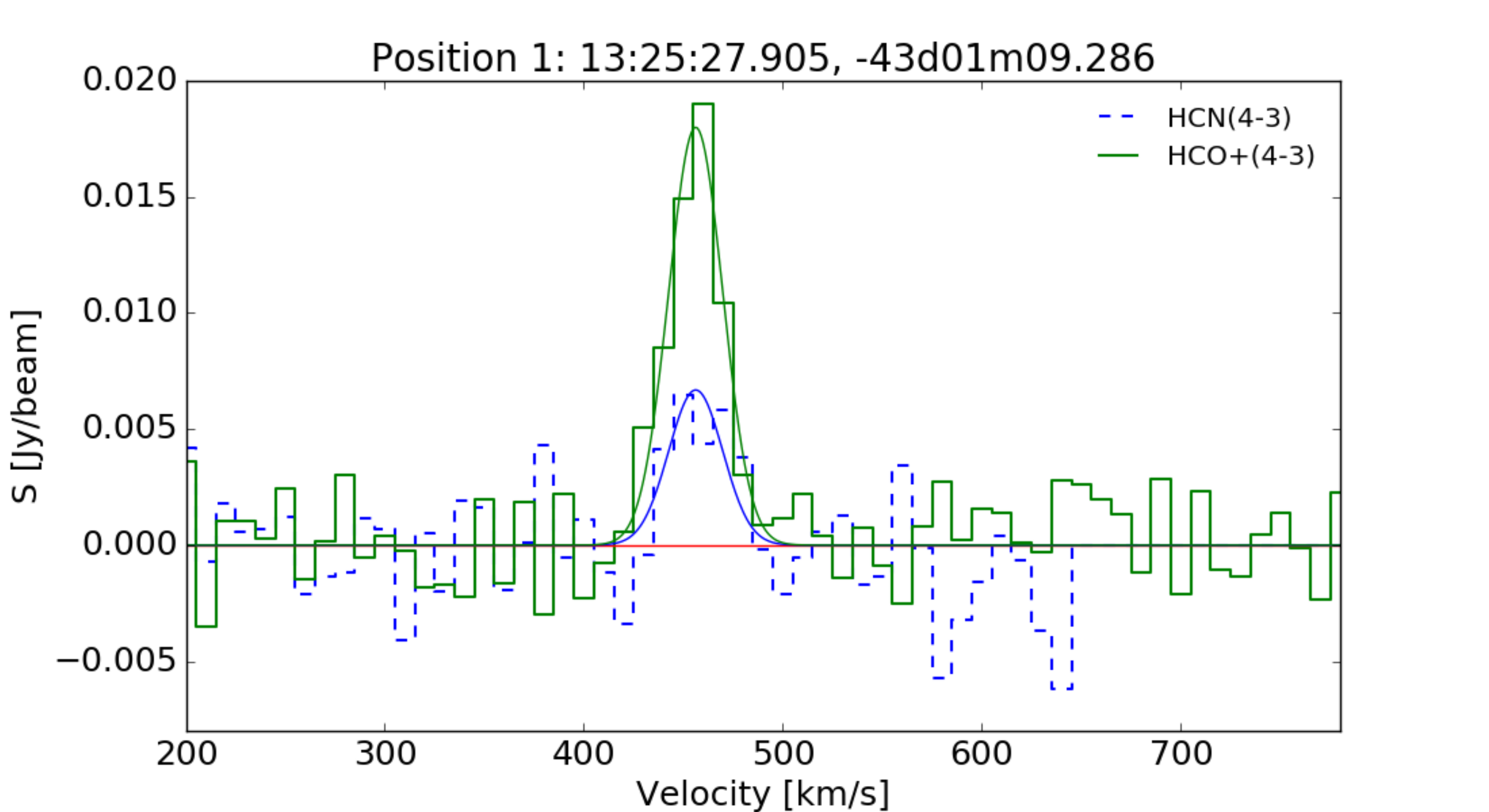}
\includegraphics[width=8cm]{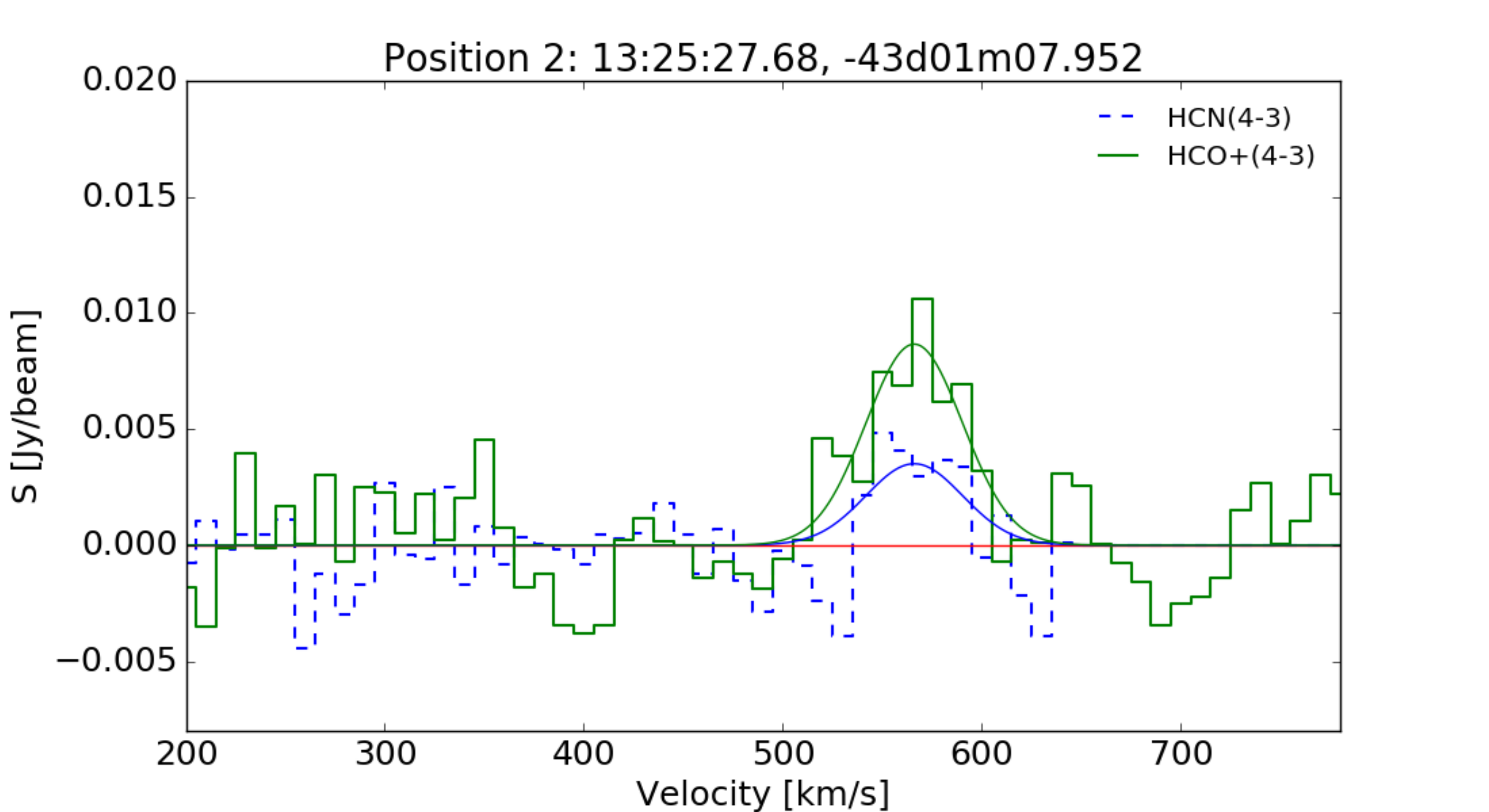}\\
\includegraphics[width=8cm]{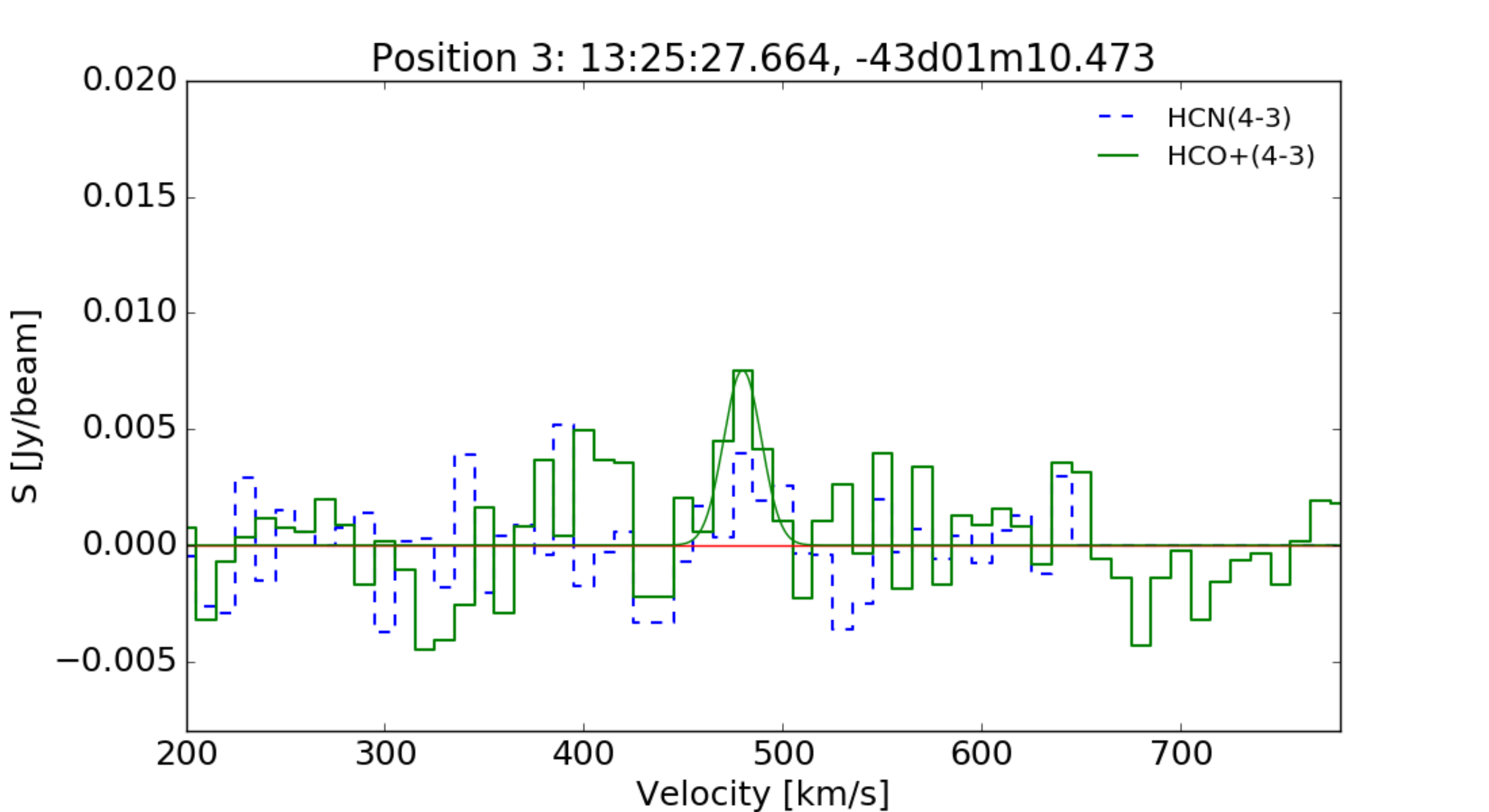}
\includegraphics[width=8cm]{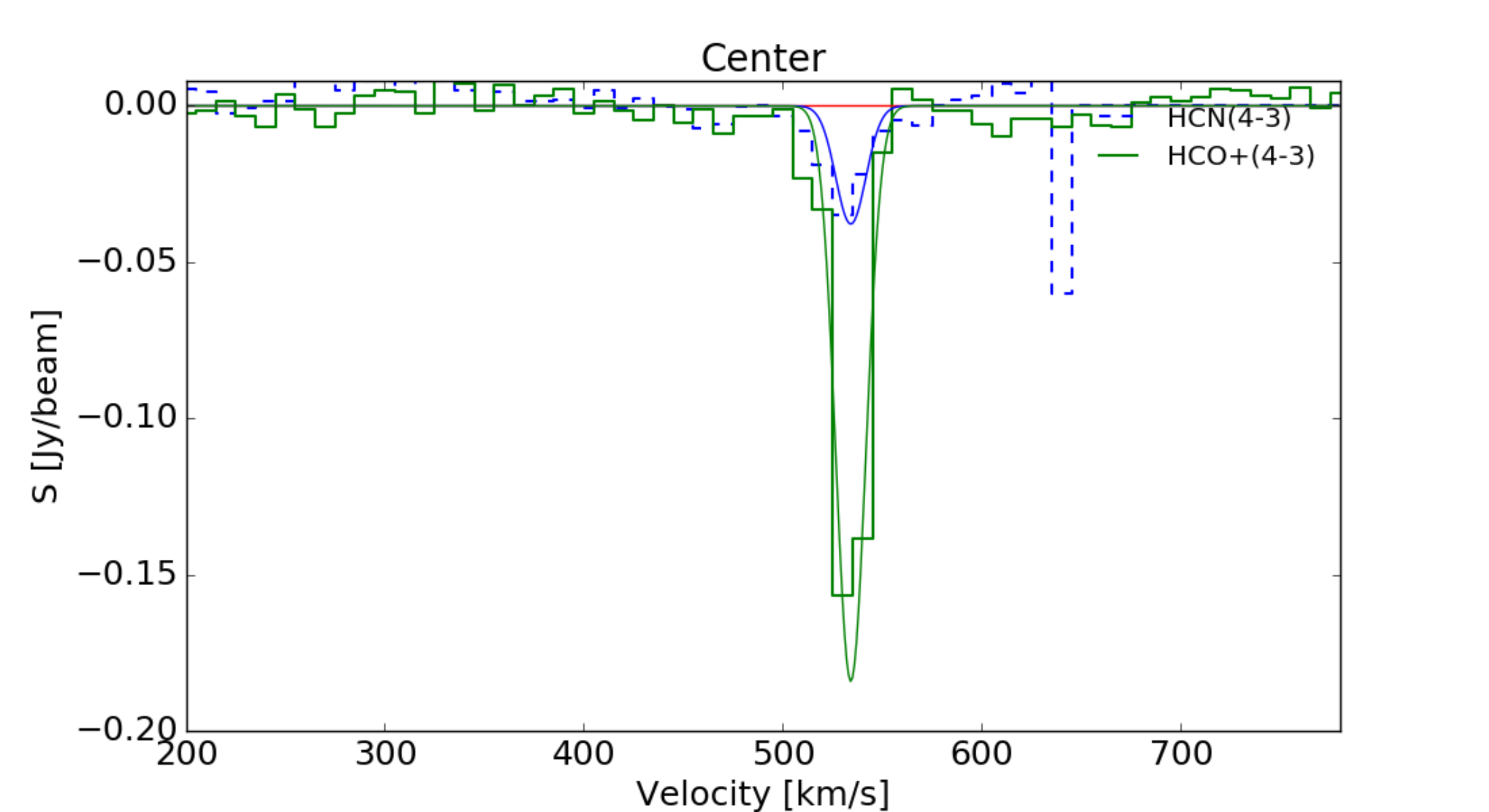}
\end{center}
\caption{HCN and HCO$^+$(4--3) spectra (blue dashed and green solid lines, respectively) towards: Position 1) $RA$\ = 13${\rm ^h}$25${\rm ^m}$27.${\rm^s}$905, $Dec$ = -43${\rm ^o}$01$\arcmin$09\farcs286, Position 2) $RA$\ = 13${\rm ^h}$25${\rm ^m}$27${\rm^s}$68, $Dec$ = -43${\rm ^o}$01$\arcmin$07\farcs952, Position 3) $RA$\ = 13${\rm ^h}$25${\rm ^m}$27${\rm^s}$.664, $Dec$ = -43${\rm ^o}$01$\arcmin$10\farcs473, and Position 4) center, at $RA$\ = 13${\rm ^h}$25${\rm ^m}$27.${\rm^s}$615 ; $Dec$ =  -43${\rm ^o}$01$\arcmin$08\farcs805 (see Figs.~\ref{fig9} and \ref{fig10}). The y-axis of the spectra are shown in units of Jy/beam from -0.008 to 0.02, except for the spectra towards the nucleus which range from -0.2 to 0.008, and the x-axis  from 200 to 780 km\,s$^{-1}$.  Gaussian fits to the profiles are also shown, and line parameters are displayed in Table\,\ref{table4}.  \label{fig20}}
\end{figure*}

Based on all the maps for the different transitions presented in \S.\,\ref{result}, we enumerate next all the distinct and main molecular gas components that we can discern from large to small scales as we go closer to the center of Cen~A. Fig.~\ref{fig4b} shows a simplified scheme that illustrates these main molecular components and the definitions we use in this Paper. A summary of the properties of each component is provided in Table\,\ref{table2} and described next.

 1) CND: The major and minor axes of  the CND are confirmed to be $20\arcsec \times 10\arcsec$, or 360\,pc $\times$ 180\,pc in linear scale (without taking into account any projection effect). The position angle is $PA$ $=155\arcdeg$. This is well in agreement with the CND size and orientation provided by \citet{2009ApJ...695..116E}, considering the coarser angular resolution of 6\arcsec . The inclination of the disk was estimated to be $i$ $\simeq$ 70\arcdeg , but with our new values for the major and minor axes the inclination assuming a circular disk is slightly smaller, $i$ $\simeq$ 60\arcdeg .  It is composed of multiple filamentary and clumpy structures, possibly arm-like features or streamers.
In Fig.~\ref{fig4b} we indicated as black lines most of the filaments that can be found within the CND
as seen from our maps.
The velocity width of the CND is $\Delta V$ $\simeq$ 475 $\pm$ 7\,\kms\ (FWZI), without correcting for inclination. With the orientation and the kinematics of the CND, the several 10--100\,pc scale filamentary structures within the CND are likely trailing.
In the external parts of this CND ($r$ $>$ 8\arcsec, or 144\,pc, from the center) the molecular gas has large velocity widths of 10\,\kms\ and up to 40\,\kms . These regions correspond to the high velocity wings of the CO(3--2) spectra. 
A double-peaked distribution is present both in CO(3--2) and CO(6--5) spectra, although it is clearer in the latter.

2) Nuclear ring: Deeper inside the CND there is a ring-like feature with a major and minor axis of 9\arcsec\  $\times$ 6\arcsec\ (162\,pc $\times$ 108\,pc ), and also at $PA$ = 155\arcdeg , which is formed by multiple filaments or streamers leading to it.  If the ring structure is coplanar and has a circular shape, then the inclination would be $i$ $\simeq$ 50\arcdeg. This ring-like structure has a velocity width of $\sim$ 260\,\kms .
This component is detected in CO(3--2) as well as partially in CO(6--5).

3) Nuclear filaments: There are two nearly parallel filamentary structures to the SE and NW of the AGN of about 2\arcsec\ in length (or $\sim$ 40\,pc)  contained within the nuclear ring-like structure, with $PA$ $\simeq$ 120\arcdeg , and with a rotational symmetry of 180\arcdeg\ around the AGN.
 These components are most prominent in CO(3--2), CO(6--5), and HCO$^+$(4--3),  and just partially detected in HCN(4--3).
The NW filament is aligned from $RA$, $Dec$ [J2000] = 13$^{\rm h}$25\arcmin27\farcs692 , -43\arcdeg01\arcmin08\farcs19 to 13$^{\rm h}$25\arcmin27\farcs407, -43\arcdeg01\arcmin06\farcs833 , and from $V$ = 555 to 695\,\kms,
and the SE filament from $RA$ , $Dec$ = 13$^{\rm h}$25\arcmin27\farcs536 , -43\arcdeg01\arcmin09\farcs65 to 13$^{\rm h}$25:27\farcs772, -43\arcdeg01\arcmin10\farcs882, and from $V$ = 455 to 555\,\kms. There is an increase in velocity of $\sim$50\,\kms\ with respect to that of the CND overall. There is an asymmetry between these two filaments. The filament to the NW is a factor of two brighter than the component to the SE. 
Although both filamentary structures are symmetric with respect to the center of the galaxy, the SE nuclear filament is curved towards the N as it extends away from the AGN and connects it to the nuclear ring structure with 9\arcsec\ diameter. Despite the asymmetry, if we add all the flux from the component to the N and W of the AGN and to the S and E in the CO(6--5) map for example, we find that the total fluxes are quite equal: 186\,Jy\,km\,s$^{-1}$ versus 180\,Jy\,km\,s$^{-1}$. 
There are well defined molecular clumps in these two structures. In the transitions reported in this paper one can see around 4-6 clumps on each side.

The connecting point between the SE nuclear filament and ring is the brightest in all probed transitions of all the components presented in this Paper, and has also a large velocity dispersion of $\sim$ 20\,\kms. 
A similar connecting point, but with slightly slower dispersion and not as prominent as the previous one, can be found at the opposite side (i.e. to the W).
These connecting points are also linked to additional filaments that are nearly perpendicular to the nuclear filaments and form part of the nuclear ring.

4) Nuclear disk:
In \S\,\ref{introduction} we introduced the few tens of parsec scale sized nuclear disk, well inside the area covered by the nuclear filaments, which contains ionized and molecular gas that presumably is the fuel feeding the nuclear massive object.  While ionized gas shows distributions elongated along the jet and is likely related to it, very warm (typically 1000 -- 2000\,K) molecular hydrogen as traced by H$_2$(J=1--0) S(1) (2.122 $\mu$m) using VLT/SINFONI in
  the inner 3$\arcsec$ ($\sim$50 pc) follows a rotating nuclear disk  
\citep{2007ApJ...671.1329N}, although the distribution shows that this disk might be quite irregular. Part of the gas in the nuclear H$_2$ disk may have been impacted by the jet \citep{2013ApJ...766...36B} and excited by shocks \citep{2017A&A...599A..53I}.

\begin{figure*}
\includegraphics[width=16cm]{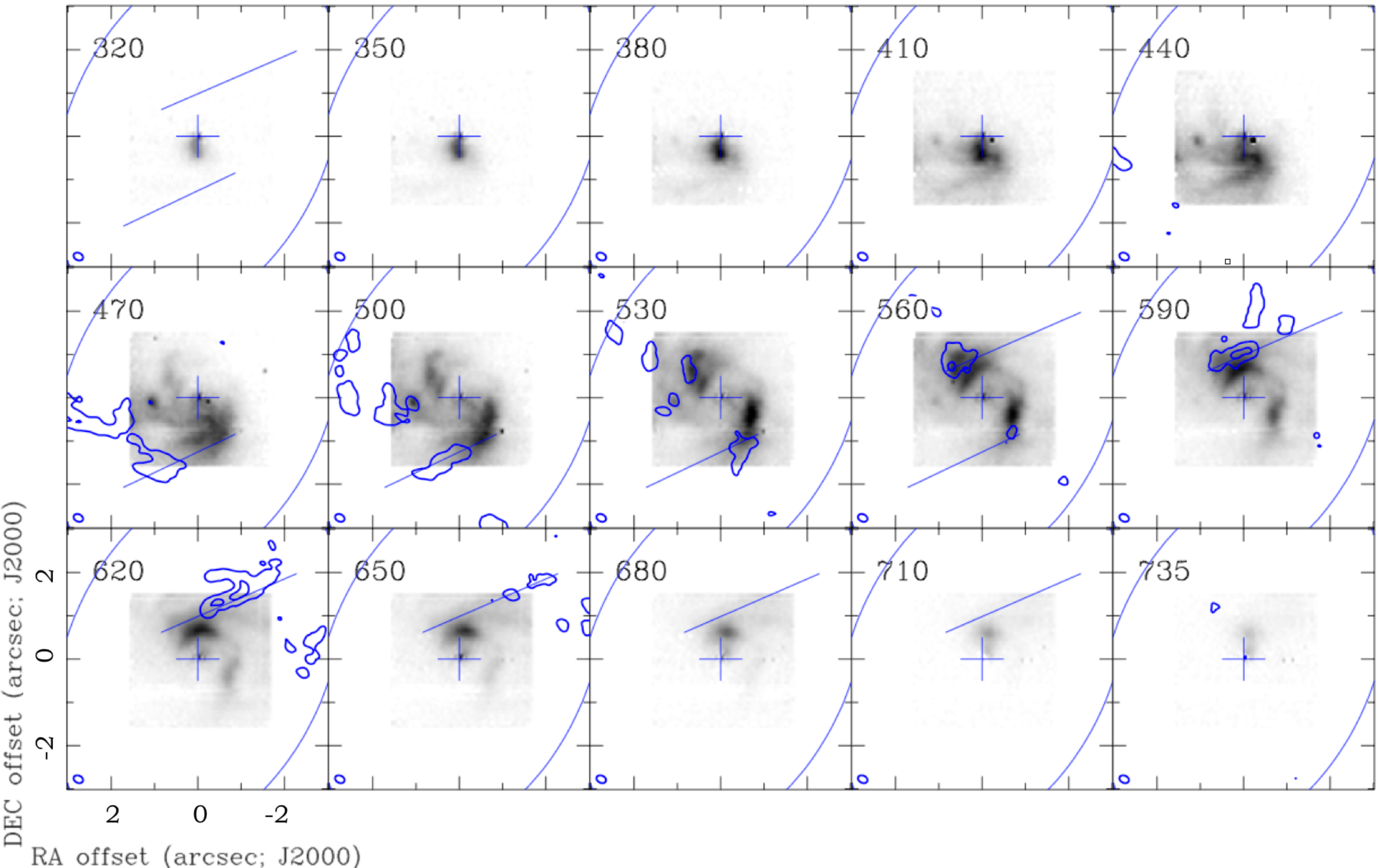}
\caption{Comparison of the channel maps of the H$_2$(J=1--0) S(1) (grey scale) and CO(6--5) (contours at 3.5 and 7$\sigma$) lines in the inner 6\arcsec\ of the CND of Cen~A in the LSR velocity range V = $320 - 740$\,\kms\ in 30\,km\,s$^{-1}$ bins. The H$_2$ map covers the inner 3\arcsec\ and the resolution of the H$_2$ map is 0.12\arcsec\ \citep{2007ApJ...671.1329N} . The velocities are shown in the upper left corner and the synthesized beam of the CO(6--5) observations at the bottom left corner of each panel.  
The cross sign shows the position of the AGN: $RA$\ = 13${\rm ^h}$25${\rm ^m}$27.${\rm^s}$615 ; $Dec$ =  --43${\rm ^o}$01$\arcmin$08\farcs805. 
 See Fig.~\ref{fig4b} for a description of the symbols representing the main molecular components of the CND of Cen\,A.
\label{fig21}}
\end{figure*}

 Fig.\,\ref{fig21} shows the comparison of channel maps of the CO(6--5) and the H$_2$ lines. 
Although most of the warm and dense gas close to the nucleus ($r$ $<$ 20\,pc) is not detected as traced by the transitions investigated in this Paper, we observe that the endings of the two nuclear filaments and a third weaker component approaching the nucleus from the E of the AGN (clearly seen in CO(3--2) and CO(6--5) lower contours in the channel maps) are counterparts of the molecular gas traced by the H$_2$ line within the nuclear disk. 
The CO filaments abruptly end in the probed transitions at $\sim$ 20\,pc from the AGN, but the H$_2$ maps show that these continue in a warmer gas phase, probably shock excited, and then wind up in the form of nuclear spiral arms (see also Fig.\,\ref{fig-1}b). 
Although with sizes a factor of 10 larger, these filaments/arms are reminiscent of the Galactic Center, where molecular and ionized components (circumnuclear disk and mini spirals) are seen within the inner few pc of Sgr\,A* \citep[e.g.][]{1983A&A...122..143E,1983Natur.306..647L,2009ApJ...699..186Z,2012A&A...539A..29M,2016PASJ...68L...7T}.

Overall these nuclear spiral features within the nuclear disk of Cen~A create a distribution that remind us of another structure that resembles a ring-like feature, but this time at $r$ = 0.5\arcsec\ (10\,pc). 
Further inside this 10\,pc ring-like feature, two additional 180\arcdeg\ rotationally symmetric regions are found along the N-S direction of about 10\,pc in length and with even larger velocity dispersions ($\sim$ 200\,\kms\ from H$_2$) than anywhere else within the CND as probed in our CO maps.

\subsection{Molecular Gas Mass}
\label{sub:moleculargasmass}

We calculate in this subsection the mass of the different molecular gas components found in our maps.
 We use a conversion factor between integrated CO intensity and H$_2$ column density $X$
  $= N_{\rm H_2} /I _{\rm CO}$ = 4 $\times$ 10$^{20}$~cm$^{-2}$~(K~km~s$^{-1}$)$^{-1}$ for the CND \citep{2014A&A...562A..96I}, with an uncertainty of a factor of two.  This value is a factor of 10 larger than that observed in other nuclear regions of galaxies
  \citep{1988ApJ...325..389M,1995ApJ...448L..97W,1996A&A...305..421M,2001A&A...365..571W} and that previously used by \citet{2009ApJ...695..116E} to calculate the molecular gas mass of the CND. 
The masses derived here should be rescaled if a better factor for Cen~A is obtained.
We also use a line ratio between $^{12}$CO(1--0) and (3--2) fluxes of 0.1 over the whole CND, again following \citet{2014A&A...562A..96I}. 

The gas mass obtained from the CO(3--2) map is $M_{\rm gas}$ $\simeq 4.8 \times 10^7$~M$_\odot$. 
Flux loss was measured in \S\,\ref{subsect:co3-2emissionline} to be { $\sim$50\%, so the final estimate is $M_{\rm gas}$ $\simeq 9 \times 10^7$~M$_\odot$.}
This is consistent with the measurement $M_{\rm gas}$ $\simeq 8 \times 10^7$~M$_\odot$ 
inside a projected distance of $r$ $<$ 200 pc (12\arcsec) in \citet{2009ApJ...695..116E} if we apply the X factor given by \citet{2014A&A...562A..96I}.
This is also in agreement with the mass of the circumnuclear gas using single dish measurements for many CO transitions and an LVG analysis, which yielded $M_{\rm gas}$ =  8.4 $\times$ 10$^7$ M$_\odot$, also assuming a 35\% mass contribution by helium \citep{2014A&A...562A..96I}. 

Table~\ref{table2} exhibits the derived main parameters of the circumnuclear gas disk, nuclear ring, and nuclear filaments (NE and SW) in CO(3--2) and CO(6--5). These parameters include peak flux densities, velocity ranges, total CO(3--2) and CO(6--5) fluxes and the corresponding molecular gas masses by using the CO(3--2) fluxes.

\section{Warped disk or Non-Circular Motions? Mechanisms Feeding the AGN from Kiloparsec to Parsec Scales}
\label{secfeeding}

\subsection{A Warped Disk?}

In the past a warped and thin disk model had been used to reproduce the observations both at a large scale \citep[e.g.][]{2006ApJ...645.1092Q} as well as at a few parsec scale using the H$_2$ line \citep{2007ApJ...671.1329N}.  \citet{2007ApJ...671.1329N} show by applying the Kinemetry analysis \citep{2006MNRAS.366..787K} on the velocity field map of the nuclear disk that it is characterized by a mean inclination angle of $i$ = 45$\arcdeg$ and a $PA$ = 155$\arcdeg$
  assuming a warped disk model to describe its gas kinematics.
  \citet{2006ApJ...645.1092Q} used the warped disk model to reproduce the morphology of the parallelogram feature seen in mid-IR (Spitzer), on top of previous studies.  
 \citet{2010PASA...27..396Q} compiled inclinations and position angles from the literature and identified that there might be (assuming that a warped disk model applies) three major changes of inclination and $PA$ at about 1.3\,kpc (1\farcm2), at about 600\,pc (33\arcsec), and a third kink at a radius of about 100\,pc (5\arcsec).
 The field of view and angular resolution in the ALMA observations fill the missing gap from tens of parsec to the $r$ = 200\,pc scale within the CND. The existence of the ring and the nuclear filaments are likely related to the third kink.

We also fitted the CO(3--2) velocity field using Kinemetry. The Kinemetry method performs harmonic expansion of 2D maps such as surface brightness, velocity or velocity dispersion, along the best fitting ellipses, in order to detect morphological and kinematic components. We use the fitting over the line of sight velocity distribution. If we assume circular orbits in a warped disk, then one can extract the inclination and position angle of each ring as a function of radius. We follow this approach to fulfill the following three aims: 1) identify general properties of the circumnuclear gas assuming a warped disk model is valid, 2) compare the results with previous fits using the same assumption but for other linear scales, and 3) quantify deviations from this assumption and identify regions that are significantly different.

\begin{figure*}
\begin{center}
\includegraphics[width=10cm]{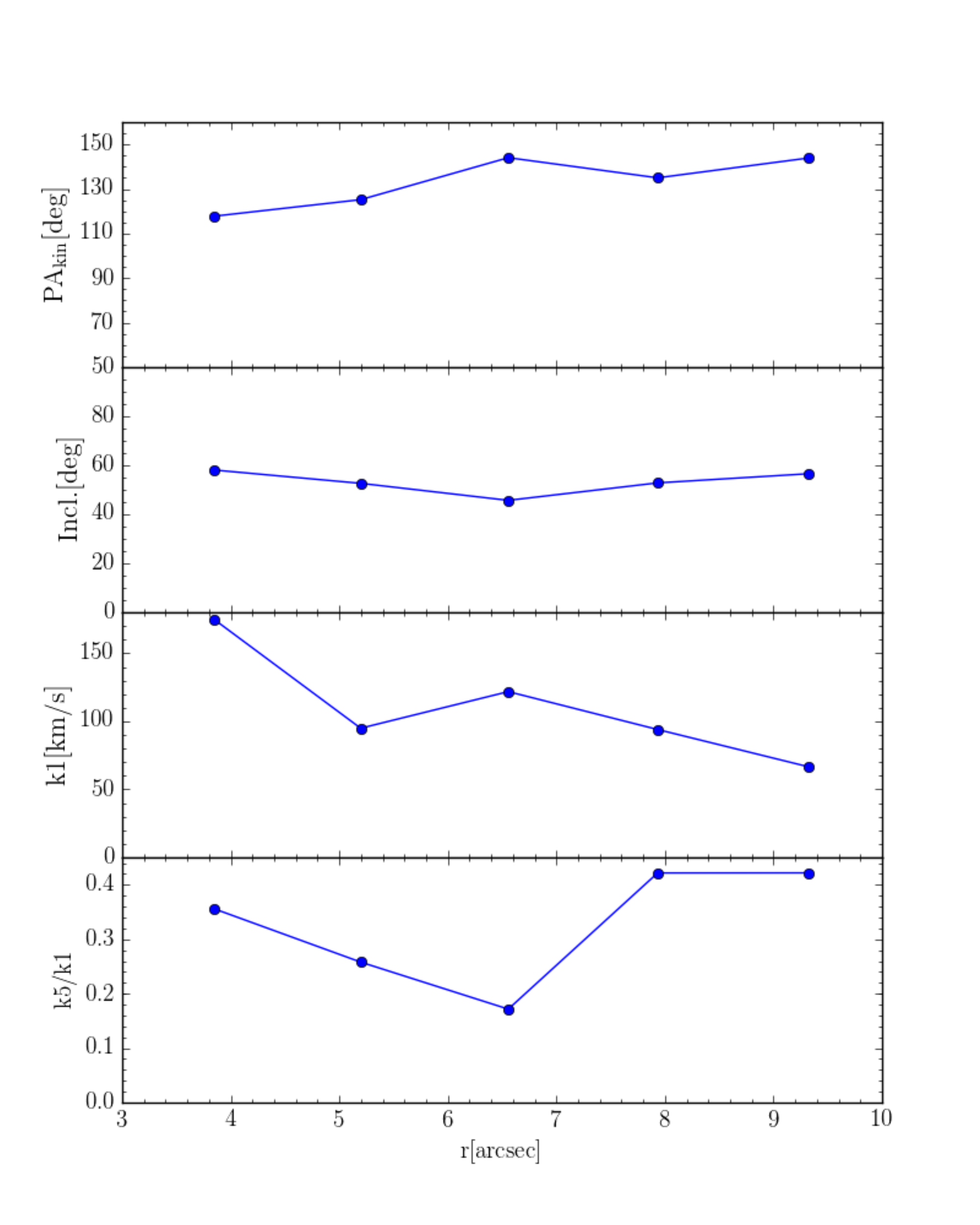}
\caption{Kinemetry analysis \citep{2006MNRAS.366..787K} using the ALMA CO(3--2) velocity field with bins of 1\farcs2: a) the kinematic PA, b) the inclination (calculated from $q$, the axial ratio or flattening of the ellipse, assuming circular orbits), c) $k1$ the amplitude of bulk velocities, and d) $k5/k1$, the ratio between harmonics $k1$ and $k5$, where $k5$ represents deviations from simple rotation. See Table~\ref{table3} for details.
\label{fig15}}
\end{center}
\end{figure*}

Several kinematic parameter profiles as a function of radius were obtained from the CO(3--2) velocity field at a scale of 1\farcs2 (Fig.~\ref{fig15} and Table~\ref{table3}): a) the kinematic $PA$, or orientation of the maximum velocity, b) the inclination (assuming circular orbits) obtained from  the axial ratio $q$ (i.e. flattening of the ellipse, where $q =$ cos$(i)$), c) $k1$, the amplitude of bulk motions (rotation curve), and d) $k5/k1$, the ratio between harmonics $k1$ and $k5$, where $k5$ is the higher-order term which is not fitted, and represents deviations from simple rotation and points to complex kinematical components.
The plots show the trends in these parameters up to a radius of 10\arcsec.

A ring width of 1\farcs2 was used and the following trends are present in the results. Note that other widths were also used showing similar results.
In the plots values smaller than 3\arcsec\ are uncertain due to a lack of CO(3--2) emission there and are not presented. 
Within the inner 5\arcsec\ the distribution and kinematics appear peculiar. Overall the mean kinematic $PA$ is 145\arcdeg, with values close to 120\arcdeg\ up to 5\arcsec , likely due to the effect of the nuclear filaments. The kinematic $PA$ is close but offset from the photometric $PA$ of the disk, which was estimated to be 155\arcdeg. Since the kinematic $PA$ and the photometric $PA$ of the disk are different, it indicates that the distribution/kinematics are not axisymmetric.  The inclination
varies from 58\arcdeg\ to 46\arcdeg , assuming circular orbits. On average the inclination within the CND is 50 -- 60\arcdeg\ and agrees well with the inclination obtained for the CND and nuclear ring.
$k1$ is very high at small radii because of the two nuclear filaments, but it is $\sim$ 100\,km\,s$^{-1}$ within most of the disk. Above 8\arcsec\ it decreases considerably and it cannot be considered to be representative of the rotation speed. This is because of the large velocity dispersion component at the edge of the CND, and probably also because of the extended molecular gas emission located at large radii ($r$ $>$ 1\,kpc) and seen in projection.
$k5/k1$ is very large ($\sim$ 0.4) above 9\arcsec\ , partly due to multiple velocity components and the low values of $k1$ as a result of some contamination with extended emission. The average from 2\arcsec\ to 9\arcsec\ is very large in comparison with other studies on nearby galaxies \citep[e.g.][]{2016ApJ...823...51B}, where $k5/k1$ is typically $\lesssim$ 0.05. This indicates that there are large deviations from circular motion due to complex kinematics of the molecular gas and/or multiple components along the line of sight.  

\begin{figure*}
\begin{center}
\includegraphics[width=10cm]{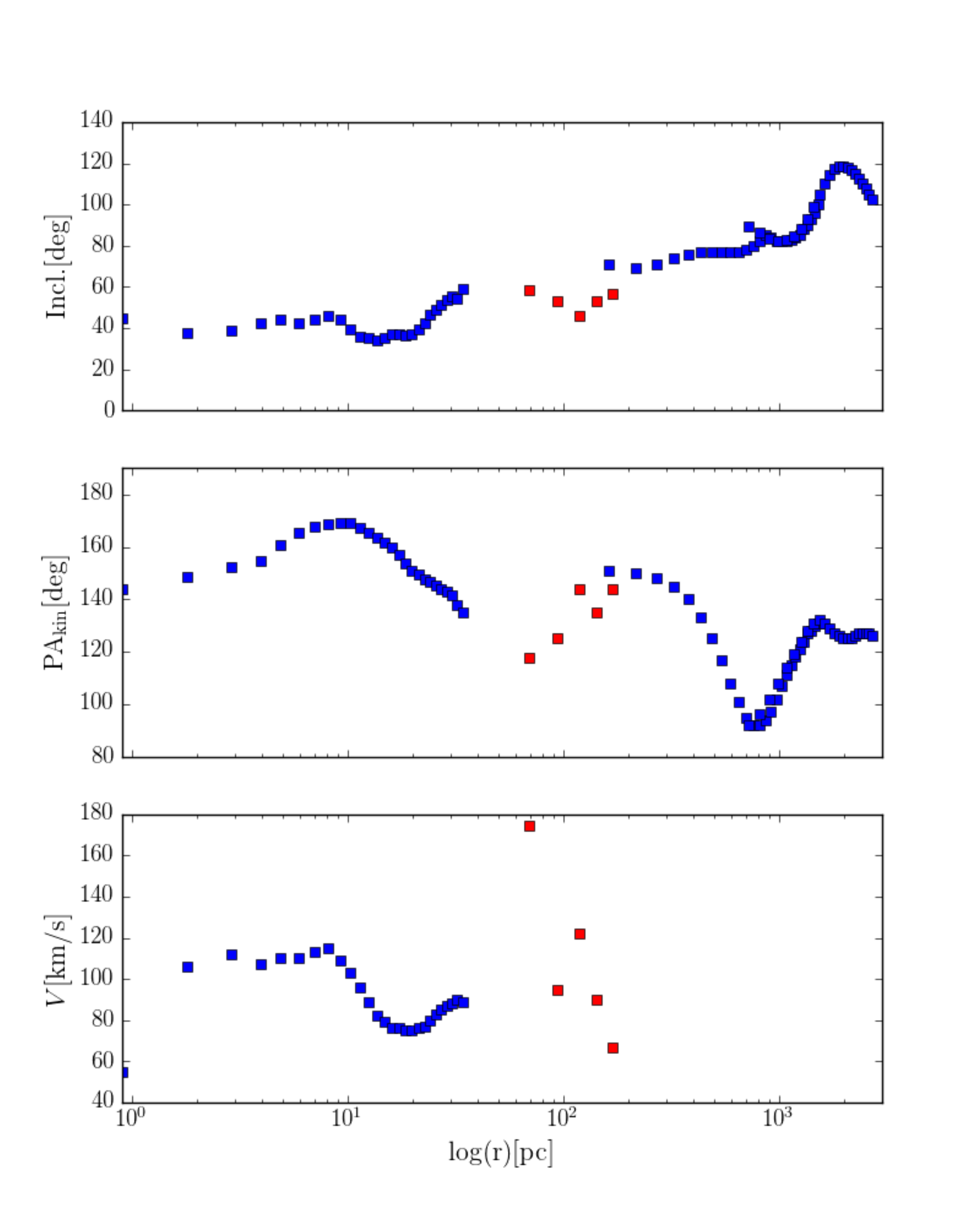}
\caption{Inclination, PA, and molecular gas bulk velocities (not corrected by inclination, projected rotational velocity if circular motions) as a function of radius from the CO(3--2) data presented in this Paper (red square symbols), to be compared with the compilation carried out by \citet[][and references therein]{2010PASA...27..396Q} (blue square symbols). Note that rotational velocities are from \citet{2007ApJ...671.1329N} and this Paper. The compilation by \citet[][]{2010PASA...27..396Q} does not include velocities.
\label{fig16}}
\end{center}
\end{figure*}

\begin{figure*}
\begin{center}
\includegraphics[width=8cm]{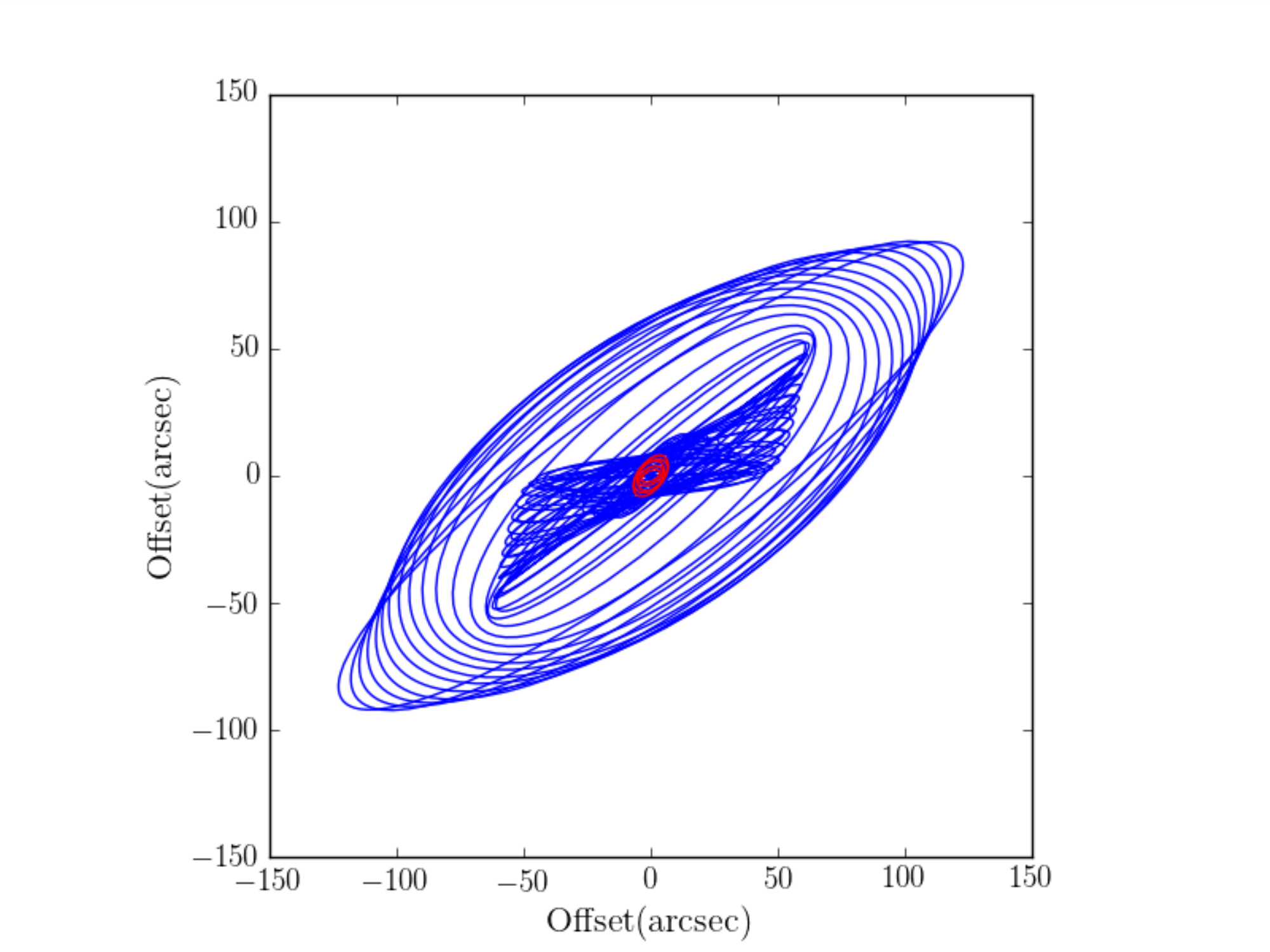}
\includegraphics[width=8cm]{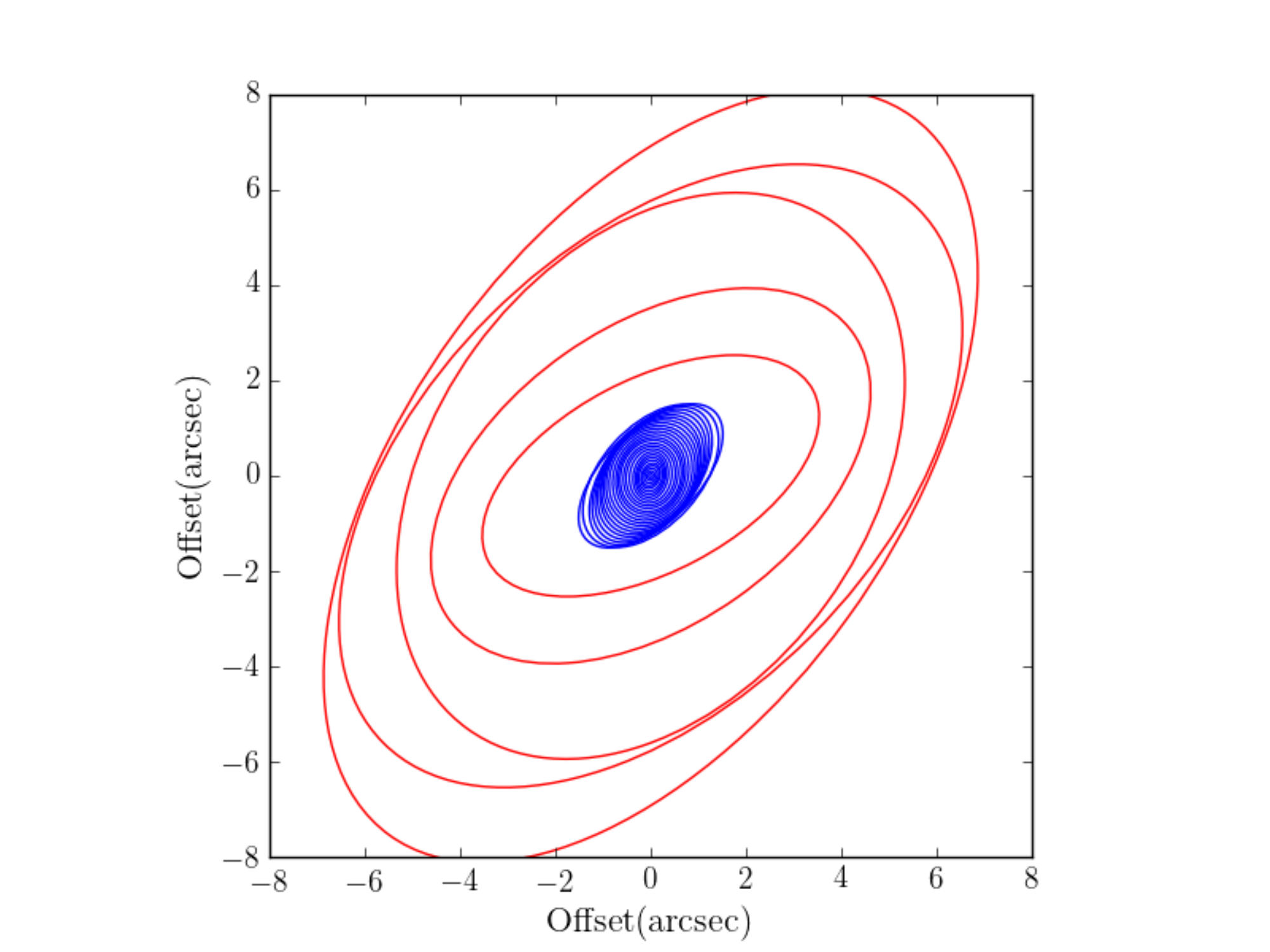}
\caption{Ellipses (red) using the position and inclination angles as a function of radius fitted by Kinemetry using the CO(3--2) data, as presented in this Paper (see Fig.~\ref{fig15}), and compared with other published fit results (blue ellipses): \citet{2006ApJ...645.1092Q} in an inner region of 300\arcsec\ (left panel) and \citet{2007ApJ...671.1329N} in an inner region of 16\arcsec\ (right panel). 
\label{fig17}}
\end{center}
\end{figure*}

Fig.~\ref{fig16} shows the inclinations and position angles as a function of radius compiled by \citet{2010PASA...27..396Q}, using data from  \citet{2006ApJ...645.1092Q}, \citet{2007ApJ...671.1329N} and \citet{2009ApJ...695..116E}
as well as those presented in this Paper for comparison. The data presented here fills for the first time the gap between the several hundred to tens of parsec scale and the obtained position angles and inclinations seem to naturally follow previously published results.
Finally, in Fig.~\ref{fig17} we plotted all ellipses fitted by Kinemetry from large to small scales. The panel on the left shows the fits for large scales, i.e. the inner 300\arcsec, while the one on the right shows the small scales, i.e. the inner 16\arcsec. The red ellipses in both panels correspond to the fits performed in this Paper using the CO(3--2) data, and in blue the fits obtained by \citet{2009ApJ...695..116E} and \citet{2007ApJ...671.1329N}.

In summary, although we tried to fit the best warped disk model possible from kinematic information, deviations from simple rotation and complex kinematic components are apparent. Using this simple warped disk model it would not be possible to reproduce the distribution.
Therefore we invoke non-circular motions as one of the main ingredients to explain the observed distribution and kinematics.

\subsection{Non-Circular Motions}
In \S~\ref{subsect:components} we described the main molecular gas components that we could discern within the CND of Cen\,A. There are large deviations from axi-symmetry in the distribution and kinematics and non-circular motions are needed to explain the observations.  
We favor the scenario where non-circular motions play a major role due to the existence of well aligned nuclear filaments and a nuclear ring. 

Non-circular motions in the gas disk of Cen\,A had been invoked before, although at a larger scale.
\citet{2009ApJ...695..116E} provided evidence that non-circular motions in the molecular gas may be present at least at kiloparsec scales. It was argued that the contribution of a weak non-axisymmetric potential (together with a warp as assumed in previous work) is able to reproduce the CO line distribution and kinematics. In particular it could well reproduce the curved emission resembling spiral arms at kiloparsec scales \citep{2012ApJ...756L..10E}, the formation of the CND, the connection of this CND to gas at larger radii and the lack of emission along the E-W direction in projection within the parallelogram filaments imaged in the mid-infrared \citep{2006ApJ...645.1092Q}. The latter is interpreted as a gap of emission at 200\,pc~$<$~$r$~$<$~800\,pc \citep{2006ApJ...645.1092Q}.  
A possible gap of CO emission in the inner $r$ $<$ 80\,pc of the CND was suggested by \citet{2009ApJ...695..116E} and \citet{2013ASPC..476...69E}, which we now identify as the area inside the nuclear ring. The lack of high angular resolution and high sensitivity observations had prevented  the identification of the nuclear filaments until now.

Since the gas is dissipative (i.e. it will shock at orbit crossings), the kinematic effects on the molecular gas is important when a non-axisymmetric potential is introduced and the resulting large non-radial hydrodynamical (pressure) forces can exert torques on the gas which alter its orbital motion. The gas flows down the shocks and "sprays" back out to large radii, then encountering another shock at the opposite region and repeating the pattern, which makes bars important agents for the gas in spiral galaxies to lose part of its angular momentum and form substantial gas concentrations in their central regions \citep[e.g.][]{1992MNRAS.259..345A,1999ApJS..124..403S}.

Additional mechanisms are invoked for driving gas past the inner Lindblad resonance to feed supermassive black holes that power AGN and nuclear SBs. \citet{1989Natur.338...45S} and \citet{1990Natur.345..679S} proposed the \emph{bars within bars} model, in which a primary (mostly stellar) bar would lead gas to the center where it there would become unstable and form a gaseous bar, further depositing gas into a nuclear disk of tens of parsec scale. This inspired many theoretical and observational works.
\citet{1993A&A...277...27F} showed that bars within bars can form in 3D self-consistent simulations with stars and gas.
The gaseous component was shown to be essential for the decoupling of the nuclear bar. The large scale stellar bar can then be rapidly destroyed by the central mass concentration.
\citet{2004ApJ...617L.115E} investigated the mechanism of formation and dynamical decoupling of bars within bars with gaseous nuclear bars encompassing the full size of the galactic disk and hosting a double inner Lindblad resonance. As these structures become massive and self-gravitating, the  nuclear bars lose internal angular momentum to the primary bars, increase their strength, diminish the nuclear bar size and increase the nuclear bar pattern speed. The viscosity of the gas is an important parameter for the decoupling of nested bars. 

Observationally, approximately one-third of barred galaxies host a secondary nuclear bar in addition to the primary large-scale stellar bar with pattern speeds exceeding those of the large-scale (primary) stellar bars \citep[e.g.][and references therein]{1990ApJ...363..391P,1993A&A...277...27F,1995A&AS..111..115W,1997A&AS..125..479J,2004A&A...415..941E}. 
An increasing number of publications show streaming gas flows down to about tens of pc from the nucleus of galaxies as a result of (stellar) nested bars (e.g., \citealt{2006ApJ...641L..25F}, \citealt{2013ApJ...770L..27F} for NGC\,1097, \citealt{2006ApJ...649..181S,2007A&A...462L..27S} for NGC\,6946, \citealt{2008ApJ...675..281M} and \citealt{2012ApJ...755..104M} for Maffei\,2).
A secondary bar in our Galaxy was reported by \citet{2001A&A...379L..44A},
and simulations to illustrate the putative nested bars were presented by \citet{2009ApJ...691.1525N}. 
In the small inner bar models with sizes of 200\,pc, straight shocks are formed within the inner bar, leading to a nuclear gas disk  formed at the center with size of 15\,pc and mass of $\sim$ 10$^7$ M$_\odot$. All of these cases are suggested to be caused by a nuclear stellar bar. 
Moreover, all nested bar observations and simulations so far have focused on spiral galaxies, but not on elliptical galaxies whose ISM is replenished by external gas.

 The inner ring and nuclear filaments seen in Cen~A with the help of the ALMA data and VLT/SINFONI data are likely due to non-circular motions. In this scenario these nuclear filaments are the loci of shocks. This is reasonable because of the peculiar observed distribution of the CO and H$_2$ nuclear filaments, i.e., they are nearly straight and are characterized by a 180$\arcdeg$ rotational symmetry. This is further reinforced by the distinct kinematics of the nuclear filaments as revealed from the P-V diagrams. Fig.\,\ref{fig-pv}$c$) shows a clearly steeper slope in the P-V diagram in both nuclear filaments, and Fig.\,\ref{fig-pv}$a$) (also Fig.\,\ref{fig-pv2} and Fig.\,\ref{fig-pv3} for CO(6--5)) that the P--V curves of the nuclear filaments systematically avoid the location of the AGN and $V_{sys}$, therefore these components are likely at forbidden velocities \citep{1991ARA&A..29..195C}.  Overall the distribution and kinematics are in agreement with existing simulations of gas under barred potentials causing non-circular motions \citep[e.g.][]{1994ApJ...437L.123W,2009ApJ...691.1525N}.
As for the nuclear disk, there is further evidence that the H$_2$ and ionized line emission of
the Cen~A center is likely dominated by shocks \citep[][and references therein]{2017A&A...599A..53I}. Based on H$_2$ line ratios, the warm molecular gas is likely excited by modest shock velocities 5-20\,\kms\ \citep{2010ApJ...724.1193O,2017A&A...599A..53I}, which might be partly explained by non-circular motions.

Within the CND significant self-gravitation ($M_{\rm gas}$ $>$ 0.3 $M_{\rm dyn}$) would be required to trigger a gaseous bar instability, but under these conditions, due to star forming gas, self gravity would decrease quickly and a very short life time of the gaseous bar would result \citep{1998IAUS..184..269F}. In \S\,\ref{sub:moleculargasmass} we estimated the molecular gas mass enclosed within the CND as $M_{\rm gas}$ $\simeq$ 9 $\times$ 10$^7$ M$_{\odot}$.
The velocity dispersion can be assumed to be small with respect to the orbital velocity within the extent of the CND, and then the dynamical mass at a radius $r$ is $M_{\rm dyn}$ = V$^2$ $r$ / $G$, where $V$ is the rotation velocity and G is the gravitational constant. At $r$ = 200\,pc, and assuming an average velocity of 100\,\kms\ and an inclination of the CND of 60\arcdeg, M$_{dyn}$  = 6 $\times$ 10$^8$ M$_\odot$. 
Therefore the {  $M_{\rm gas}$/$M_{\rm dyn}$ $\simeq$ 0.14}, which indicates that the CND is currently not an entity dominated by self-gravitation.
Although this structure may have been generated as a result of a spontaneous gaseous bar, it is currently not self-gravitating.

In the case of Cen~A neither large scale nor nuclear bars have been reported. The task of identifying such structures is complicated due to the large extinction towards the dust lane along the minor axis of this elliptical galaxy.
The stellar component of Cen A appears very round, but it is likely that it is intrinsically flattened. \citet{1986MNRAS.218..297W} and \citet{1995ApJ...449..592H} support this scenario and find axis ratios of 1: 0.98 : 0.55 and 1: 0.92 : 0.79, respectively.
The stellar component shows little rotation within one effective radius, with 
 a maximum rotation of around 40\,\kms\ roughly along the direction of the major axis at $PA_{\rm kin}$ $\simeq$ 35$\arcdeg$  (almost perpendicular to the dust lane) out to 100\arcsec\ from the nucleus (the N side is approaching and the S receding) \citep{1986MNRAS.218..297W}. \citet{2005AJ....130..406S} found that the stellar rotation is slower, 20\,\kms , at smaller radii of about 40\arcsec .  
 The velocity dispersion is $\sim$135\,\kms\ on both axes for most of the radius range from 2\arcsec\ to 4\arcsec\ \citep{2005AJ....130..406S}.

There is also an S-shape twist in the (stellar and planetary nebulae) kinematics \citep{1995ApJ...449..592H,2004ApJ...602..685P}. Rotation at smaller scales of tens of parsec is  actually $PA_{\rm kin}$ $\simeq$ 165\arcdeg , similar to that of the CND (i.e. $\sim$ 155\arcdeg), 
but the nuclear stellar rotation is counter-rotating relative to the gas \citep{2009MNRAS.394..660C}.

Counter-rotation between stars and gas might be an important aspect to explain the distribution and kinematics. The gas that was likely accreted from an \ion{H}{1}-rich dwarf companion galaxy was probably not rotationally supported and then it went into orbits at a smaller distance from the center in a few orbital time scales \citep{2010PASA...27..463M}.
 From the obtained circular velocities within the CND, $V_c$ $\sim$ 200\,\kms. Therefore at the nuclear ring ($r$ = 4\farcs5, or 80\,pc), the characteristic orbital period is $T$ $\sim$ 1\,Myr. The very short orbital period in the central regions suggests that although the gas should have already reached an equilibrium configuration, there are apparent asymmetries probably as a result of gas having recently reached the central region or being misaligned with respect to the gravitational potential. 
In the external regions further out the characteristic orbital period is larger, being $\sim$ 2 $\times$ 10$^7$\,yr at a radius of 1\,kpc. 
Numerical simulations of the evolution of counterrotating gaseous disks in disk galaxies is such that there is a transitory ($\sim$ 500\,Myr) main $m$ = 1 dynamical instability mode consisting of one-arm spirals which would be on the leading side with respect to the most massive disk, then transform into a more stationary phase with a composite of $m$ = 1 and $m$ = 2 features trailing and forming a ring \citep[e.g.][]{2000A&A...363..869G}.

Another piece of information is that the nuclear filaments (i.e. loci of shocks) are straight, and parallel to each other, although curved at large radii. From the inclination of the CND, the loci of shocks would be on the leading side of the putative nuclear bar if any.
From the morphology of the loci of shocks one can infer the properties of the non-axisymmetric potential  \citep[e.g.][]{1992MNRAS.259..345A}.
Strong bars tend to have straight loci of shocks, while ovals or weak bars should have more curved shapes. Therefore we can infer that somehow the gravitational potential would likely resemble that of a strong stellar bar.

Detailed hydrodynamical simulations would be needed to try to reproduce the molecular gas properties under such a complex stellar configuration and are beyond the scope of this Paper. No matter what the origin is,  the existence of the nuclear ring and shock loci leads to m = 2 instabilities, which may play a major role in feeding AGNs of powerful radio sources.

\section{Geometry of the CND and Orientation of the Jet}
\label{secjet}

In this section we revisit the question of whether the CND is perpendicular to the jet, which is expected 
if there is a physical connection between the two.
 The relativistic jet is at a $PA$ $\simeq 51\arcdeg$ and pointing towards us \citep[e.g.][]{1998AJ....115..960T}, which is nearly perpendicular, at least in projection, to the CND, $PA$ $\simeq 155\arcdeg$. 

Here, we derived an inclination for the CND and nuclear ring of $i$ $\simeq 60\arcdeg$  (near side is to the S) and $i$ $\simeq 50\arcdeg$, respectively (see \S~\ref{subsect:components}). These measurements are improvements over the inclination of $\simeq$ 70\arcdeg\ provided by \citet{2009ApJ...695..116E}.
The CND extent is 20\arcsec , while that of the nuclear ring is half of that, so within 80 $<$ $r$ $<$ 200\,pc it seems likely that the gas is nearly coplanar.

The inclinations are compatible with smaller scale ($<$ 54\,pc) estimates by \citet{2007ApJ...671.1329N} using VLT/SINFONI molecular hydrogen line data. The inclinations of the tilted rings in their warped disk model range from $i$ = 38$\arcdeg$  at 2\,pc to $i$ = 59$\arcdeg$ at $r$ $>$ 34 pc.
However, given that non-circular motions might be playing a major role as in the field of view probed by our ALMA observations (see \S\,\ref{secfeeding}), it is not clear how justified it is to model the molecular kinematics just using a pure warped disk model.
In that case, maybe the average inclination of all the concentric rings, $i$ = 45$\arcdeg$, would be a better estimate for the inclination of the nuclear disk.
Finally, at an even smaller scale of 7 -- 15\,mas (0.13 -- 0.25\,pc) and at 12.5$\mu$m, \citet{2010PASA...27..490B}, using the MID-infrared Interferometric instrument (MIDI) at the Very Large Telescope Interferometer (VLTI), provided an orientation of dust emission with a position angle (PA) of $\sim$10\arcdeg , and an inclination of $\sim$63\arcdeg, which overall does not seem compatible with the other measurements and may reflect more chaotic distribution at these small scales.

As for the jet, various estimates of its orientation have been published so far. \citet{1998AJ....115..960T}, using VLBI observations, found a range for the axis of the nuclear jet with respect to the line of sight of $\theta$ $\simeq$ 50\arcdeg - 80\arcdeg . Other estimates in the literature typically range from $\theta$ = 60\arcdeg\ to 70\arcdeg\ \citep{1979AJ.....84..284D,1979ApJ...232...60G,1994ApJ...426L..23S,1996ApJ...466L..63J}. These are in agreement with the inclinations that we find for the CND. VLBI observations by \citet{2014A&A...569A.115M} recently constrained the angle to the line of sight using the jet-to-counterjet brightness ratio, in combination with proper motions, to $\theta$ $\sim$ 12\arcdeg\ -- 45\arcdeg . A value close to $\sim$50\arcdeg\ would satisfy most constraints for the nuclear jet, and in that case the jet would be perpendicular to the circumnuclear and nuclear disks. The larger scale (hundred parsec) radio jet using VLA seems to be compatible with smaller angles ($\sim$15\arcdeg), contrasting with the pc-scale jet \citep{2003ApJ...593..169H}.

Nuclear radio jets are seen to align well with the rotation axis of megamaser disks, even though the subparsec-scale accretion disks do not align with the kiloparsec-scale galactic disks \citep[e.g.][]{2013ApJ...771..121G}.
Possible causes for these misalignments are, among others, radiation pressure \citep[e.g.][]{1997MNRAS.292..136P}, and accretion events \citep[e.g.][]{2010MNRAS.407.1529H}. 
Therefore a similar misalignment could also be present in Cen~A, i.e. a warp of few tens of degrees.
In such a case \citet{2012MNRAS.422.2547N} find that disk misalignments boost accretion rate and angular momentum is dissipated in the interface where the two disks meet.

\section{Comparison with Numerical Simulations of the multi-phase ISM around SMBHs}
 \label{secnumerical}

In this section, we compare the properties of the molecular gas in Cen~A derived from our observations with recent numerical simulations studying feedback and the onset of thick tori in the center of disk galaxies. {
 Numerical simulations have shown that the (energy-driven) jet feedback may create a cavity mostly devoid of dense gas in the nuclear region which quenches star formation, although the final distribution of dense gas depends on the initial configuration  \citep[e.g.][]{2013ApJ...763L..18W}. While negative feedback is most effective if clouds are tenuous and distributed spherically, positive feedback is most effective if clouds are in a disk-like configuration.  }

The jet of Cen~A is very well collimated, with an
opening angle of less than 12\arcdeg\ on scales of 0.3\,pc \citep{2011A&A...530L..11M}.
There is no clear evidence in our maps of positive feedback such as gas compressed by the jet or negative feedback, where outflows may take gas away from the nuclear region \citep[e.g.][]{2012ARA&A..50..455F,2015ARA&A..53..115K,2007A&A...468L..49M,2015ApJ...799...26M}.
In the CO maps we could not discern any large velocity component that would resemble an outflowing component. However, the outflow might be more extended and recovering the whole flux in the maps could be important. Recently, \citet{2017A&A...599A..53I} reported the signature of an outflow in this target as seen in [\ion{C}{1}] and CO(4--3) using the single dish telescope APEX, as well as [\ion{C}{2}], [\ion{N}{2}] and [\ion{N}{3}] using Herschel/PACS, although angular resolution is limited. Probably some of the most direct evidence of interaction between jet and surrounding gas is given by the fact that other ionized tracers are distributed almost along the direction of the jet (Br$\gamma$, [\ion{Fe}{2}], [\ion{Si}{4}] , [\ion{O}{4}] (25.9\,$\mu$m), [\ion{Ne}{5}] (24.3\,$\mu$m))  \citep[e.g.][]{2007ApJ...671.1329N,2008MNRAS.384.1469Q}.  However, from arguments related to the geometry and kinematics, \citet{2007ApJ...671.1329N} noted that ionized emission such as [\ion{Si}{6}] must indicate an inflow rather than outflow, and it is interpreted as backflow of gas along the side of the jet's cocoon.
Also, although farther apart, \citet{2016A&A...595A..65S}  and \citet{2016A&A...592L...9M} reported the detection of molecular gas 15\,kpc NE
from Cen~A's centre (i.e. in the region of the outer filament), where there is an ongoing interaction between the radio jet and gas clouds.

Three-dimensional radiative hydrodynamical models of the circumnuclear gas in galaxies predict that a clumpy torus-like structure is naturally produced on scales of tens of parsec around an SMBH, due to radiative pressure from the AGN resulting in vertical circulation of gas (or fountains) \citep{2012ApJ...758...66W}, supernova-driven turbulence \citep{2002ApJ...577..197W,2009ApJ...702...63W,2009MNRAS.393..759S}, as well as due to dynamical arguments related to galactic inflows \citep{2012MNRAS.420..320H}.
As a result of the energy feedback from SNe and AGN the circumnuclear disks possess a: i) highly inhomogeneous internal density and temperature structure, ii) filamentary structures such as nuclear spiral arms, and iii) a large scale height in the outer regions at galactocentric radii of a few tens of parsec, resembling the shape of a torus and being characterized by large internal turbulent motion.
 \citet{2016ApJ...828L..19W} modelled the multi-phase properties of the ISM around a SMBH of $\sim$ 10$^6$ M$_\odot$ including the effect of AGN, SNe, and XDR conditions. While the molecular hydrogen co-exists at high (T\,$>$\,1000\,K) and cold (T\,$<$\,100\,K) temperatures, the dense and cold molecular ISM is expected to be distributed preferentially close to the equatorial plane.

Our observations agree in general with these numerical simulations. Our maps cover the central 400\,pc down to 10\,pc scale.
We observe a cold disk with filamentary and clumpy structure at scales of tens to hundreds of parsec.
Several long filamentary structures are seen, and the general distribution is quite irregular and inhomogeneous. 
The velocity dispersion is larger at the edges of the CND, where multiple regions have velocity widths of 20\,\kms\ up to 40\,\kms .

A measure of the molecular gas disk scale height can be estimated from the rotational velocity at a given radius, ranging roughly from 170\,\kms\ in the center of the disk to $\sim$100\,\kms\ at its edges { (see Fig.~\ref{fig15})}, and the velocity dispersion, which ranges from 10 to 40\,\kms\  also from the center to its edges.
We assume hydrostatic equilibrium of the molecular gas disk, an axisymmetric potential, and follow $h~=~R~q~\sigma~/~V$ (Eq. 3.1 of \citealt{1992ApJ...391..121Q}), where $V$ is velocity at radius $R$, $h$ is the scale height and $q$ is the axis ratio of the potential which is assumed to be $q$ $\sim$ 1. If $R$ = 100 - 200\,pc, $\sigma$ = 10 - 40\,\kms\ (although the larger velocity dispersion limit could be due to non-circular motions), and $V~\simeq $ 200\,\kms, then $h$ = 5 -- 40\,pc from the inner to the outer radius. This increase in molecular gas scale height is not as large as in the scenario proposed by \citet{2009ApJ...702...63W} and \citet{2012ApJ...758...66W}, although it is unclear what the scale height should be for the different phases of the ISM for an object like Cen~A, and in particular for the cold molecular gas.

There are other differences with respect to these numerical simulations. First, although filamentary and clumpy structures are present in the simulations, the loci of shocks and a ring structure are not well reproduced, which may reflect that a non-axisymmetric potential (i.e. m = 2 mode instability) would be needed in those simulations. Second, in the inner $20$\,pc simulations predict a relatively cold and thin molecular disk to be formed, but such molecular gas component is not found as traced by CO(3--2) or CO(6--5) to our sensitivity limit. A warm (T$\sim$1000\,K) molecular disk is instead present in the inner tens of parsec.

\section{Physical and chemical properties of the molecular gas around the Cen\,A nucleus}
\label{secratio}

\subsection{CO Line Ratios}
\label{co}

Analysis of single dish data for many CO transitions has shown that the CND of Cen~A has an emission ladder of CO transitions quite different from those of either starburst galaxies or (Seyfert) AGNs \citep{2014A&A...562A..96I}.

While the average over the 20\arcsec\ CND is $R_{65/32}$ $\simeq$ 0.4 using $T_{mb}$ units (or 1.85 in flux units)  \citep{2014A&A...562A..96I}, 
CO(6--5) to CO(3--2) line ratios in the inner half of the CND, as obtained from our ALMA data (Fig.~\ref{lineratio}), are characterized by a larger average value,  $R_{65/32}$ = 0.9 (or 3.5 in flux units). 
 Nearby star-forming galaxies, mostly classified as LIRGs, exhibit these high R$_{65/32}$ ratios globally, and even higher values are found in powerful radio-galaxies such as 3C\,293 \citep{2010ApJ...715..775P}.

The values that we find in the CND of Cen~A are similar to those in the Seyfert\,2 galaxy NGC\,1068 at similar spatial scales.
\citet{2014A&A...567A.125G} show that the average R$_{65/32}$ ratio in the CND of NGC\,1068 is $\sim$0.8, with values ranging  from $\sim$0.7 to 2.0 (using T$_{mb}$ units) close to the AGN. \citet{2017arXiv170205458A} also finds an average of 1.2 (also using $T_{mb}$ units) in the nuclear regions of the lenticular galaxy NGC\,1377, which is probably hosting a radio-quiet, obscured AGN \citep{2016A&A...594A.114C}.

In \S\,\ref{colineratio}, we showed that there is a gradient of the CO(6--5) to CO(3--2) flux density ratios, $R_{65/32}$, along the nuclear filaments.
There is gradual increase towards the AGN by a factor of 2--3, reaching the highest values closest to the nucleus (i.e. $\sim$ 20\,pc), where maxima of $R_{65/32}$ $\simeq$ 1.3  using $T_{mb}$ units (or 5 if flux units) are found. These high values should imply high density ($>$10$^4$\,cm$^{-3}$) and high kinematic temperatures ($T_{kin}~>$ 100\,K), similar to the centers of other galaxies observed on similar spatial scales \citep[][]{1998ApJ...495..267M,2004ApJ...616L..55M,2008ApJ...683...70H,2013PASJ...65..100I,2014A&A...567A.125G,2017arXiv170205458A}. Mechanical heating from shocks as well as illumination from X-ray and UV photons might be playing a role in the formation of such a gradient.

The higher ratio in the area probed by our ALMA observations with respect to the average in the CND suggests that CO emission will be more luminous with increasing transition preferentially towards the central regions.

\subsection{ A Lack of CO/HCO$^+$/HCN Emission in the Inner 30\,pc?}

 An interesting difference in contrast with other galaxies also observed with matching spatial scales of $\sim$5\,pc is that there is not much molecular emission as traced by CO transitions in the center of Cen~A. For example, NGC\,1068 hosts in its center a nuclear torus 
 \citep{2016ApJ...823L..12G,2016ApJ...829L...7G}, while NGC\,1377 shows a hot compact core dominated by gas along the jet  \citep{2017arXiv170205458A}, which is extremely radio-quiet in contrast to the
one in Cen~A. 
A question that remains open is whether this difference reflects different steps in the evolutionary sequence or just simply different scenarios and morphologies.

A caveat in the Cen~A data is that towards the AGN we may find absorption lines. However, the continuum source is unresolved, so at least we would have expected to find emission from 10 to 30\,pc if the molecular distribution were centrally concentrated.
The lack of detected emission is likely due to a large gradient in temperature and differences in the chemical properties of the ISM caused by the energetics from the AGN, because there exists molecular gas traced by H$_2$ \citep[][see also \S\,\ref{introduction} and \ref{subsect:components}]{2007ApJ...671.1329N}. In this very nuclear region the molecules probed in this Paper may have been mostly dissociated and ionized.

A consequence of this is that the molecular transitions such as those probed in this Paper may not be good tracers of the molecular gas that obscures the AGN under the special conditions in the nuclear regions of objects like Cen~A, and that using H$_2$ lines probing warmer gas might be more appropriate. This interpretation is in agreement with \citet{2009ApJ...696..448H}, who argued that the molecular gas as traced by the H$_2$ line provides sufficient obscuring column densities toward a sample of local AGNs.

\subsection{HCN/HCO$^{+}$ Line Ratios}

 The dense gas tracers HCN and HCO$^+$ are detected in emission in the J = 4--3 transitions towards several positions in the imaged region as well as in absorption towards the nucleus.
Here we discuss the low HCN/HCO$^+$ line ratios that are systematically found in the inner circumnuclear regions of Cen~A, as mentioned in \S\,\ref{hcohcn}.

For years, the HCN/HCO$^+$ ratio, R$_{HCN/HCO^+}$, was found observationally to be $>1$ in the presence of an AGN and was claimed to be enhanced due to X-ray irradiation \citep[e.g.][]{2001ASPC..249..672K,2008ApJ...677..262K}.  Theoretically, however, HCN and HCO$^+$ are generally of the same order in PDRs, but in XDRs HCN is not as bright as HCO$^+$ lines under the majority of conditions \citep{2005A&A...436..397M}. 
 Three-dimensional radiative transfer simulations of HCN to HCO$^{+}$ line diagnostics in AGN molecular tori by \citet{2007ApJ...671...73Y} also show that under a wide range of molecular abundances R$_{HCN/HCO^+}$ $\lesssim$ 1. 

Recently the favored scenario for the enhancement of HCN around AGNs in observational data is high temperature chemistry likely due to mechanical heating \citep{2010ApJ...721.1570H,2013PASJ...65..100I,2015ApJ...799...26M}. The increasing number of sources detected in these species seem to show in average an enhancement of the HCN/HCO$^+$ ratio in the presence of an AGN \citep{2015ApJ...814...39P,2016ApJ...818...42I,2016AJ....152..218I}. However, the scatter is large among the ratios observed in both galaxies hosting AGNs and starbursts \citep{2015ApJ...814...39P}, which makes it necessary to study individual cases at high resolution to understand the origin of such dispersion. In addition, recent high resolution observations with ALMA have shown that this enhancement may occur in the surroundings and not towards the AGN itself \citep{2014A&A...567A.125G,2015A&A...573A.116M,2016ApJ...818...42I}. 

In the regions where these dense tracers are detected in Cen~A, all R$_{HCN/HCO^+}$ ratios lie well below those of the AGN dominated galaxies \citep{2016AJ....152..218I} and within the range found for starburst galaxies \citep{2016ApJ...818...42I}. Given our low R$_{HCN/HCO^+}$ ratios we can discard that conditions as in other AGN dominated galaxies apply in the circumnuclear environments of Cen~A. 
Given the shocked regions identified in this Paper, possible interaction with the jet, and the elliptical nature of Cen~A, one may expect that in these regions mechanical heating is also important, and thus that high temperatures chemistry also apply producing high HCN to HCO$^+$ ratios.


 A possible explanation is that in Cen~A XDR conditions dominate over mechanical heating. 
The bolometric luminosity of the nuclear source of Cen\,A is estimated to be 2$\times$10$^{43}$ erg s$^{-1}$, half of it at high energies \citep{1998A&ARv...8..237I}.
 Detailed high angular resolution observations were obtained with Chandra, revealing nuclear emission, a jet, and more extended emission \citep{2002ApJ...569...54K}.
The 2 -- 10\,keV unabsorbed luminosity of the nucleus of Cen\,A is estimated to be as high as $\sim$ 5 $\times$ 10$^{41}$\,ergs\,s$^{-1}$ from Chandra and XMM-Newton, after correcting for the pile-up problem and adding extended emission \citep{2004ApJ...612..786E}. 
\citet{2016ApJ...819..150F} use NuSTAR and XMM-Newton data and calculated an unabsorbed 3 -- 50 keV luminosity of 3.4 $\times$ 10$^{42}$ erg\,s$^{-1}$.
The source is not resolved (the point-spread function of NuSTAR has a half-power diameter of 60\arcsec), but it can be argued that the jet is mainly visible in the soft X-rays and most of the high energy radiation is coming from the nucleus itself. At even higher energies, in the 3 -- 1000\,keV range, the luminosity is L$_X$ = 2.0 $\times$ 10$^{43}$ erg s$^{-1}$ as seen by INTEGRAL \citep{2011A&A...531A..70B}.
However, the X-ray luminosity of Cen~A is in between other low luminosity AGNs such as NGC\,1097 \citep[e.g.][]{2006ApJ...643..652N, 2013PASJ...65..100I}, and NGC\,7469, and in these two cases the obtained $R_{HCN/HCO^+}$ ratios towards their nuclei are $>1$ \citep{2015ApJ...811...39I}. Cosmic-ray ionization rates might be higher in Cen\,A producing even further an over-density of HCO$^+$ over HCN. Examples of low ratios can be found in M\,82 and NGC\,1614, where
enhanced cosmic-ray ionization rate due to frequent SNe would increase  the abundance of HCO$^+$ even in a dense molecular cloud \citep[e.g.][]{2011MNRAS.414.1583B,2011A&A...525A.119M,2013A&A...549A..39A}. However, it is likely that the number of SNe in Cen~A is not that high. In the case of Cen\,A, the AGN itself may contribute \citep{2009ApJ...695L..40A}.
Another possible contribution to the low ratios is that HCN(4--3) is more sub-thermally excited than HCO$^+$(4--3) because of the difference in critical densities (\S\,\ref{introduction}). Also, as Cen~A may have acquired the gas through an HI-rich dwarf galaxy, maybe the gas is less processed (i.e. it shows a lack of products of mostly secondary elements like N) and HCN would be underabundant with respect to HCO$^+$. 

Another variable to take into account is time. The HCN/HCO$^+$ abundance ratio is seen to be highly time-dependent \citep{2013ApJ...765..108H,2014A&A...567A.125G}, in the sense that although HCN might be overabundant in early times after the onset of chemical reactions, of the order of 10$^{4-5}$\,yr ($t_{cross}$, the shock crossing time), in steady state this might not be true.
The age of AGN activity in Cen~A is considerably longer than that time scale, of the order of 10$^{7-8}$\,yr \citep[e.g.][and references therein]{2010PASA...27..463M}, so no overabundance of HCN would be expected then.

\section{Summary and conclusions}
\label{conclusion}

We have observed the molecular gas as traced by the CO(3--2), HCO$^+$(4--3), HCN(4--3), and CO(6--5) lines in the circumnuclear disk (CND) of the nearby elliptical and powerful radio galaxy Cen~A.  The angular resolution achieved using ALMA is $\sim$5\,pc (0.3\arcsec). Our main findings are summarized as follows:

\begin{itemize}

\item The CND's (unprojected) extent is 400\,pc $\times$ 200\,pc along $PA$ = 155\arcdeg\ and is composed of multiple filamentary structures or streamers, which form a nuclear ring-like structure with an unprojected diameter of 9\arcsec\ $\times$ 6\arcsec\ (162\,pc $\times$ 108\,pc), and with a similar $PA$ as that of the CND as a whole. The kinematic $PA$ along the CND is slightly offset from 155\arcdeg\, ranging from 110\arcdeg\ to 150\arcdeg .

\item Inside the nuclear ring-like feature, there are two nearly parallel filamentary structures with lengths of about 2--3\arcsec\ (or 36 -- 54\,pc) at a $PA$ = 120$^{\rm o}$, and with a rotational symmetry of 180\arcdeg\ with respect to each other and centered at the AGN. Although we confirm that CO emission is not bright in the inner $\sim$ 80\,pc  inside the nuclear ring as previously suggested \citep{2009ApJ...695..116E}, we reveal for the first time that inside there are nuclear filamentary structures. The line widths along these filamentary structures are as high as $\sim$ 100\,\kms, and are $\sim$ 50\,\kms\ larger than those within the CND. We interpret that these nuclear filaments are shock regions caused by non-circular motions because of their distribution and distinct kinematics as revealed by the P-V diagrams. 

\item The velocity field within the CND presents an S-shape feature, indicative of non-circular motions and/or a warp in the disk.
The CND, with a gradient of velocity from 315\,\kms\, to 795\,\kms , can be separated using P--V diagrams into three well differentiated parts: 1) At  -8\arcsec\ and 8\arcsec\ there is a large velocity dispersion component of $\sim$100\,\kms , 2) $\Delta v / \Delta r$ = 1.8\,km~s$^{-1}$~pc$^{-1}$ in the inner parts of the CND, and 3) the nuclear filaments have a distinct velocity gradient, even steeper, with $\Delta v / \Delta r$ = 3.4\,km~s$^{-1}$~pc$^{-1}$. 

\item  Large velocity dispersions as high as 40\,\kms\ are found towards the edges of the CND. If circular motions dominate at these radii and for hydrostatic equilibrium conditions, this can be interpreted as larger scale heights (up to $\sim$ 40\,pc) as a function of radius. This increase of scale height is probably not as large as predicted by numerical simulations of the multi-phase ISM of CNDs perturbed by SNe and AGN activities.

\item Low HCN/HCO$^+$(4--3) ratios ($<$ 0.5) are found towards several molecular clumps located at few tens of parsec from the AGN, unlike in the nuclear regions of other low luminosity AGN galaxies, where high temperature chemistry due to mechanical heating is invoked. 

\item  CO emission connects to the even warmer molecular gas probed using the H$_2$(J=1--0) S(1) transition line within the inner 50\,pc \citep{2007ApJ...671.1329N}. The gas winds up in the form of two or three main nuclear spirals as traced by the H$_2$ line. There is no substantial emission in the molecular transitions probed by us (i.e. CO(3--2), CO(6--5), HCN and HCO$^+$(4--3)) within the inner 20\,pc away from the AGN. A large gradient of excitation conditions must be present because molecular hydrogen emission in that region is prominent \citep{2007ApJ...671.1329N}. 
Within $r$ $\simeq$ 10\,pc H$_2$ emission forms an inner ring structure and again another pair of filaments of $\sim$10\,pc length, with a rotational symmetry of 180\arcdeg , along the N--S direction, and with even larger velocity dispersions ($\sim$200\,\kms) than anywhere else within the CND as probed in our CO maps, which may also indicate shocks at even smaller scales.

\item The overall molecular gas distribution and kinematics is interpreted as produced primarily by non-circular motions, similar to those in a \emph{bars within bars} scenario, being the successive shocked regions at different scales where loss of angular momentum occurs most efficiently.  Although the disk might be warped, we argue that it is not the dominating factor. 
This study reveals observationally that non-circular motions and successive shocks are important mechanisms at play in feeding powerful radio galaxies down to parsec scales.

\end{itemize}

\acknowledgements{
This paper makes use of the following ALMA data:
   ADS/JAO.ALMA\#2012.1.00225.S. ALMA is a partnership of ESO (representing its member states),
   NSF (USA) and NINS (Japan), together with NRC (Canada), NSC and ASIAA (Taiwan), and KASI
   (Republic of Korea), in cooperation with the Republic of Chile. The Joint ALMA Observatory is operated by ESO, AUI/NRAO and NAOJ.
   We thank V. Impellizzeri and T. Wiklind for fruitful discussions.
   DE was supported by the ALMA Japan Research Grant of NAOJ Chile Observatory NAOJ-ALMA-0031.
   DE was supported by JSPS KAKENHI Grant Number JP 17K14254.
   SM is supported by the Ministry of Science and Technology (MoST) of Taiwan, MoST 103-2112-M-001-032-MY3.
   Data analysis was in part carried out on the open use data analysis computer system at the Astronomy Data Center, ADC, of the National Astronomical Observatory of Japan. This research made use of Astropy, a community-developed core Python package for Astronomy (Astropy Collaboration, 2013).
}

{\it Facilities:} \facility{ALMA}

\bibliography{cena+.bib}

\clearpage

\begin{table*}
\begin{center}
\caption{Main parameters of the ALMA observations \label{table1}}
\begin{tabular}{lll}
\tableline
\\
 &  Band 7   & Band 9 \\ \tableline
Date           &  2014/07/07 &  2014/04/14 \\
Number of antennas  &  33  & 34   \\
Unprojected baselines   [m] &  20 -- 650 & 20 -- 558 \\
Time on source [min.]       &  15 & 54  \\
FWHM of primary beam [\arcsec]     &  16.9  & 8.4 \\
FWHM of synthesized beam  & $0\farcs36 \times 0\farcs29$ (6.5 $\times$ 5.2\,pc ), $PA$ = $70\arcdeg$ & $0\farcs23 \times 0\farcs16$ (4.1 $\times$ 2.9\,pc ), $PA$ = $48\arcdeg$ \\
Spectral window center freq. [GHz] & 345.302, 355.320, 356.734 & 691.473, 693.300, 705.209, 708.877 \\
Lines/rest freq. [GHz]  &CO(3--2) 345.796, HCN(4--3) 354.505, & CO(6--5) 691.473, HCN(8-7) 708.877 \\
                          &HCO$^+$(4--3) 356.734 &  \\
Bandwidth per baseband&  0.9375\,GHz (790\,\kms)  &1.875\,GHz (790\,\kms) \\ 
Spectral resolution & 244.141\,kHz  (0.4\,kms) & 488.281\,kHz  (0.4\,\kms)  \\
rms continuum& 2.2\,mJy\,beam$^{-1}$ &16.6\,mJy\,beam$^{-1}$ \\
rms in 10~\kms            &  1.3\,mJy\,beam$^{-1}$ (CO(3--2))  & 5.8\,mJy\,beam$^{-1}$ (CO(6--5))\\
            &   1.8\,mJy\,beam$^{-1}$ (HCN(4--3), HCO$^+$(4--3)) & \\
Calibrators (ampl. / bandpass / phase) & Titan / 3C279 / J1427-4206  &Titan / 3C279 / J1427-4206 \\
\\
\tableline
\end{tabular}
\end{center}
\end{table*}

\begin{table*}
\caption{Derived parameters of the different regions detected in CO(3--2) and CO(6--5)  \label{table2}}
\begin{center}
\begin{tabular}{l l l l r r}
\tableline
\small Component   &\small  Peak$^{a)}$  &\small Velocity range &\small ${S_{\rm  ^{12}CO(3-2)}}^{b)}$ &\small      ${S_{\rm  ^{12}CO(6-5)}}^{c)}$ & $M_{\rm gas}$ $^{d)}$ \\
          &\small [Jy~beam$^{-1}$ km s$^{-1}$]  &\small [km s$^{-1}$] &\small  [Jy~km~s$^{-1}$] &\small  [Jy~km~s$^{-1}$] & [10$^6$ M$_\odot$] \\
\tableline
\\
Filament NW   &  3.7 $\pm$ 0.2 / 5.9 $\pm$ 0.8 & 550 - 660 &   45 $\pm$ 1 (46)     & 151 $\pm$ 6  (176) & 1.42\\
Filament SE    &  2.4 $\pm$ 0.2 / 4.4 $\pm$ 0.8 & 410 - 570 &   23 $\pm$ 1 (23)     &  93 $\pm$ 5 (106)  & 0.71\\
Nuclear ring$^{e}$    &  6.8 $\pm$ 0.2 / 8.4 $\pm$ 0.8 & 420 - 700  &   339 $\pm$ 3 (364) & 888 $\pm$ 20 (1259) & 11.25\\
CND &  6.8 $\pm$ 0.2  /  8.4 $\pm$ 0.8 & 290 - 820 & 1102 $\pm$ 6  (1552) & -   & 47.95 \\
\\
\tableline
\end{tabular}
\tablecomments{
a) Peak flux densities are derived from the primary beam corrected maps. Two values are given, the first referring to CO(3--2) and the second to CO(6--5).
b) and c) The total fluxes for CO(3--2) and CO(6--5). Numbers in parenthesis indicate primary beam corrected values.
d) The gas mass $M_{\rm gas}$ is calculated using the CO(3--2) flux. We adopt a galactic CO-to-H$_2$ conversion factor  $X = 4 \times 10^{20}$~cm$^{-2}$~(K~\kms)$^{-1}$ for the circumnuclear gas \citep{2014A&A...562A..96I}. The molecular gas mass, $M_{\rm gas}$, is calculated as $M_{\rm H_{2}[M_\odot]}$ = 1.18 $\times$ 10$^4$ $\times$ D[Mpc]$^2$ S$_{CO(1-0)}$[Jy km s$^{-1}$] $\times$ [$X$ / 3.0 $\times$ 10$^{20}$ cm$^{-2}$ (K km s$^{-1}$)$^{-1}$, a conversion from CO(1--0) to CO(3--2) fluxes of $S_{CO(1-0)}$ / $S_{CO(3-2)}$ $\simeq$ 0.1 following the fluxes normalized to a half power beam size of 22\arcsec\ and decomposed between CND and extended component \citep{2014A&A...562A..96I}, and a factor of 1.36 to account for elements other than hydrogen \citep{2000asqu.book.....C}.
e) Values for the nuclear ring contain CO(3--2) and CO(6--5) emission in the inner 8\farcs5.
}
\end{center}
\end{table*}

\begin{table*}
\caption{ Gaussian fits to HCO$^+$ and HCN(4--3) spectra in brightest detected positions \label{table4}}
\begin{center}
\begin{tabular}{l l l l l l l}
\tableline

Position $^{a}$ & Distance &  Line  & Flux     &      Velocity  &            Width     &        Peak \\
               & [pc] &    &  [Jy\,km\,s$^{-1}$]       &     [km\,s$^{-1}$]  &            [km\,s$^{-1}$]     &        [Jy/beam] \\\tableline
\\
 1          &58   & HCO$^+$(4--3)    &0.62    $\pm$ 0.05  & 461.2  $\pm$ 1.3   & 32 $\pm$ 3 & 0.018 \\
              &      & HCN(4--3)    & 0.23    $\pm$ 0.04   & 461.2   & 32.2  & 0.006     \\
              \\
 2           &20   &HCO$^+$(4--3)   & 0.53  $\pm$ 0.08 &  571 $\pm$ 4   & 57 $\pm$ 10  & 0.009 \\
                    & &HCN(4--3)     & 0.21  $\pm$ 0.04    &571   & 57  &0.004    \\
                    \\
 3           &32  & HCO$^+$(4--3)   & 0.18  $\pm$  0.05  & 485 $\pm$ 3   & 22 $\pm$ 7 & 0.008   \\
                    & &HCN(4--3)     &$<$0.1     \\    
                    \\                                                                 
 Center       &    &  HCO$^+$(4--3)  &  -3.49     $\pm$  0.12 &  539 $\pm$ 1  &  17.8 $\pm$  0.8 & -0.18    \\
                   & &HCN(4--3)    & -0.72   $\pm$ 0.09  & 539 &  17.8 & -0.04 \\
\tableline
\end{tabular}
\tablecomments{a) The coordinates for each position is 1) $RA$\ = 13${\rm ^h}$25${\rm ^m}$27.${\rm^s}$905, $Dec$ = -43${\rm ^o}$01$\arcmin$09\farcs286, 2) $RA$\ = 13${\rm ^h}$25${\rm ^m}$27${\rm^s}$68, $Dec$ = -43${\rm ^o}$01$\arcmin$07\farcs952, 3) $RA$\ = 13${\rm ^h}$25${\rm ^m}$27${\rm^s}$.664, $Dec$ = -43${\rm ^o}$01$\arcmin$10\farcs473, Center: $RA$\ = 13${\rm ^h}$25${\rm ^m}$27.${\rm^s}$615 ; $Dec$ =  --43${\rm ^o}$01$\arcmin$08\farcs805
}
\end{center}
\end{table*}

\begin{table*}
\caption{Kinemetry parameter fits to the CO(3--2) velocity field  \label{table3}}
\begin{center}
\begin{tabular}{l l l l r r}
\tableline
\small r             &\small Inclination  &\small $PA$     &\small     $k_1$  & $k_1/k_5$ \\
         \arcsec   &\small [deg]         &\small [deg]   &\small [\kms ]  &  \\
\tableline
\\

3.85	& 58 &	117 &174 & 0.35 & \\
5.19	& 53 &	125  &94 & 0.25 & \\
6.55	& 46 &	144 &121&  0.17 &\\
7.93	& 52 &	135  &90 & 0.42 &\\
9.32	& 57 &	143 & 66 &0.42 &\\
\\
\tableline
\end{tabular}
\tablecomments{a) the radial range considered, b) the kinematic PA, b) $q$ the axial ratio or flattening of the ellipse, c) $k1$ the amplitude of bulk velocities, and d) $k5/k1$ the ratio between harmonics $k1$ and $k5$, where $k5$ represents deviations from simple rotation.
}
\end{center}
\end{table*}

\end{document}